\documentstyle[12pt,epsf]{article}

\def\hybrid{\topmargin 0pt      \oddsidemargin 0pt
	\headheight 0pt \headsep 0pt
	\textwidth 6.25in       % A4 paper
        \textheight 9.5in       % A4 paper
	\marginparwidth .875in
	\parskip 5pt plus 1pt   \jot = 1.5ex}

\catcode`\@=11
\def\marginnote#1{}

\newcount\hour
\newcount\minute
\newtoks\amorpm
\hour=\time\divide\hour by60
\minute=\time{\multiply\hour by60 \global\advance\minute by-\hour}
\edef\standardtime{{\ifnum\hour<12 \global\amorpm={am}%
	\else\global\amorpm={pm}\advance\hour by-12 \fi
	\ifnum\hour=0 \hour=12 \fi
	\number\hour:\ifnum\minute<10 0\fi\number\minute\the\amorpm}}
\edef\militarytime{\number\hour:\ifnum\minute<10 0\fi\number\minute}

\def\draftlabel#1{{\@bsphack\if@filesw {\let\thepage\relax
   \xdef\@gtempa{\write\@auxout{\string
      \newlabel{#1}{{\@currentlabel}{\thepage}}}}}\@gtempa
   \if@nobreak \ifvmode\nobreak\fi\fi\fi\@esphack}
	\gdef\@eqnlabel{#1}}
\def\@eqnlabel{}
\def\@vacuum{}
\def\draftmarginnote#1{\marginpar{\raggedright\scriptsize\tt#1}}

\def\draft{\oddsidemargin -.5truein
	\def\@oddfoot{\sl preliminary draft \hfil
	\rm\thepage\hfil\sl\today\quad\militarytime}
	\let\@evenfoot\@oddfoot \overfullrule 3pt
	\let\label=\draftlabel
	\let\marginnote=\draftmarginnote
   \def\@eqnnum{(\theequation)\rlap{\kern\marginparsep\tt\@eqnlabel}%
\global\let\@eqnlabel\@vacuum}  }

\def\numberbysection{\@addtoreset{equation}{section}
	\def\theequation{\thesection.\arabic{equation}}}

\catcode`@=12
\relax

\def\ve{\varepsilon}
\def\nn{\nonumber}
\def\beq{\begin{equation}}
\def\eeq{\end{equation}}
\def\bea{\begin{eqnarray}}
\def\eea{\end{eqnarray}}
\def\ep{\epsilon}
\def\si{\sigma}
\relax
\numberbysection
\hybrid
\begin{document}
\begin{titlepage}
\begin{center}
December~1998 \hfill    PAR--LPTHE 98/56 \\[.3in]
{\large\bf Coupled Potts models: Self-duality and fixed point
structure}\\[.2in] 
        {\bf Vladimir Dotsenko (1), Jesper Lykke Jacobsen (2), Marc-Andr{\'e}
             Lewis(1) and Marco Picco (1)} \\ 
	{\bf (1)} {\it LPTHE\/}\footnote{Unit{\'e} Mixte de Recherche CNRS
UMR 7589.},\\ 
        {\it  Universit{\'e} Pierre et Marie Curie, PARIS VI\\
              Universit{\'e} Denis Diderot, PARIS VII\\
	      Bo\^{\i}te 126, Tour 16, 1$^{\it er}$ {\'e}tage \\
	      4 place Jussieu,
	      F-75252 Paris CEDEX 05, FRANCE}\\
%              and\\
	{\bf (2)} {\it Laboratoire de Physique Statistique%
             \footnote{Laboratoire associ{\'e} aux universit{\'e}s
                       Paris 6, Paris 7 et au CNRS.},\\
             Ecole Normale Sup{\'e}rieure,\\
             24 rue Lhomond,
             F-75231 Paris CEDEX 05, FRANCE \\}
\end{center}
\vskip .15in
\centerline{\bf ABSTRACT}
\begin{quotation}
{\small We consider $q$-state Potts models coupled by their energy
 operators. Restricting our study to self-dual couplings, numerical
 simulations demonstrate the existence of non-trivial fixed points for
 $2 \le q \le 4$. These fixed points were first predicted by perturbative
 renormalisation group calculations. Accurate values for the central
 charge and the multiscaling exponents of the spin and energy
 operators are calculated using a series of novel transfer matrix
 algorithms employing clusters and loops. These results compare well
 with those of the perturbative expansion, in the range of parameter
 values where the latter is valid. The criticality of the fixed-point
 models is independently verified by examining higher eigenvalues in
 the even sector, and by demonstrating the existence of scaling laws
 from Monte Carlo simulations. This 
 might be a first step towards the identification of the conformal
 field theories describing the critical behaviour of this class of
 models.}
\end{quotation}
\end{titlepage}

\newpage

\section{Introduction}

It is well-established that many spin models of statistical physics
with random-valued nearest-neighbour couplings possess distinct
critical points, their critical properties being different from those
of the corresponding fixed-coupling models. To ensure the existence of
a ferromagnetic-paramagnetic phase transition one usually restricts
the study, within this class of models, to those where the random
exchange couplings are exclusively ferromagnetic. This restriction
may seem simplistic, but in fact it turns out that many models
exhibiting site or bond dilution (systems where some couplings are
zero), or even asymmetric distributions of positive and negative
couplings, belong to the same critical universality class. Sometimes
these statistical models are called ``weakly disordered'' in order to
distinguish them from the strongly disordered models encountered in
spin-glass theory, but in fact this nomenclature is somehow
inappropriate, since the disorder (whose strength is here given by the
spread of the distribution of random-valued couplings) could either be
weak or strong. 
It is often observed that models with varying disorder strength
display identical critical properties, a situation reminiscent of the
well-known universality of the second-order phase transitions in
non-disordered systems. Besides being of fundamental theoretical
interest this universality is often of great practical importance:
Whilst in calculations it is often simpler to consider weak
randomness, in numerical studies stonger randomness is preferred, in order
to reach more easily the new critical regime induced by the disorder.

An efficient laboratory for the study of disordered systems is
provided by two-dimen\-sional spin models. These models are
non-trivial, critical, universal (unlike their one-dimensional
counterparts) and easier to analyse, both analytically and
numerically, than three-dimensional systems.

In this broad playground many results have been obtained over the
years. Analytic results were mainly derived from perturbative
renormalisation group calculations \cite{dot-dot,ludwig,dpp1} whilst numerical
results were obtained using Monte Carlo simulations and numerical
diagonalisation of transfer matrices \cite{cfl,mp,cj,cb98}. Much less
has been found out concerning the conformal field theories (CFTs)
which should describe exactly the random systems at their critical
points. This is deceiving, considering the success of CFT in
describing the critical behaviour of the corresponding pure models
(without disorder). Our study of coupled Potts models is mainly
motivated by our interest to progress towards the identification of
these conformal theories.

The relation between the two problems, coupled and random, becomes
clear when one considers the replica approach to random systems. For
instance, the partition function of Ising-type models with random
nearest-neighbour spin couplings can be cast in the form
\beq
  Z(\beta) = \sum_{\{\sigma\}}
    \mbox{exp}\left\{\beta\sum_{x,\alpha}J_{x,\alpha}\sigma_{x}\sigma_{x+\alpha}\right\}
  = \sum_{\{\sigma\}}\mbox{exp}\left\{-\beta\sum_{x,\alpha}
    J_{x,\alpha}\ve_{x,\alpha}\right\}.
  \label{fpart}
\eeq
We can consider, for simplicity, the model on a square lattice. Then
$x$ runs over the sites of the lattice and the unit vector $\alpha$
over the neighbours, $\alpha= {\bf e}_1, {\bf e}_2$.
The couplings $J_{x,\alpha}$ assigned to links of the lattice take
random values, independently for each link, with some specified
distribution. The product of neighbouring spins
$-\sigma_{x}\sigma_{x+\alpha}$ in Eq.~(\ref{fpart}) is the energy
operator $\ve_{x,\alpha}$ assigned to a link, which becomes a
local energy operator $\ve(x)$ in the continuum limit near the
critical point. In this limit the partition function (\ref{fpart}) can
symbolically be represented as
\beq
  Z(\beta)\propto \mbox{Tr}\,\,\exp\left\{-{\cal H}_{0}-\int
  d^{2}x\,m(x)\ve(x)\right\}, 
\eeq
${\cal H}_{0}$ representing the Hamiltonian (or Euclidian action, in the
field theoretic terminology) of the corresponding pure CFT if it is known.
\beq
  m(x)\propto \frac{\beta J_{x,\alpha}-
  \beta_{\rm c}J_{0}}{\beta_{\rm c}J_{0}},
  \label{mass}
\eeq
is a mass-type coupling, which replaces $\beta J_{x,\alpha}$ in the
continuum limit: the randomness of $J_{x,\alpha}$, whose average value
is $J_0$, translates into the randomness of $m(x)$. The trace
$\mbox{Tr}$ in Eq.~(\ref{mass}) is assumed to represent, in the
continuum limit theory, the summation over the spin configurations in
the lattice model (\ref{fpart}).

In quenched disordered systems, averaging over
the randomness should be done at the level of the free energy
$F(\beta) \propto \log Z(\beta)$.  To this end one introduces
replicas, that is, $N$ copies of the same system
\beq
  (Z(\beta))^N \propto {\mbox{Tr}}\,\,
    \exp\left\{-\sum^N_{a=1}{\cal H}^{(a)}_{0}-
    \int d^{2}x\,m(x)\sum^N_{a=1} \ve_{a}(x)\right\}
  \label{replica}
\eeq
and averages over the randomness:
\beq
  \overline{(Z(\beta))^{N}}.
  \label{rand-av}
\eeq
By taking the limit $N \to 0$ one recovers the average of the free energy
for a single system:
\beq
  \overline{F(\beta)}\propto\overline{\log Z(\beta)}\propto
  \lim_{N\rightarrow 0}\frac{\overline{(Z(\beta))^{N}}-1}{N}.
\eeq
If the distribution of the couplings $\{J_{x,\alpha}\}$ in
Eq.~(\ref{fpart}) is gaussian, and consequently that of $m(x)$ in
Eq.~(\ref{replica}), the disorder-averaged $(Z(\beta))^{N}$ in
(\ref{rand-av}) is equivalent to a system of $N$ models coupled by
their energy operators:
\beq
  \overline{(Z(\beta))^N}\propto
  \mbox{Tr}\exp\left\{-\sum^{N}_{a=1}{\cal H}^{(a)}_{0}-m_{0}\int d^{2}x
  \sum^{N}_{a=1}\ve_{a}(x) +
  g_{0}\int d^{2}x\sum_{a\neq b}\ve_{a}(x)\ve_{b}(x)\right\}.
  \label{averaged}
\eeq
Here $g_0$ is the width of the distribution of $m(x)$, and $m_0$
is its average value. At the critical point $m_0$ should vanish.

In cases where the distribution of $\{J_{x,\alpha}\}$, or of $m(x)$,
is not gaussian, the resulting theory in Eq.~(\ref{averaged}) contains
higher-order coupling terms, like
\beq
  \sum_{a,b,c}\ve_{a}(x)\ve_{b}(x)\ve_{c}(x) \qquad
  \mbox{and} \qquad
  \sum_{a,b,c,d}\ve_{a}(x)\ve_{b}(x)\ve_{c}(x)
    \ve_{d}(x),
\eeq
but these are either irrelevant (in the renormalisation group
sense) or do not modify the fixed point structure and its stability.
We can thus limit our study to second order energy-energy couplings. 

On the lattice, before taking the continuum limit, a similar procedure
applied to $Z(\beta)$ in Eq.~(\ref{fpart})
produces
\beq
  \overline{(Z(\beta))^{N}}=
   \sum_{\{\sigma_{a}\}}\mbox{exp}\left\{-\beta
   J_{0}\sum_{x,\alpha}\sum^{N}_{a=1} 
   \ve^{(a)}_{x,\alpha}+
   Ag_{0}\sum_{x,\alpha}\sum_{a\neq b}\ve^{(a)}_{x,\alpha}
   \ve^{(b)}_{x,a}\right\},
  \label{averlattice}
\eeq
where $\ve^{(a)}_{x,\alpha}=-\sigma^{(a)}_{x}\sigma^{(a)}_{x+\alpha}$,
and $A$ is some constant coefficient.

Thus, in order to solve the random coupling problem, one first has to
solve the theory of $N$ coupled homogeneous models,
Eqs.~(\ref{averaged}) and (\ref{averlattice}), ultimately taking the
limit $N\rightarrow 0$. A complete solution to the problem would be
the identification of the exact conformal field theory associated with
the critical point.

When looking for possible candidates to this conformal theory, an
important issue arising is the non-unitarity of the $N\rightarrow 0$
limit theory, as can be seen using perturbative renormalisation group (RG).
For instance, in a non-unitary theory the central charge {\em
increases} along the RG flow, in contradistinction to what is the case
for a unitary ({\em i.e.}~with reflection positivity) theory
\cite{zamo}. As a first step, in order to avoid the problems of
non-unitarity and work with a well-defined problem, we suggest to
consider the problem of $N$ (positive, integer) coupled models,
Eqs.~(\ref{averaged}) and (\ref{averlattice}), and to examine the
critical properties of such systems. In the conformal theory language
this amounts to studying $N$ minimal models, coupled two by two by the
operators 
$\Phi_{1,2}$ which are the energy operators in the case of Ising or
Potts models. This problem is unitary. The $\beta$-function of the
renormalisation group, to second order in perturbation, is given by
\beq
  \frac{dg}{d\xi}=\beta(g)\propto(N-2)g^{2}+{\cal O}(g^{3})
  \label{betai}
\eeq
in the case of Ising \cite{dot-dot}, and
\beq
  \frac{dg}{d\xi}=\beta(g)\propto 3\epsilon g+(N-2)g^{2}+
  {\cal O}(\epsilon^2,g^{3})
  \label{betap}
\eeq
in the case of $q$-state Potts models \cite{ludwig,dpp1}, where
$\epsilon\propto q-2$. By the usual analysis, for $N<2$ (and
eventually for $N\rightarrow 0$) the theory (\ref{betai}) runs into
zero coupling along the RG flow, and the theory (\ref{betap}) runs
into a new non-trivial fixed point with
\beq
 g_{\rm FP} \equiv g_{\ast}=\frac{3\epsilon}{2-N}+{\cal O}(\epsilon^{2}).
 \label{fpcoup}
\eeq
Thus, the random Potts model possesses a new non-trivial fixed point,
and it is therefore of interest to look for the associated conformal
theory. 

These considerations hold true when $g$ is initially positive, that is
$g_{0}>0$, which is the case for the random model. If now, in order to
attain unitarity, we replace $N<2$ by $N>2$ ($N$ integer), the
coefficients in Eqs.~(\ref{betai}) and (\ref{betap}) change sign and
the theory runs into a strong coupling regime, which is not controlled
by pertubative RG and believed to be massive. This means a finite
correlation length, and thus a non-critical theory.

To avoid this problem, when passing from $N<2$ to $N>2$ one should
simultaneously change the sign of the initial coupling $g_{0}$. 
We expect some sort of similarity, or duality, of the domains $N<2,
g_{0}>0$ and $N>2, g_{0}<0$. In fact, if multiplied by $(2-N)$,
Eq.~(\ref{betap}) takes the form:
\beq
  \frac{d\lambda}{d\xi}\propto 3\epsilon\lambda - \lambda^{2}+
  {\cal O}(\lambda^{3}),
\eeq
where we have defined a new coupling $\lambda=(2-N)g$. In other words,
the RG flows depend on $g$ in the combination $(2-N)g$, so that the
regions $N<2, g_{0}>0$ and $N>2, g_{0}<0$ are similar.

In this way one can gain unitarity whilst still staying critical. Note
that the critical coupling $g_{\ast}$, Eq.~(\ref{fpcoup}), depends
only on $N$ and $\epsilon(q)$. It is not obvious, but we can hope that
the unitary $N>2$ problem being solved, that is the exact conformal
field theory being found for general $N$, one could analytically
continue into the domain $N<2$ and extract exact information about the
random model, $N\rightarrow 0$.

Following this approach, we first have to find the exact conformal theory
of the non-trivial fixed point $g_{\ast}$ of $N$ coupled Potts models,
$N=3,4,5,...$. The first step is to study three coupled models,
expecting to be able to extend the result to general $N$ later. It
should be observed at this point that the model of $N$ coupled Potts
lattices (minimal conformal theories) is interesting on its own right,
independently of its relation to the random problem. This makes the
project doubly interesting.

The system composed of three coupled $q$-state Potts models can be
studied by the usual methods, in particular numerically, either by Monte
Carlo simulations or by diagonalising the transfer matrix in a strip
geometry. Our first objective is to get some confidence in the
existence of the fixed points predicted by perturbative RG, and
second, to get fairly accurate numerical values for the central charge
and dimensions of operators like spin, energy and their moments
$\sigma_{1}(x)\sigma_{2}(x),\,\sigma_{1}(x)\sigma_{2}(x)\sigma_{3}(x)\,$,
etc. This should be useful when searching for the corresponding
conformal theory.

To put the numerical study on a firm basis we need a properly defined
model on the lattice, or rather on three lattices which are coupled,
link to link, by their energy operators; see Eq.~(\ref{averlattice}).
For the Potts model the energy operator is of the form
\beq
  \ve_{x,\alpha}=-\delta_{\sigma_{x},\sigma_{x+\alpha}},
\eeq
where $\delta_{\sigma,\sigma'}=1$ for $\sigma=\sigma'$ and $0$ otherwise,
and the spin operator $\sigma_{x}$ at the lattice site $x$ takes $q$
different values. The partition function of three coupled models is of
the form
\beq
  \sum_{\{\sigma,\tau,\eta\}}\exp\left(\tilde{\beta}\sum_{x,\alpha}
    (\delta_{\sigma,\sigma'}+\delta_{\tau,\tau'}+\delta_{\eta,\eta'})
  +\tilde{g}_{0}\sum_{x,\alpha}(\delta_{\sigma,\sigma'}\delta_{\tau,\tau'}
  +\delta_{\tau,\tau'}\delta_{\eta,\eta'}+\delta_{\sigma,\sigma'}
   \delta_{\eta,\eta'})\right)
  \label{lat3}
\eeq
with $\sigma=\sigma_{x}$ and $\sigma'=\sigma_{x+\alpha}$ {\em etc}. In this way
we have simply copied Eq.~(\ref{averlattice}) for the particular case
of $N=3$ and $a,b=1,2,3$, absorbing the coupling constant $J_{0}$ into
a redefinition of $\tilde{\beta}$.

Next we have to look for critical points of Eq.~(\ref{lat3}), for some
$\tilde{\beta}$ and negative $\tilde{g}_{0}$. If the position of the
critical point is not known analytically and has to be determined
numerically for a system as complicated as three coupled Potts
lattices, with $q^3$ degrees of freedom at each site, accuracy is
likely to be very poor, leading to unprecise or absurd measurements of
critical properties. Thus, our numerical studies would greatly benefit
from an exact determination of the position of the critical
point. This can be done if the model is self-dual, like the single
Ising or Potts models. For three coupled models, the existence of
duality relations requires the inclusion of a three-energy
interaction term
\beq
  \sum_{x,\alpha}\delta_{\sigma,\sigma'}\delta_{\tau,\tau'}
    \delta_{\eta,\eta'}.
  \label{3int}
\eeq
In the continuum limit this becomes
\beq
  \int d^{2}x \, \ve_{1}(x)\ve_{2}(x)\ve_{3}(x).
\eeq
It is straightforwardly shown that such a term does not modify
the fixed point structure nor its stability.  This means that adding
(\ref{3int}) to 
the lattice Hamiltonian should not modify the critical properies of
the model, the continuum limit theory being the same.  In this way we
finally arrive at a lattice model with partition function
\beq
\sum_{\{\sigma,\tau,\eta\}}\exp\left(\sum_{x,\alpha}
   \left[a\,(\delta_{\sigma,\sigma'}+\delta_{\tau,\tau'}
+\delta_{\eta,\eta'})+b\,(\delta_{\sigma,\sigma'}\delta_{\tau,\tau'}
+\delta_{\tau,\tau'}\delta_{\eta,\eta'}
+\delta_{\sigma,\sigma'}\delta_{\eta,\eta'}) 
+c\,\delta_{\sigma,\sigma'}\delta_{\tau,\tau'}\delta_{\eta,\eta'}\right]\right).
  \label{3modpf}
\eeq
The coupling constants, $a$, $b$ and $c$, can be chosen so as to
render this model self-dual. In Sec.~\ref{sec:duality} we shall
establish the corresponding duality transformations. It will turn out
that the model possesses not a point, but a line in parameter space
$(a,b,c)$ on which it is self-dual. On symmetry grounds we should
expect that its critical points belong to this line. In fact, it will
be shown that generically there exists three self-dual fixed points:
\begin{enumerate}
\item That of a single Potts model with $q^3$ states of the spin
  variable. At this point $a=b=0$ and $c\neq 0$.
  Whenever the energy-energy coupling between the models is relevant
  ({\em i.e.}, for $q > 2$), we have $q^3 \gg  4$. The phase transition is
  thus first order, and the model is not critical.
  \label{point-1}
\item That of three decoupled $q$-state Potts models. At this point
  $b=c=0$ and $a \neq 0$. The phase transition is second order if $q \leq 4$.
  \label{point-2}
\item The non-trivial fixed point of three coupled models. The
  transition is here second order, and the model is critical. The study
  of this point is the principal subject of this paper.
  \label{point-3}
\end{enumerate}

To make a better connection with the continuum limit theory, the
energy operators of the lattice model will be redefined in the next
section to take the form
\beq
E_{\sigma,\sigma'}=1-\delta_{\sigma,\sigma'}.
\eeq
The principal coupling term then becomes
\beq
  -g_{0}\sum_{x,\alpha}\left(E_{\sigma,\sigma'}E_{\tau,\tau'}+
    E_{\tau,\tau'}E_{\eta,\eta'}+E_{\sigma,\sigma'}E_{\eta,\eta'}\right),
\eeq
where the parameter $g_0$ is a linear combination of the parameters
$b,c$ in Eq.~(\ref{3modpf}):
\beq
  g_0=b+c.
\eeq
With respect to $g_{0}$ the critical point \ref{point-1}),
corresponding to a first order transition, has positive $g_0$, as it
should have been expected. The decoupling point \ref{point-2}), has
$g_{0}=0$, whilst the fixed point \ref{point-3}) is found for finite
negative $g_{0}$.

As will be evident from the subsequent sections, to locate this last
fixed point we have relied on the $c$-theorem of Zamolodchikov
\cite{zamo}, which states that the effective central charge of the
theory decreases along the RG flow. The effective central charge has
been measured using the strip geometry, as will be explained in detail
in Sec.~\ref{sec:numerics}. We assumed on symmetry grounds, and
verified numerically, that the RG flow from the decoupling point to
the non-trivial coupling point goes along the line of self-duality
that we have found. To locate the point \ref{point-3}) we stayed on
the self-dual line, on the negative $g_{0}$ side, and followed, using
transfer matrices on a strip, the evolution of the effective central
charge along the line. This led us to a particular limiting point on
the line of self-duality, actually its end-point, at which the exact
values of the couplings are known.  We then used transfer matrices and
Monte Carlo simulations, with the couplings tuned exactly to their
end-point values, to check scaling laws and measure critical
dimensions of various operators, comparing the result with those
obtained by perturbative CFT.

The paper is laid out as follows. In section \ref{sec:duality}, we
present duality transformations and identify the self-dual lines for
two and three coupled models. Section \ref{sec:RG} is devoted to the
computation of the central charge and the critical exponents of
physical operators. To this end we employ perturbative CFT
techniques. Section \ref{sec:numerics} introduces the various transfer
matrix algorithms we used to numerically compute the critical
properties. Since these algorithms are interesting on their own right,
some of them being more efficient than those previously described in
the Potts model literature, they are presented extensively. Section
\ref{sec:results} presents the numerical results obtained using these
newly introduced algorithms. A Monte Carlo study of scaling laws is
also undertaken.  Finally, section \ref{sec:discussion} is devoted to
concluding remarks and to a brief summary of the obtained results.

\section{Self-duality and criticality}

\label{sec:duality}
Duality relations, {\em i.e.} maps between sets of coupling constants
that lead to the same partition functions, are central objects in the
study of critical systems. These relations map one part of the
coupling phase space to another one, a self-dual manifold separating
the two.  If the fixed points we are looking for are unique, they
should be self-dual points in the phase space.  If this were not the
case, then fixed points would arise in dual pairs, which implies dual
RG flows.  Either these flows would never cross the self-dual line or
they would both cross it at a given point. We are not able to picture
in which way our system could behave like that. Numerical results
presented later in this paper strongly support the conjecture that the
fixed points of interest, for our models, are self-dual.

\subsection{Duality relations}

Let us first outline the construction of a global duality transformation
from which the self-dual lines will be extracted. Following
Ref.~\cite{Wu}, we shall restrict our attention to Hamiltonians of the
form
\beq
  {\cal H} = -\sum_{\langle i, j \rangle}
  J(\xi^{(1)}_{ij},\ldots,\xi^{(N)}_{ij}),
\eeq
where $\xi^{(k)}_i=1,\ldots,q$ are the spin variables of the $k$th model
and $\xi^{(k)}_{ij}\equiv |\xi^{(k)}_i-\xi^{(k)}_j|$. Energy-energy
coupled Potts models naturally have Hamiltonians of this form. For
generic $J$, the partition function can be written as
\beq 
  Z = \sum_{\vec\xi_i} \prod_{\langle i, j \rangle} u(\vec\xi_{ij}),
  \label{sum1}
\eeq
where $\vec\xi_{ij} = (\xi^{(1)}_{ij},\ldots \xi^{(N)}_{ij})$ and the
local Boltzmann weights $u(\vec\xi)$ are
\beq 
  u(\vec\xi) = \exp\left(\beta J(\vec\xi)\right).
\eeq
The matrix $U$, whose elements are $u^{(k)}(\xi_{ij})$, is a $q\times q$
cyclic matrix. It is rather obvious that the partition function can be
rewritten as
\beq
  Z = q^N \sum_{\vec\xi_{ij}} \prod_{\langle i, j \rangle} u(\vec\xi_{ij}),
\eeq 
where a factor of $q^N$ has been pulled out in front, since after
setting all the $\xi_{ij}$s, one still needs to fix the absolute value
of one spin in each model in order to completely specify the
configuration. Using Fourier transform the partition function can be
defined on the dual lattice. Since it is translationally invariant,
the eigenvalues of the matrix $U$ are given by
\beq
  \lambda(\vec\eta) = \sum_{\vec\xi} \exp \left( \frac{2\pi {\rm i}
  \vec\xi\cdot \vec\eta}{q} \right) u(\vec\xi),
  \label{duality} 
\eeq
which implies, using inverse Fourier transform, that
\beq
  u(\vec\xi_{ij}) = \sum_{\vec\eta} T(\vec\xi_i,\vec\eta) \lambda(\vec\eta)
  T^*(\vec\xi_j,\vec\eta),
\eeq
where
\beq
  T(\vec\xi,\vec\eta) = q^{-\frac{1}{2}} \exp \left(\frac{2\pi {\rm i}
  \vec\xi\cdot \vec\eta}{q} \right).
\eeq
This can be inserted in Eq.~(\ref{sum1}), and leads directly to
\beq
  Z= q^{1+N-N_{\rm D}} \sum_{\vec\eta_{ij}=1}^q \prod_{\langle i, j \rangle}
  \lambda(\vec\eta_{ij})
  \label{sum2}, 
\eeq
where $N_{\rm D}$ is the number of sites on the dual lattice.
The additional factor of $q$ comes from the fact that once all the
$\vec\eta_{ij}$ are set, one still has to define the absolute value on
one site. On a self-dual lattice, such as the square, the number of
spin sites on the direct lattice and on its dual are equal (neglecting
boundary effects), and one thus gets
\beq
  Z(u) = q Z(\lambda).
\eeq
The transformations (\ref{duality}) are well-defined global duality
transformations. From these, self-dual solutions can be extracted. Let
us treat the cases of two and three coupled models separately.

\subsubsection{Two models}

Consider the case of two coupled models, with Hamiltonian
\beq
{\cal H} = \sum_{\langle i,j\rangle} {\cal H}_{ij},
\eeq
\beq
{\cal H}_{ij} = -a\,(\delta_{\si_i,\si_j}+\delta_{\tau_i,\tau_j})
- b\,\,\delta_{\si_i,\si_j}\delta_{\tau_i,\tau_j}.
\eeq
The choice of sign for the coupling $b$ conforms with existing
computations for the random model.

The duality relations, mapping the couplings $(a,b)$ to $(a^*,b^*)$,
are given by
\bea 
  {\rm e}^{2a^*+b^*} &=& \frac{{\rm e}^{2a+b} + 2(q-1) {\rm e}^a +
                         (q-1)^2}{{\rm e}^{2a+b} -2 {\rm e}^a + 1}, \\ 
  {\rm e}^{a^*} &=& \frac{{\rm e}^{2a+b} + (q-2){\rm e}^a -
                         (q-1)}{{\rm e}^{2a+b} -2 {\rm e}^a + 1}.  
\eea
The denominator is introduced to ensure that configurations of zero
energy remain as such under duality. The self-duality condition
constrains couplings $a$ and $b$ to satisfy
\beq
  {\rm e}^b = {2 {\rm e}^a + (q-1)\over {\rm e}^{2a}}.
  \label{sd2m}
\eeq

Two points of direct physical interpretation can be found on this
line. First, the decoupled point $b=0$ for which  $a =
\ln(1+\sqrt{q})$, which is the usual critical temperature $\beta_{\rm c}(q)$
of the decoupled models. The second, at $a=0$ and $b = \ln(1+q)$,
corresponds to a $q^2$-state Potts model, for which
$\beta_{\rm c}(q^2) = \ln(1+q)$.  
We remark that the case of two coupled models was previously
considered by Domany and Riedel \cite{domany79}.

\subsubsection{Three models}

The Hamiltonian associated with the case of three coupled models is
\beq
  {\cal H} = \sum_{\langle i,j\rangle} {\cal H}_{ij},
\eeq
\bea
  \label{dua3}
  {\cal H}_{ij} &=& -a\,(\delta_{\si_i,\si_j} +\delta_{\tau_i,\tau_j}
  +\delta_{\eta_i,\eta_j} ) 
  - b\,(\delta_{\si_i,\si_j}\delta_{\tau_i,\tau_j}
  +\delta_{\si_i,\si_j}\delta_{\eta_i,\eta_j}
  +\delta_{\tau_i,\tau_j}\delta_{\eta_i,\eta_j}) \nn\\
  &-& c\,\,\delta_{\si_i,\si_j} \delta_{\tau_i,\tau_j}
  \delta_{\eta_i,\eta_j}.
\eea
The introduction of a three-coupling term is necessary to produce
self-dual solutions, since it is generated by applying the duality
relations to our original Hamiltonian.  The duality transformations
are given by
\bea
  {\rm e}^{3a^*+3b^*+c^*} &=& \frac{{\rm e}^{3a+3b+c} + 3(q-1)
  {\rm e}^{2a+b} + 3(q-1)^2 {\rm e}^a + 
  (q-1)^3}{{\rm e}^{3a+3b+c} -3 {\rm e}^{2a+b} + 3 {\rm e}^a -1}, \\  
  {\rm e}^{2a^*+b^*} &=& \frac{{\rm e}^{3a+3b+c} + (2q-3)
  {\rm e}^{2a+b} + (q^2-4q+3) {\rm e}^a - 
  (q-1)^2}{{\rm e}^{3a+3b+c} -3 {\rm e}^{2a+b} + 3 {\rm e}^a -1}, \\  
  {\rm e}^{a^*} &=& \frac{{\rm e}^{3a+3b+c} + (q-3){\rm e}^{2a+b} +
  (3-2q) {\rm e}^a + 
  (q-1)}{{\rm e}^{3a+3b+c} -3 {\rm e}^{2a+b} + 3 {\rm e}^a -1}.  
\eea
Self-duality solutions are found to be
\bea
  \label{sd3m}
  {\rm e}^{b}&=&{(2 + \sqrt{q}) {\rm e}^a -( \sqrt{q}+1)\over
  {\rm e}^{2a}}, \nn \\ 
  {\rm e}^c&=& {\rm e}^{3a} {3 ({\rm e}^a-1)(\sqrt{q} +1) + q\sqrt{q}
  +1 \over ({\rm e}^a  
  (2+\sqrt{q}) -(1+\sqrt{q}))^3 }.
\eea
The two trivial points found in the two-models case also arise here;
the decoupling point, for which $b=c=0$ and $a = \ln(1+\sqrt{q})$, and
the $q^3$-state Potts model, with $a=b=0$ and
$c = \beta_{\rm c}(q^3) = \ln(1+q^{3/2})$.

\subsection{A convenient reparametrisation}
\label{sec:repara}

There is a certain arbitrariness in the way we choose to parametrise a
point on the self-dual line. Ideally, the couplings entering the
lattice Hamiltonian would in some way be comparable to those of the
field theory describing its continuum limit. Evidently, decoupling
points should be associated with null couplings in both schemes, but
this still leaves plenty of room for a reparametrisation. We propose
such a reparametrisation which makes closer contact with the physical
considerations put forward in the Introduction, and, at the same time,
facilitates the implementation of numerical methods. 

\subsubsection{General case}

Introducing the parameter $x$ ($x \in [1,+\infty[$) in
Eq.~(\ref{dua3}), the self-duality relations can be rewritten as
\bea
  {\rm e}^a &=& \frac{\sqrt{q}+1}{\sqrt{q}+2} x, \\
  {\rm e}^b &=& \frac{(\sqrt{q}+2)^2}{\sqrt{q}+1} \frac{x-1}{x^2}, \\
  {\rm e}^c &=& \frac{3(\sqrt{q}+1)^2}{(\sqrt{q}+2)^4}
                \frac{x^3(x-\gamma)}{(x-1)^3}, 
\eea
where $\gamma\equiv \frac{4-q}{3}$. Since the decoupled models we
study have ferromagnetic ground states, it is more convenient to trade
the $\delta_{\sigma_i,\sigma_j}$ for the operators 
\beq 
  E_{\si_i,\si_j} = 1-\delta_{\si_i,\si_j}.
  \label{enlat}
\eeq 
Doing so, the Hamiltonian becomes 
\bea 
  {\cal H}_{ij} &=& -(3a+3b+c) + (a+2b+c)\,(E_{\si_i,\si_j} +E_{\tau_i,\tau_j} 
   +E_{\eta_i,\eta_j} ) \nn\\ &-& (b+c) \,(E_{\si_i,\si_j}E_{\tau_i,\tau_j}
   +E_{\si_i,\si_j}E_{\eta_i,\eta_j}
   +E_{\tau_i,\tau_j}E_{\eta_i,\eta_j})\nn\\
  &+& c\,\,E_{\si_i,\si_j} E_{\tau_i,\tau_j} E_{\eta_i,\eta_j}.
  \label{H3-repara}
\eea
The constant term, being the ground state energy, can be gauged out.
Now define the new one- and two-energy coupling constants
\bea 
  \beta &=& a+2b+c, \nn\\
  g     &=& b+c.
\eea 
With this set of couplings the Hamiltonian is turned into
\bea 
  {\cal H}_{ij} &=& \beta\,(E_{\si_i,\si_j} +E_{\tau_i,\tau_j} 
   +E_{\eta_i,\eta_j} ) \nn\\ &-& g\,(E_{\si_i,\si_j}E_{\tau_i,\tau_j}
   +E_{\si_i,\si_j}E_{\eta_i,\eta_j}
   +E_{\tau_i,\tau_j}E_{\eta_i,\eta_j})\nn\\
  &+& c\,\,E_{\si_i,\si_j} E_{\tau_i,\tau_j} E_{\eta_i,\eta_j}.
\eea 
On the self-dual line the couplings $\beta$ and $g$ are parametrised by 
\bea
 {\rm e}^{-\beta} &=& \frac{\sqrt{q}+2}{3(\sqrt{q}+1)} \frac{x-1}{x-\gamma},
 \\ 
 {\rm e}^{g} &=&
 \frac{3(\sqrt{q}+1)}{(\sqrt{q}+2)^2}\frac{x(x-\gamma)}{(x-1)^2}.   
\eea

A similar reparametrisation for the case of two models leads to the
Hamiltonian
\beq
  {\cal H}_{ij} = \beta (E_{\si_i,\si_j} +E_{\tau_i,\tau_j})
  - g(E_{\si_i,\si_j}E_{\tau_i,\tau_j}),
  \label{H2-repara}
\eeq
with
\bea
  {\rm e}^{-\beta} &=& \frac{x^2}{2x^2 + x(q-1)}, \nn \\
  {\rm e}^{g}      &=& \frac{2x + (q-1)}{x^2}.
\eea

This particular parametrisation has some definite advantages over the
original one. First, all Boltzmann weights remain finite along the
self-dual line, something useful both in numerical studies and in a
comparison with perturbative CFT. Second, as we shall see in
Sec.~\ref{sec:RG}, along the self-dual line, the coupling constant $g$
has a sign in agreeement with perturbative computations.  Finally, in the
three-models case, the three-coupling term becomes infinite when
$x \to \infty$, leading to a null Boltzmann weight. This implies some
simplifications of numerical studies, as we now show.

\subsubsection{Limits on the self-dual lines}

As we just said, the limit $x \to \infty$ will be of special interest
in the following sections. There we can simplify numerical
computations (both using transfer matrices and Monte Carlo
simulations) by suppressing some of the Boltzmann weights attached to
the links.  The local Boltzmann weight $W(L_{ij})$ for the degrees of
freedom coupling spin sites $i$ and $j$ is completely determined by
specifying the number $L_{ij}$ of layers having different spin values at
the two ends of the bond ($ij$). The total Boltzmann weight of a spin
configuration is then given by
\beq
  W = \prod_{\langle i,j \rangle} W(L_{ij}).
  \label{local-Boltz}
\eeq

If we consider the case of two coupled models, Eq.~(\ref{H2-repara}),
three possible weights $W(L)$ arise:
\beq
  W(0)= 1, \ \ \ \ W(1)= {\rm e}^{-\beta}, \ \ \ \ W(2)={\rm e}^{-2\beta+g}. 
\eeq
Using the self-dual parametrised solutions, we see that in the limit
$x \to \infty$ this becomes
\bea
\label{esd2}
  W(0)=1, \ \ \ \ W(1)=1/2, \ \ \ \ W(2)=0.
\eea
For the case of three models, Eq.~(\ref{H3-repara}), there are four
possible weights
\beq
  W(0)=1, \ \ \ \ W(1)={\rm e}^{-\beta}, \ \ \ \
  W(2)={\rm e}^{-2\beta+g}, \ \ \ \ W(3)={\rm e}^{-3\beta+3g-c}.
\eeq
At $x\rightarrow \infty$ this becomes
\beq
  W(0)= 1, \ \ \ \ W(1)=\frac{\sqrt{q}+2}{3(\sqrt{q}+1)}, \ \ \ \
  W(2)= \frac{1}{3(\sqrt{q}+1)}, \ \ \ \ W(3)= 0,   
  \label{esd3r}
\eeq
We shall use this limit extensively later on. The fact that
configurations with one or more $L_{ij}$ equal to three are forbidden
greatly simplifies the problem. It even gives hope to solve exactly
the lattice model at the critical point. In Sec.~4.5 below we shall
see that it is possible to reformulate the Hamiltonians
(\ref{H2-repara}) and (\ref{H3-repara}) so as to obtain even more null
Boltzmann weights, by passing on to a random cluster picture.

\section{Renormalisation group study of \\ coupled Potts models}
\label{sec:RG}

Coupled Potts models have already been studied in details using
renormalisation group techniques.  Details can be found in references
\cite{ludwig,dpp1,cardy,pujol,simon,ls}.  We shall only present here a
summary of these results, which are exposed extensively in the given
references.

The continuum limit of the models under consideration is
defined by 
\beq
  {\cal H} = \displaystyle\sum_{i=1}^N {\cal H}_{0,i} - g \int d^2x
  \displaystyle\sum_{i \neq j}^N \ve_{i}(x)\ve_{j}(x),
  \label{nonrandham}
\eeq 
where $\ve_{i}(x)$ is the continuum limit of $E_{\sigma_i,\sigma_j}$
[see Eq.~(\ref{enlat})] and corresponds to the energy operator, and
${\cal H}_{0,i}$ are the Hamiltonians of the decoupled Potts models in
the continuum limit. We shall restrict the discussion to $q$-state
Potts models with $2\leq q \leq 4$, for which the dimension of the
energy operator varies between $1$ (for $q=2$) and $1/2$ (for $q=4$).
For such values of $q$, the continuum limit of the models are
conformal field theories.  The term $g\int \ve\ve $ is considered as a
perturbation, which is relevant by dimensional analysis (except for
$q=2$, where it is marginal).  The computation of the $\beta$-function
to two loops was done in Refs.~\cite{ludwig,dpp1} and is given by  
\beq
  \beta(g) \equiv {dg\over d\log (r)} = 3\ep g(r) + 4 \pi (N-2) g^2(r)
   -16 \pi^2 (N-2) g^3(r) + {\cal O}(g^4(r)). 
\eeq 
Here $\ep$ is the perturbation parameter, and it corresponds to the
dimension of the perturbation:
$\ep = \frac{2}{3}\left(1 - \Delta_{\ve}\right)$ where $\Delta_{\ve}$
is the dimension of the energy operator of the decoupled models. For
$q=2$ we have $\ep =0$, whilst for $q=3$, $\ep = 2/15$ and for $q=4$,
$\ep=1/3$. The parameter $r$ is an infrared cut-off.

In order to compute critical exponents we need to compute the matrices 
$(Z_\ve)_{ij}$ and $(Z_\si)_{ij}$ which are respectively the
renormalisation constants of the energy and spin operators. These have
to be understood in the following way:
\beq 
  \vec{\cal O}_{\rm R} = (Z_{\cal O})\vec{\cal O},
\eeq
where the $i$th element of vector $\vec{\cal O}$ is the operator
$\cal O$ of model $i$, and ${\cal O}_{\rm R}$ is the
corresponding renormalised operator. These matrices are, again to
second order in the perturbation, given by \cite{dpp1}
\begin{equation}
  \frac{d \log (Z_\ve(r))_{ij}}{dr} = \left(-4\pi(N-1)g - 8\pi^2(N-1)
  g^2 + {\cal O}(g^3)\right)(1-\delta_{ij}),
\end{equation}
\begin{equation}
  \frac{d \log (Z_\sigma(r))_{ij}}{dr} = \left(3(N-1)g^2
  \pi^2\epsilon\,{\cal F} + 4(N-1)(N-2)\pi^3g^3 +
  {\cal O}(g^4)\right)\delta_{ij}, 
\end{equation}      
with
\begin{equation}
  {\cal F} = 2 \, \frac{\Gamma^2(-\frac{2}{3}) \Gamma^2(\frac{1}{6})}
  {\Gamma^2(-\frac{1}{3})\Gamma^2(-\frac{1}{6})}. \label{calF}
\end{equation}
The fact that the energy matrix is not diagonal was observed in
Refs.~\cite{ludwig,ls}. This implies that the energy operators for
each individual layer are no longer the proper ones to study critical
behaviour. Instead, the eigenvectors of the matrix turn out to be the
ones we observe in numerical studies. We shall come back to this point
when we compute critical exponents in the next sub-section.

The possible behaviours of our coupled systems can be summarised as
follows \cite{pujol}.
\begin{itemize}
\item For $q=2$ the model corresponds to the $N$-colour Ashkin-Teller
  model. For $N=2$ it is integrable \cite{baxwukad}. For $N>2$ the
  sign of $g$ is determinant for the large-scale behaviour: When $g>0$
  a second order phase transition of the Ising type is observed,
  whilst for $g<0$ the scenario is that of a fluctuation-driven first
  order phase transition \cite{fradkin,shankar}.
\item For $q>2$ and $N=2$ the model is still integrable, but now presents
  a mass generation \cite{vays}. This again indicates a first order
  phase transition.
\item For the case $N>2$ and $g>0$ the coupling constants flow far
  from our perturbative region. Even if a definite proof is not given, a
  comparison with the case $q=2$ seems to tell us that a mass gap is
  dynamically generated, indicating a first order phase transition.

  The situation is completely different for $g<0$. In that case there is
  a non-trivial infrared fixed point for
  \beq
  \label{fixed1}
    g_* = -{3\epsilon \over 4
    \pi (N-2)} + {9\epsilon^2 \over 4\pi (N-2)^2} + {\cal O}(\epsilon^3)
  \eeq
  The critical exponents associated with the energy and spin
  operators for this fixed point, along with the values of the central
  charge, are computed below. 
\end{itemize}

\subsection{Critical exponents for $q>2,N>2$}

We now compute the critical exponents of our coupled models at the
fixed points of the renormalisation group. For the decoupling point,
the critical exponents are evidently the ones of the pure models.  We
shall therefore concentrate on the non-trivial fixed point identified
above. Since the renormalisation matrix for the spin operator is
diagonal and the coupling is invariant under permutation, any linear
combination of the spin operators has a well-defined critical exponent
which is found to be \cite{dpp1}
\bea
  \Delta'_{\sigma} = \Delta_{\sigma} -
  \frac{27}{16}\frac{N-1}{(N-2)^2}\,
  {\cal F} \epsilon^3 + {\cal O}(\epsilon^4),
\eea
where ${\cal F}$ was defined in Eq.~(\ref{calF}).

Defining critical exponents for the energy operators is somehow more
tricky. Correlation functions between renormalised operators (at a
given cut-off $R$) are of the following form
\begin{equation}
  \langle \ve_i(0)\ve_j(R) \rangle \sim
  \sum_{k \ne i}\sum_{l\ne j} (Z_\ve)_{ik}(Z_\ve)_{jl}
  \frac{1}{R^{2\Delta_\ve}} \langle
  \ve_k(0)\ve_l(1)\rangle, 
\end{equation}
mixing correlation functions of the $N$ different layers. One way to
extract unambiguous critical exponents is to diagonalise the
renormalisation matrix, thus introducing a new basis of operators.
This diagonalisation is exact, in the sense that eigenvectors for the
one-loop renormalisation matrix remain eigenvectors to all order in
the perturbation. For three models the eigenvectors are 
\beq
  \ve_1+\ve_2+\ve_3, \qquad
  \ve_1-\ve_2, \qquad
  \ve_2-\ve_3. 
\eeq
When using this basis, the computation of the different exponents is
straightforward. We have \cite{ls}
\beq
  \Delta_{\ve_1+\ve_2+\ve_3} =
  \Delta_\ve+6\epsilon-9\epsilon^2+{\cal O}(\epsilon^3)
  \label{symm-energy}
\eeq
and
\beq
  \Delta_{\ve_1-\ve_2} =
  \Delta_\ve-3\epsilon+\frac{9}{2}\epsilon^2+{\cal O}(\epsilon^3).
\eeq
Some explicit values of the different critical exponents for the case
of three coupled model are provided by Table \ref{tablecrit}.  The
appearance of negative magnetic exponents for sufficiently large $q$
shows the limitations of the perturbative expansion. It is also
possible to compute the critical exponents for higher moments of the
spin and energy operators. The physically significant operators are
found to be
\bea
&&  \sigma_{1}\sigma_{2}, \qquad \sigma_{1}\sigma_{3}, \qquad
  \sigma_{2}\sigma_{3}, \qquad \sigma_1\sigma_2\sigma_3; \\
&&  \ve_{1}\ve_{2}+\ve_{2}\ve_{3}+\ve_{3}\ve_{1}, \qquad
  \ve_{1}\ve_{2}- \ve_{2}\ve_{3}, \qquad
  \ve_{1}\ve_{2}- \ve_{3}\ve_{1}, \qquad
  \ve_{1}\ve_{2}\ve_{3}.
\eea
As was shown for the random Potts model \cite{MAL}, perturbative
computations for higher moments of the spin and energy operators are
much less precise than for the operators themselves, and one should
keep in mind that they eventually become absurd for sufficiently high
moments (for example, the third magnetic moment has negative exponent
for $q>3.7$). For the spin operators, the second order perturbative
computations lead to the following exponents \cite{ddp,MAL}
\beq
  \Delta_{\sigma_1\sigma_2} = 2\Delta_\sigma(\epsilon) +
  \frac{3\epsilon}{4(N-2)} \left(1-\frac{3\epsilon}{N-2}\left((N-2)\log 2 +
  \frac{11}{12}\right)\right) + {\cal O}(\epsilon^3),
\eeq
\beq
  \Delta_{\sigma_1\sigma_2\sigma_3} = 3\Delta_\sigma(\epsilon) +
  \frac{9\epsilon}{4(N-2)}\left(1-\frac{3\epsilon}{N-2}\left((N-2)\log 2 +
  \frac{11}{12}+\frac{\alpha}{24}\right)\right) + {\cal O}(\epsilon^3),
\eeq
with
\beq
\alpha=33-\frac{29\sqrt{3}\pi}{3}.
\eeq
For the energy operators, critical exponents for the diagonal
operators are given by
($\ve_{\rm S}^2\equiv \ve_{1}\ve_{2}+\ve_{2}\ve_{3}+\ve_{3}\ve_{1}$ and
 $\ve_{\rm A}^2\equiv \ve_{1}\ve_{2}- \ve_{2}\ve_{3}$)
\bea 
  \Delta_{\ve_{\rm S}^2} &=& 2\Delta_\ve(\epsilon) + 3\epsilon +
                       \frac{9}{2}\epsilon^2 +{\cal O}(\epsilon^3) \\
  \Delta_{\ve_{\rm A}^2} &=& 2\Delta_\ve(\epsilon) - \frac{3}{2}\epsilon -
                       9\epsilon^2 + {\cal O}(\epsilon^3) \\
  \Delta_{\ve_1\ve_2\ve_3}
                   &=& 3\Delta_\ve(\epsilon) -
                       \frac{27}{4}\epsilon^2 + {\cal O}(\epsilon^3)
\eea
The numerical values for these operators are given in Table 2.
\begin{table}
\begin{center}
\begin{tabular}{|l||l|l|l|} \hline
  $q$  & $\Delta_{\ve_1+\ve_2+\ve_3}$ & $\Delta_{\ve_1-\ve_2}$ &
         $\Delta_{\sigma_1}$\\ \hline
 2.00  & 1.0000 & 1.0000 & 0.12500 \\ \hline
 2.25  & 1.1447 & 0.8499 & 0.12789 \\ \hline
 2.50  & 1.2639 & 0.7154 & 0.12964 \\ \hline
 2.75  & 1.3615 & 0.5930 & 0.12985 \\ \hline
 3.00  & 1.4400 & 0.4800 & 0.12805 \\ \hline
 3.25  & 1.5006 & 0.3737 & 0.12353 \\ \hline
 3.50  & 1.5429 & 0.2710 & 0.11501 \\ \hline
 3.75  & 1.5624 & 0.1656 & 0.09926 \\ \hline
 4.00  & 1.5000 & 0.0000 & 0.04238 \\ \hline
\end{tabular}
\end{center}
\protect\caption{\label{tablecrit}Perturbative critical exponents of
  physically significant energy and spin operators for three coupled
  $q$-state Potts models.}
\end{table}

\begin{table}
\begin{center}
\begin{tabular}{|l||l|l|l|l|l|} \hline
  $q$  & $\Delta_{\ve_1\ve_2+\ve_2\ve_3+\ve_3\ve_1}$ &
         $\Delta_{\ve_1\ve_2-\ve_2\ve_3}$
       & $\Delta_{\ve_1\ve_2\ve_3}$ & $\Delta_{\sigma_1\sigma_2}$ &
         $\Delta_{\sigma_1\sigma_2\sigma_3}$ \\ \hline
 2.00  & 2.000 &  2.000 & 3.000 & 0.2500 &  0.3750 \\ \hline
 2.25  & 2.005 &  1.834 & 2.837 & 0.2775 &  0.4553 \\ \hline
 2.50  & 2.021 &  1.653 & 2.664 & 0.2949 &  0.5190 \\ \hline
 2.75  & 2.046 &  1.456 & 2.479 & 0.3030 &  0.5685 \\ \hline
 3.00  & 2.080 &  1.240 & 2.280 & 0.3023 &  0.6048 \\ \hline
 3.25  & 2.126 &  0.997 & 2.060 & 0.2921 &  0.6283 \\ \hline
 3.50  & 2.186 &  0.713 & 1.806 & 0.2703 &  0.6375 \\ \hline
 3.75  & 2.272 &  0.350 & 1.486 & 0.2303 &  0.6268 \\ \hline
 4.00  & 2.500 & -0.500 & 0.750 & 0.0975 &  0.5301 \\ \hline
\end{tabular}
\end{center}
\protect\caption{\label{tablemom}Perturbative critical exponents of
  physically significant energy and spin operator moments for three
  coupled $q$-state Potts models.}
\end{table}

\subsection{Central charge and the $c$-theorem for $q>2,N>2$}

One of the most important tools of CFT, Zamolodchikov's $c$-theorem
\cite{zamo}, provides a simple way of computing the central charge for
perturbed theories.  Moreover, it gives us crucial information:
\begin{itemize}
\item The $c$-function, to be defined below, is decreasing along the
  renormalisation flow.
\item If the flow has a fixed point, the field theory at the fixed
  point is conformally invariant and its central charge is the value of
  the $c$-function at that point. 
\end{itemize}

With our conventions for the $\beta$-function taken into account, the
$c$-function is defined as
\beq
  c(g')= c_{\rm pure} - 24 \int_{0}^{g'} \beta(g) \mbox{d}g,
  \label{c-function}
\eeq
with $c_{\rm pure}$ being the total central charge of the decoupled
models. The central charge deviation from the decoupling point value
is thus easily computed. Substituting $g'=g_*$, the fixed point value
of the couplings, we get the following correction:
\bea
  \Delta c &=& -24\int_{0}^{g_*} \beta(g) \mbox{d}g \\
           &=& -\frac{27}{8}\frac{N(N-1)}{(N-2)^2}
               \left(\epsilon^3 - \frac{9}{2(N-2)}\epsilon^4\right) +
               {\cal O}(\epsilon^5) 
  \label{c-correction}
\eea
Table \ref{tablecharge} presents the pure and perturbative values of
the central charge. Clearly, for large enough $q$ the latter are not
to be trusted, since the perturbation theory eventually breaks
down. This is witnessed by the change of sign in
Eq.~(\ref{c-correction}) when $\epsilon > \frac29$.

\begin{table}
\begin{center}
\begin{tabular}{|l||l|l|} \hline
  $q$  & $c_{\rm pure}$ & $c_{\rm FP}$ \\ \hline
 2.00  & 1.5000 & 1.5000 \\ \hline
 2.25  & 1.7627 & 1.7620 \\ \hline
 2.50  & 1.9975 & 1.9931 \\ \hline
 2.75  & 2.2089 & 2.1976 \\ \hline
 3.00  & 2.4000 & 2.3808 \\ \hline
 3.25  & 2.5734 & 2.5500 \\ \hline
 3.50  & 2.7309 & 2.7164 \\ \hline
 3.75  & 2.8734 & 2.9054 \\ \hline
 4.00  & 3.0000 & 3.3750 \\ \hline
\end{tabular}
\end{center}
\protect\caption{\label{tablecharge}Comparison of the central charge
  values at the pure and the non-trivial fixed points.}
\end{table}

\section{The transfer matrices}
\label{sec:numerics}

Systems of several coupled $q$-state Potts models possess an enormous
number of states, and one should think that accurate numerical results
cannot be obtained by the transfer matrix technique, since only very
narrow strips can be accessed. However, there are several
possibilities for dramatically reducing the size of the state space,
as we show in the present Section. In particular we shall
see that it is possible to numerically study the $q$-state
Potts model for {\em any} real $q$ with less computational effort than
is required in the traditional transfer matrix approach to the Ising
model.

We have optimised the algorithm described in Ref.~\cite{mp}, and
adapted it to the case of three coupled models, by taking into account
all known symmetries of the Boltzmann weights in the spin
basis. Further progress can be made by trading the spin variables for
clusters or loops. The cluster algorithm of Ref.~\cite{Blote82} has
been adapted to the problem of coupled models, and we also describe, for the
first time in the literature, the practical implementation of the
associated loop algorithm. The latter algorithm allows us to address
the case of three coupled models on strips of width $L_{\rm max}=12$.

\subsection{The four algorithms}

Consider a system of coupled Potts models defined on a cylinder of
circumference $L$ and length $M$, measured in units of the lattice
constant. The imposition of periodic boundary conditions in the
transversal direction is understood throughout. The transfer matrix
can be viewed as a linear operator ${\bf T}$ that acts on the
partition function of the $M$ row system, where fixed boundary
conditions have been imposed on the degrees of freedom of the 
$M$th row, so as to generate the corresponding quantity for a system
with $M+1$ rows. To fix terminology, we shall refer to the
specification of the boundary condition on the last row as the {\em
state} of the system. Thus, symbolically, 
\begin{equation}
  Z_{\alpha}^{(M+1)} = \sum_{\beta} T_{\alpha \beta}
                       Z_{\beta}^{(M)},
  \label{add-a-row}
\end{equation}
where the Greek subscripts label the possible states.
The matrix elements $T_{\alpha \beta}$ are nothing but the Boltzmann
factors $\exp[-{\cal H}(\alpha,\beta)]$ arising from the interaction between
the degrees of freedom in the $M$th and the $(M+1)$th row, when these
are in the fixed states $\beta$ and $\alpha$ respectively.

It is well-known that crucial information about the free energy, the
central charge, and the scaling dimensions of various physical
operators can be extracted from the asymptotic scaling (with strip
width $L$) of the eigenvalue spectrum of the transfer matrix, and so
the question arises what is the most efficient way of diagonalising
it. A quite common trick is to decompose it as a product of sparse
matrices which each add a single degree of freedom to the $M$th row,
rather than an entire new row at once. But even more important is the
realisation that the total number of states, and thus the
dimensionality of the transfer matrix, can be reduced by taking into
account the various symmetries of the interactions and possibly by
rewriting the model in terms of new degrees of freedom other than
simply the Potts spins. We shall refer to any such collection of states
as a {\em basis} of the transfer matrix, and the cardinality of the
basis as its {\em size}. The accuracy of the finite-size data, of
course, improves significantly with increasing $L$, and since the
size is a very rapidly (typically exponentially) increasing function of
$L$ we must face the highly non-trivial algorithmic problem of
identifying and implementing the most efficient basis.

These considerations have led to the development of four different
algorithms for the coupled Potts models problem. We list them here in
order of increasing efficiency and, roughly, increasing algorithmic
ingenuity. Details on the implementations will be given in the
following sub-sections. As a rough measure of their respective
efficiency we indicate how large $L$ can be attained in practice%
\footnote{The largest sparse matrices that we could numerically
diagonalise using a reasonable amount of computation time had size
$\sim 10^7$.} for the three-layered model.
\begin{itemize}
 \item {\tt alg1} uses the trivial basis of Potts spins and can access
       $L=5$ for $q=3$ and $L=4$ for $q=4$. Although hopelessly
       inefficient in the general case this algorithm is still of some
       use for small, integer $q$ since it allows
       for a straightforward access to magnetic properties.
 \item {\tt alg2} still works in the spin basis, but the $Z_q$
       symmetry of the Potts spins has been factored out. In this
       process the magnetic properties are lost. The size is now
       independent of $q$, but for $q<L$ a further reduction of the
       size occurs due to a
       truncation of the admissible states. For $q=3$ this algorithm
       can access $L=7$, but for general $q$ only $L=6$.
 \item {\tt alg3} utilises a mapping to the random cluster model, and
       the basis consists of the topologically allowed (`well-nested')
       connectivities with respect to clusters of spins that are in
       the same Potts state. Magnetic properties are lost, but can be
       restored through the inclusion of a ghost site at the expense
       of increasing the number of basis states. Since $q$ enters only
       as a (continuous) parameter, this algorithm is particularly
       convenient for comparing with analytic results obtained by
       perturbatively expanding in powers of $(q-2)$. In the
       non-magnetic sector $L=7$ can be accessed.
 \item {\tt alg4} works in a mixed representation of random clusters
       and their surrounding loops on the medial lattice. Magnetic
       properties can be addressed through the imposition of twisted boundary
       conditions, which by duality are equivalent to the topological
       constraint that clusters and loops do not wrap around the cylinder.
       Again $q$ is a continuous parameter. Due to the definition of
       the medial lattice, only even $L$ are allowed. This algorithm
       can access $L=12$ in the non-magnetic sector and $L=10$ in the
       magnetic one.
\end{itemize}

Before turning our attention to a detailed description of these
algorithms we briefly recall how physically interesting quantities may
be extracted from the transfer matrix spectra.

\subsection{Extraction of physical quantities}

The reduced free energy per spin in the limit $M \to \infty$ of an
infinitely long cylinder is given by
\begin{equation}
  f_0(L) = - \lim_{M \to \infty} \frac{1}{LM} \ln {\rm Tr} Z^{(M)}
         = - \frac{1}{L} \ln \lambda_0(L),
\end{equation}
where $\lambda_0$ is the largest eigenvalue of ${\bf T}$. Starting
from some arbitrary inititial vector of unit norm $|{\bf v}_0 \rangle$,
this can be found by simply iterating the transfer matrix \cite{Furstenberg}
\begin{equation}
 \label{iterate}
 \lambda_0(L) = \lim_{M \to \infty} \frac{1}{M}
                \ln \left \| {\bf T}^M |{\bf v}_0 \rangle \right \|.
\end{equation}
Higher eigenvalues $\lambda_k(L)$ are found by iterating a set of $n$
vectors $\{ |{\bf v}_k \rangle \}_{k=0}^{n-1}$, where a given
$|{\bf v}_k \rangle$ is orthogonalised to the set
$\{ |{\bf v}_l \rangle \}_{l=0}^{k-1}$ after each
multiplication by ${\bf T}$ \cite{Benettin}.

In general, of course, there exists more expedient methods for
diagonalising a square matrix, but since each of our four algorithms
allows for factorising the transfer matrix as a product of {\em sparse}
matrices this simple iteration method is superior. The advantage of using
sparse-matrix factorisation is that it reduces the number of elementary
operations from $(C_L)^2$ to $L C_L$ per iterated row
\cite{nightingale}, where $C_L$ is the size of the basis.

It is well-known from conformal field theory how to relate the central
charge to the finite-size scaling of the free energy \cite{bcn,affleck}
\begin{equation}
 \label{fc}
 f_0(L) = f_0(\infty) - \frac{\pi c}{6 L^2} + \cdots.
\end{equation}
Similarly, the gaps of the eigenvalue spectrum fix the scaling
dimensions $x_i$ of physical operators%
\footnote{The notational discrepancy with Sec.~\ref{sec:RG} is
intentional. Indeed, the identification between the scaling dimensions
of physical operators and numerically measured quantities is the
object of Sec.~\ref{sec:results}.}~\cite{Cardy83}
\begin{equation}
 \label{fx}
 f_i(L) - f_0(L) = \frac{2\pi x_i}{L^2} + \cdots.
\end{equation}

In general one may construct several {\em sectors} of the transfer
matrix by identifying the various irreducible representations of the
symmetry group of the microscopic (spin) degrees of freedom. As a
familiar example consider just one single Ising model ($q=2$). Here
there are two sectors (even and odd) corresponding to the possible
transformations of a state vector
$|{\bf v} \rangle \to \pm |{\bf v} \rangle$ under a global spin flip
\begin{equation}
  \sigma \to (\sigma+1) \mbox{ mod } q \; .
  \label{shift}
\end{equation}
This generalises straightforwardly to the $q$-state Potts model for which
we must consider transformations
$|{\bf v} \rangle \to {\rm e}^{2i\pi{j\over q}} |{\bf v} \rangle$,
$j=0,1,\ldots,q-1$.
The thermal and magnetic scaling dimensions, $x_t$ and $x_h$, are then
extracted by applying Eq.~(\ref{fx}) to $f_1^{\rm even} - f_0^{\rm even}$
and $f_0^{\rm odd} - f_0^{\rm even}$ respectively.
In the ordinary spin basis (as used by {\tt alg1}) the choice between
the even and odd sector can be very easily accomplished by iterating
initial vectors (cfr.~Eq.~(\ref{iterate})) that are either even or
odd under the transformation (\ref{shift}). When the $Z_q$ symmetry
has been factored out (as in {\tt alg3} and {\tt alg4}) the odd sector
has to be constructed explicitly by means of a ghost site or a seam
(see below for details).

When considering several coupled Potts models more than one magnetic
exponent may be defined. Namely, for each individual model one may
independently choose between the even (e) and odd (o) sector. At the
decoupling point, of course, one has
\begin{equation}
 \frac13 x_h^{\rm ooo} = \frac12 x_h^{\rm eoo} = x_h^{\rm eeo},
\end{equation}
but at a general critical fixed point the magnetic exponents thus
defined are independent.

When using Eqs.~(\ref{fc}) and (\ref{fx}) to obtain finite-size
estimates for $c$ and $x_i$ the convergence properties can be
considerably improved by explicitly including higher-order terms in $1/L$.
In an extensive study of the $q$-state Potts model Bl{\"o}te and
Nightingale \cite{bn} numerically showed that the sub-leading correction
to the free energy takes the form of an additional
$1/L^4$ term on the right-hand side of Eq.~(\ref{fc}). Not surprisingly an
analogous statement can be made about Eq.~(\ref{fx}) for the scaling
dimensions. From the
viewpoint of conformal field theory such a non-universal correction to
scaling can be rationalised as arising from the operator $T \overline{T}$,
where $T$ denotes the stress-energy tensor \cite{cj}. This
operator of dimension 4 is present in any theory that is conformally
invariant \cite{Cardy-CFT}.

Finite-size estimates $c(L,L')$ can then be extracted either from
two-point fits for $f_0(L)$ and $f_0(L' \equiv L+1)$ using
Eq.~(\ref{fc}) or from three-point fits for $f_0(L)$, $f_0(L+1)$ and
$f_0(L' \equiv L+2)$ using 
\begin{equation}
  f_0(L) = f_0(\infty) - \frac{\pi c}{6 L^2} + \frac{A}{L^4} + \cdots.
  \label{fc2}
\end{equation}
Similarly, for the scaling dimensions the one-point estimates $x_i(L)$
obtained from Eq.~(\ref{fx}) can be improved by extracting two-point
estimates $x_i(L,L+1)$ from fits of the form
\begin{equation}
 \label{fx2}
 f_i(L) - f_0(L) = \frac{2\pi x_i}{L^2} + \frac{B}{L^4} + \cdots.
\end{equation}

\subsection{Algorithm {\tt alg1}}

In the trivial spin basis each state is specified by the values of the
$L$ Potts spins in row $M$. The size of the transfer matrix for $N$
coupled models is therefore
\begin{equation}
  N_{q,N}^{\tt alg1}(L)= q^{LN}.
\end{equation}
These numbers increase very rapidly as a function of strip width $L$,
and they do not have a $q$-independent upper bound. To highlight the
merits of the more sophisticated algorithms that we are about to
develop we present some explicit values of the $N_{q,N}^{\tt alg1}$ in
Table~\ref{table1}.

\begin{table}
\begin{center}
\begin{tabular}{|l||r|r||r|r||r|r||} \hline
$N_{q,N}^{\tt alg1}(L)$
     & $q=2$ &          & $q=3$   &          & $q=4$    &           \\
\hline
$N$  &      2&         3&        2&         3&         2&         3 \\
\hline
$L=2$&     16&        64&       81&       729&       256&     4,096 \\
\hline
$L=3$&     64&       512&      729&    19,683&     4,096&   262,144 \\
\hline
$L=4$&    256&     4,096&    6,561&   531,441&    65,536&16,777,216 \\
\hline
$L=5$&  1,024&    32,768&   59,049&14,348,907& 1,048,576&           \\
\hline
$L=6$&  4,096&   262,144&  531,441&          &16,777,216&           \\
\hline
$L=7$& 16,384& 2,097,152&4,782,969&          &          &           \\
\hline
$L=8$& 65,536&16,777,216&         &          &          &           \\
\hline
$L=9$&262,144&          &         &          &          &           \\
\hline
\end{tabular}
\end{center}
\protect\caption{\label{table1}Number $N_{q,N}^{\tt alg1}(L)$ of
  basis states in {\tt alg1} as a function of strip width $L$ and the
  number $N$ of coupled Potts models. We indicate here these numbers
  up to the largest strip width for which we were able to diagonalise
  the transfer matrix.}
\end{table}

The reason that the trivial spin basis is still of some use is that it
does not explicitly break the $Z_q$ symmetry of the Potts
spins. Therefore, ${\bf T}$ contains both the even and the odd
sectors, and the various magnetic scaling exponents can easily be
extracted. Furthermore, {\tt alg1} has served as a check of the
results obtained from the optimised algorithms presented below.

In the case of the Ising model ($q=2$), states which are even and odd
under a global spin flip [cfr.~Eq.~(\ref{shift})] are given by
\begin{equation}
  |{\rm e}\rangle = \frac{1}{\sqrt{2}}
                    (|00 \cdots 0> + |11 \cdots 1>), \ \ \ \
  |{\rm o}\rangle = \frac{1}{\sqrt{2}}
                    (|00 \cdots 0> - |11 \cdots 1>).
\end{equation}
This is easily generalised to several layers of spins by simply
forming direct product states. For example, for two coupled models
we can define the states
\begin{equation}
  |{\rm eo}\rangle = |{\rm e}\rangle \otimes |{\rm o}\rangle, \ \ \ \
  |{\rm oo}\rangle = |{\rm o}\rangle \otimes |{\rm o}\rangle,
\label{prods}
\end{equation}
so that the gap between the largest eigenvalue in each of these sectors
and the largest eigenvalue in the totally symmetric
($|{\rm ee}\rangle = |{\rm e}\rangle \otimes |{\rm e}\rangle$) sector
defines two magnetic exponents, $x_h^{\rm eo}$ and $x_h^{\rm oo}$, which
are in general independent.

For the $q$-state Potts model the appropriate prescription reads 
\bea
  |{\rm e}\rangle &=& \frac{1}{\sqrt{q}}
    \left( |00 \cdots 0\rangle + |11 \cdots 1\rangle+\cdots +
           |(q-1)(q-1) \cdots (q-1)\rangle \right),\\
  |{\rm o}\rangle &=& \frac{1}{\sqrt{q}}
    \left( |00 \cdots 0\rangle +{\rm e}^{2i\pi\over q} |11 \cdots 1\rangle +
    \cdots+{\rm e}^{2i\pi{q-1\over q}} |(q-1)(q-1) \cdots (q-1)\rangle\right),
\nn
\eea
where the physical state is obtained from Eq.~(\ref{prods}) by taking the
real part.

\subsection{Algorithm {\tt alg2}}
\label{sec:alg2}

When the $Z_q$ symmetry of the Potts spins is factored out the size of
the basis can be dramatically reduced. The drawback of this approach,
however, is that the magnetic properties are lost. In other words, the
resulting transfer matrix does only have an even sector, but this is
still enough to extract finite-size estimates for the central charge
and the thermal exponent. (On the other hand, we shall see in the
following sub-sections that in the case of the algorithms {\tt alg3}
and {\tt alg4} it is possible explicitly to reconstruct the odd sector
from topological considerations.)

The algorithm {\tt alg2} was already used in Ref.~\cite{mp} to study
the random-bond Potts model, and before adapting it to the
case of several coupled Potts models we briefly recall its application
to the single-layered model.

The basic idea is that the $\delta_{\sigma_i,\sigma_j}$-type interactions
do not depend on the explicit values of the spins $\sigma_i$ and
$\sigma_j$. What matters is whether the two spins take identical or
different values. Therefore, the possible number of states for a row of $L$
spins is equal to the number of ways $b_L$ in which $L$ objects can be
partitioned into indistinguishable parts \cite{Wu97b}. 
With $m_{\nu}$ parts of $\nu$ objects each ($\nu = 1,2,\ldots$) 
this can be rewritten as
\begin{equation}
  b_L = \sum_{m_{\nu}=0}^{\infty} {'}
        \prod_{\nu=1}^{\infty}
        \frac{L!}{(\nu!)^{m_{\nu}} \, m_{\nu}!},
\end{equation}
where the primed summation is constrained by the condition
$\sum_{\nu=1}^{\infty} \nu m_{\nu} = L$. Alternatively one can
write \cite{mp}
\begin{equation}
  b_L = \sum_{i_2=1}^2 \sum_{i_3=1}^{m_3} \sum_{i_4=1}^{m_4} \cdots
        \sum_{i_L=1}^{m_L} 1,
\end{equation}
with $m_k = 1 + \max(i_2,i_3,\ldots,i_{k-1})$.
{}From the former representation the generating function is found as
\begin{equation}
  \exp({\rm e}^t - 1) = \sum_{n=0}^{\infty} \frac{b_n t^n}{n!},
\end{equation}
whence
\begin{equation}
  b_L = \frac{1}{\rm e} \sum_{k=0}^\infty \frac{k^L}{k!}.
\end{equation}

Yet another way of interpreting the $b_L$ is to notice
\cite{cj} that they are the total number of possible $L$-point
connectivities, including the non-well nested ones \cite{bn}. We shall
come back to the notion of well-nestedness when we discuss {\tt alg3}.

A major advantage of the basis used in {\tt alg2} as compared to the
trivial spin basis is that the $b_L$ do not depend on $q$. Thus, any
integer value of $q$ can be accessed with this algorithm. However, for
any fixed $q$ the size of the transfer matrices that have $L<q$ can be
further reduced.%
\footnote{This possibility was not discussed in Ref.~\cite{mp}.}
Namely, in this case the number of permissible states is truncated due
to the fact that $L$ objects cannot be partitioned into more than $L$
indistinguishable parts! In this way we identify the number of states
$N_{q,N}^{\tt alg2}(L)$ for $N=1$ Potts layers as a sum over Stirling
numbers of the second kind \cite{Gradshteyn}
\begin{equation}
  N_{q,1}^{\rm alg2}(L) = \sum_{m=1}^q {\cal S}_L^{(m)} =
  \sum_{m=1}^q \frac{1}{m!} \sum_{k=0}^m (-1)^{m-k}
  {m \choose k} k^L.
\end{equation}
Explicit values for $q=2,3,4$ are shown in Table \ref{table2}, and we
conjecture that asymptotically $N_{q,1}^{\tt alg2}(L) \sim q^L$. We recall
that the upper limit for general $q$ is $b_L$, and this increases
as $L^{\alpha L}$, where $\alpha$  is a constant of order unity
\cite{cj}. However, by comparing Tables 
\ref{table1} and \ref{table2} we see that even though the number of
states in algorithms {\tt alg1} and {\tt alg2} exhibit the same
asymptotic growth, {\tt alg2} is much more efficient than {\tt alg1}.

\begin{table}
\begin{center}
\begin{tabular}{|l||r|r|r||r|r|r||r|r|r||} \hline
$N_{q,N}^{\tt alg2}(L)$
     & $q=2$&   &         & $q=3$&      &         & $q=4$&        &          \\
\hline
$N$  &  1&     2&        3&    1&      2&        3&    1&        2&        3 \\
\hline
$L=2$&  2&     4&        4&    2&      3&        4&    2&        3&        4 \\
\hline
$L=3$&  4&    10&       20&    5&     15&       35&    5&       15&       35 \\
\hline
$L=4$&  8&    36&      120&   14&    105&      560&   15&      120&      680 \\
\hline
$L=5$& 16&   136&      816&   41&    861&   12,341&   51&    1,326&   23,426 \\
\hline
$L=6$& 32&   528&    5,984&  122&  7,503&  310,124&  187&   17,578&1,107,414 \\
\hline
$L=7$& 64& 2,080&   45,760&  365& 66,795&8,171,255&  715&  255,970&          \\
\hline
$L=8$&128& 8,256&  357,760&1,094&598,965&         &2,795&3,907,410&          \\
\hline
$L=9$&256&32,896&2,829,056&     &       &         &     &         &          \\
\hline
\end{tabular}
\end{center}
\protect\caption{\label{table2}Number $N_{q,N}^{\tt alg2}(L)$ of
  basis states in {\tt alg2}. For $N=2,3$ coupled models these numbers
  are shown up to the largest strip width $L$ for which we were able
  to diagonalise the transfer matrix.}
\end{table}

\subsubsection{Layer indistinguishability}
\label{sec:indist}

In the general case of $N$ coupled Potts models further progress can
be made by observing that since all layers interact symmetrically there is
an additional $S_N$ permutational symmetry. If we imagine numbering
the one-layer states by an integer $i=1,2,\ldots,N_{q,1}(L)$
a general $N$-layer state can be represented by
$(i_1,i_2,\ldots,i_N)$ where $i_1 \ge i_2 \ge \ldots \ge i_N$.
The total number of states for $N$ coupled models is then
\begin{equation}
  N_{q,N} = \sum_{i_1=1}^{N_{q,1}} \sum_{i_2=1}^{i_1}
            \sum_{i_3=1}^{i_2} \cdots \sum_{i_N=1}^{i_{N-1}} 1.
\end{equation}
For $N=2,3$ this is easily found to be
\begin{eqnarray}
  N_{q,2} &=& \frac12 \left( N_{q,1}^2 + N_{q,1} \right), \nonumber \\
  N_{q,3} &=& \frac16 \left( N_{q,1}^3 + 3 N_{q,1}^2 + 2 N_{q,1} \right).
  \label{23layers}
\end{eqnarray}
Again explicit values pertaining to {\tt alg2} are shown in Table
\ref{table2}.

\subsection{Mapping to random cluster model}
\label{sec:rand-cluster}

An altogether different approach for setting up the Potts model
transfer matrices consists in rewriting the partition function in
terms of extended objects (clusters and loops) rather than the local
spins. The resulting random cluster models \cite{Kasteleyn63} and loop
gases \cite{Baxter82} have the advantage that the specification of
their states no longer depends on the value of $q$. Instead $q$ enters
only through the fugacity of the non-local objects, and hence it can
be taken as a continuously varying parameter. Such reformulations of
the problem are especially convenient for making contact with the
predictions of conformal field theory, and in particular with the
perturbative expansion in powers of $(q-2)$ which has been studied in
Sec.~\ref{sec:RG}.

The practical implementation of transfer matrices for such non-local
degrees of freedom was pioneered by Bl{\"o}te and collaborators
\cite{Blote82,Blote89}. The basic idea is here that in two dimensions
the allowed connectivities of clusters and loops are strongly
constrained by topological considerations. Accordingly the size of the
corresponding transfer matrix is drastically reduced.

In the following two sub-sections we generalise these algorithms to
the case of $N$ coupled Potts models, and we give explicit details on
the implementation for $N=1,2,3$. The cluster algorithm {\tt alg3} has
previously been described for $N=1$ by Bl{\"o}te and Nightingale
\cite{bn}. It was discussed in more detail in Ref.~\cite{cj},
where it was also shown how to adapt it to the case of bond
randomness. The loop algorithm {\tt alg4} has to our knowledge not
previously been used to study the Potts model.%
\footnote{Details on the implementation of related but more
  complicated loop models can be found in Refs.~\cite{Blote89,jj_npb98}.}
It is however the most efficient algorithm that we know of, and in
fact it performs much faster than the spin basis algorithm for the
Ising model!

Before focusing on the concrete implementation of {\tt alg3} and
{\tt alg4}, which is the object of the following two sub-sections, we
dedicate this sub-section to developing the appropriate mappings of
the partition function. We need to consider the cases of two and three
coupled Potts models in turn, but it should be clear that the results
generalise straightforwardly to any number of models $N$.

\subsubsection{Two models}

The partition function for two coupled Potts models can be written as
\begin{equation}
  Z = \sum_{\{\sigma,\tau\}} \prod_{\langle ij \rangle}
      \exp({\cal H}_{ij}),
  \label{Z2}
\end{equation}
where ${\cal H}_{ij} = a (\delta_{\sigma_i,\sigma_j} + \delta_{\tau_i,\tau_j})
                     + b \delta_{\sigma_i,\sigma_j} \delta_{\tau_i,\tau_j}$.
Using the standard Fortuin-Kasteleyn trick \cite{Kasteleyn63} the
exponential of ${\cal H}_{ij}$ is turned into
\begin{equation}
  \exp({\cal H}_{ij}) = (1 + u_a \delta_{\sigma_i,\sigma_j})
    (1 + u_a \delta_{\tau_i,\tau_j}) 
    (1 + u_b \delta_{\sigma_i,\sigma_j} \delta_{\tau_i,\tau_j}), 
\end{equation}
where we have defined $u_a = {\rm e}^a - 1$ and $u_b = {\rm e}^b - 1$.
After some straightforward algebra we obtain
\begin{equation}
  \exp({\cal H}_{ij}) = k_0 +
    k_1 (\delta_{\sigma_i,\sigma_j} + \delta_{\tau_i,\tau_j}) +
    k_2 \delta_{\sigma_i,\sigma_j} \delta_{\tau_i,\tau_j},
  \label{expH2}
\end{equation}
where
\begin{equation}
  k_0 = 1, \ \ \ \ \
  k_1 = u_a, \ \ \ \ \
  k_2 = u_a^2 + u_b (1+u_a)^2.
  \label{weights2}
\end{equation}

Now imagine expanding the $\prod_{\langle ij \rangle}$ product in
Eq.~(\ref{Z2}), using Eq.~(\ref{expH2}). In this way we obtain a total of
$3^E$ terms ($E = {|\langle ij \rangle|}$ being the number of lattice
edges), each consisting of a product of $E$ factors of
$k_i$ ($i=0,1,2$) multiplying a
product of Kronecker deltas. Define ${\cal L}$ to be the graph
consisting of two copies (`layers') of the lattice on which the Potts
model is defined, one copy placed on top of the other. To each of the
terms in the expansion we associate a graphical representation ${\cal G}$ 
on ${\cal L}$ by colouring the lattice edges for which the
corresponding Kronecker deltas occur in the product. In other words,
${\cal G}$ takes the form of a two-layered bond percolation graph.

To reproduce the partition function (\ref{Z2}) we need to perform the
sum $\sum_{\{\sigma,\tau\}}$ over the spin degrees of freedom. Because
of the Kronecker deltas this is trivially done for each term ${\cal G}$,
yielding simply a factor of $q$ for each of the $C$ connected components
(`clusters') in ${\cal G}$.%
\footnote{Note that a single isolated site is to be counted as a
  cluster on its own.}
The factors of $k_i$ are easily accounted for by writing
$E = E_0 + E_1 + E_2$, where $E_j$ is the number of occurences
in ${\cal G}$ of a situation where precisely $j=0,1,2$ edges placed on top
of one another have been simultaneously coloured. The partition
function then takes the form
\begin{equation}
  Z = \sum_{\cal G} q^C k_0^{E_0} k_1^{E_1} k_2^{E_2}.
  \label{Z2new}
\end{equation}

On the self-dual line (\ref{sd2m}) the edge weights (\ref{weights2})
assume the simple form
\begin{equation}
  k_0 = 1, \ \ \ \ \
  k_1 = {\rm e}^a - 1, \ \ \ \ \
  k_2 = q.
\end{equation}
As has already been mentioned, the limit $a \to \infty$ is of special
interest. In this limit we can rewrite the partition function as
$Z = \exp(Ea) \tilde{Z}$, where $\tilde{Z}$ has the same form as
(\ref{Z2new}) but with the {\em finite} edge weights
\begin{equation}
  \tilde{k_0} = 0, \ \ \ \ \
  \tilde{k_1} = 1, \ \ \ \ \
  \tilde{k_2} = 0.
\end{equation}
It is worthwhile noticing that in this limit it suffices to specify
the colouring configuration of one of the layers to deduce that of the
other layer. Namely, whenever an edge in the first layer is coloured
its counterpart in the second layer has to be uncoloured,
and {\em vice versa}. Therefore, the configuration sum
$\sum_{\cal G}$ appearing in Eq.~(\ref{Z2new}) in fact runs over one
layer only.

\subsubsection{Three models}

In the case of three coupled Potts models the nearest-neighbour
interactions take the form
${\cal H}_{ij} = a \delta_1 + b \delta_2 + c \delta_3$,
where we have introduced the short-hand notation
\begin{eqnarray}
  \delta_1 &=& \delta_{\sigma_i,\sigma_j} + \delta_{\tau_i,\tau_j} +
               \delta_{\eta_i,\eta_j}, \nonumber \\
  \delta_2 &=& \delta_{\sigma_i,\sigma_j} \delta_{\tau_i,\tau_j} +
               \delta_{\sigma_i,\sigma_j} \delta_{\eta_i,\eta_j} +
               \delta_{\tau_i,\tau_j} \delta_{\eta_i,\eta_j}, \\
  \delta_3 &=& \delta_{\sigma_i,\sigma_j} \delta_{\tau_i,\tau_j}
               \delta_{\eta_i,\eta_j}.
\end{eqnarray}
The objects $\delta_1$, $\delta_2$, $\delta_3$ are easily shown to
obey the following relations
\begin{equation}
  \begin{array}{lllll}
    \delta_1 \delta_1 =   \delta_1 + 2 \delta_2, & \ \ &
    \delta_1 \delta_2 = 2 \delta_2 + 3 \delta_3, & \ \ &
    \delta_1 \delta_3 = 3 \delta_3, \\
    \delta_2 \delta_2 =   \delta_2 + 6 \delta_3, & \ \ &
    \delta_2 \delta_3 = 3 \delta_3, & \ \ &
    \delta_3 \delta_3 =   \delta_3.
  \end{array}
\end{equation}
Simple algebraic manipulations then lead to the following result,
generalising Eq.~(\ref{expH2}),
\begin{equation}
  \exp({\cal H}_{ij}) = k_0 + k_1 \delta_1 + k_2 \delta_2 + k_3 \delta_3
  \label{expH3}
\end{equation}
with edge weights
\begin{eqnarray}
  k_0 &=& 1, \nonumber \\
  k_1 &=& u_a, \nonumber \\
  k_2 &=& u_a^2 + u_b(1+u_a)^2, \nonumber \\
  k_3 &=& u_a^3 + 3 u_a u_b (1+u_a)^2 +
          (1+u_a)^3 \left[ u_c (1+u_b)^3 + 3 u_b^2 + u_b^3 \right].
  \label{weights3}
\end{eqnarray}
We recall that the Fortuin-Kasteleyn parameters are
$u_{a,b,c} = {\rm e}^{a,b,c} - 1$.
After the original spin degrees of freedom have been summed over, the
partition function takes on the form
\begin{equation}
  Z = \sum_{\cal G} q^C k_0^{E_0} k_1^{E_1} k_2^{E_2} k_3^{E_3}
  \label{Z3new}
\end{equation}
with a notation analogous to that employed for the case of two coupled
models. The graph configurations ${\cal G}$ now consist of three bond
percolation graphs stacked on top of one another.

Along the self-dual line (\ref{sd3m}) the edge weights (\ref{weights3})
simplify to
\begin{equation}
  k_0 = 1, \ \ \ \ \
  k_1 = {\rm e}^a - 1, \ \ \ \ \
  k_2 = q^{1/2} ({\rm e}^a - 1), \ \ \ \ \
  k_3 = q^{3/2}.
\end{equation}
Finally, in the limit $a \to \infty$ the modified partition function
$\tilde{Z} = \exp(-Ea) Z$ has the {\em finite} edge weights
\begin{equation}
  k_0 = 0, \ \ \ \ \
  k_1 = 1, \ \ \ \ \
  k_2 = q^{1/2}, \ \ \ \ \
  k_3 = 0.
\end{equation}

\subsection{Algorithm {\tt alg3}}

With the mapping to a random cluster model in hand we are now ready to
discuss the implementation of {\tt alg3}. The point of depart of this
algorithm is the form (\ref{Z3new}) [or (\ref{Z2new})] of the
partition function, in which the spin degrees of freedom have been
turned into non-local clusters.

For simplicity we consider first the case of a single Potts model,
with partition function $Z = \sum_{\cal G} q^C u_a^{E_1}$
\cite{Kasteleyn63}. To construct the
transfer matrix we need to specify a basis, so that the knowledge of
the state before and after the addition of a new degree of freedom
gives us enough information to compute the appropriate Boltzmann
weights, cfr.~Eq.~(\ref{add-a-row}). As shown by Bl\"{o}te and
Nightingale \cite{Blote82} this is achieved by specifying the
{\em connectivity} (with respect to the clusters in ${\cal G}$)
of the $L$ points in the last row of the strip. Since connections can
only be mediated by the lattice edges which have
previously been added there is a very powerful topological constraint
(known as {\em well-nestedness}) on the connectivity states. Namely,
given any four consecutive points, if the first is connected to the
third and the second to the fourth, then all four points must be
connected. Consequently, the number $c_L$ of allowed (well-nested)
$L$-point connectivities is less than the total number of
connectivities $b_L$ considered in Section \ref{sec:alg2}.

Using a recursive principle the well-nested connectivities can be
enumerated \cite{Blote82}, and they turn out to be nothing but the
Catalan numbers
\begin{equation}
  c_L = \frac{(2L)!}{L!(L+1)!} \sim 4^L \mbox{ for $L \gg 1$}.
\end{equation}
Since the number of Potts states enters only as a (continuous)
parameter in the partition function this is {\em independent} of $q$.
A mapping from the connectivities to the set of consecutive integers
$1,2,\ldots,c_L$ can also be established. More details on the
practical implementation of the transfer matrices can be found in
Ref.~\cite{cj}.

For $N>1$ layers of coupled Potts models we need to simultaneously
keep track of the connectivity state of each layer in order to compute
the factors of $q$ occuring in Eq.~(\ref{Z2new}) or (\ref{Z3new}).
To specify the
state of the layered system we can however take advantage of the
indistinguishability of the layers, cfr.~Eq.~(\ref{23layers}). The
resulting number of connectivity states $N_{N}^{\tt alg3}(L)$ used in
{\tt alg3} can be found in Table~\ref{table3}.

\begin{table}
\begin{center}
\begin{tabular}{|r|r||r|r|r||} \hline
  $L^{\tt alg3}$ & $L^{\tt alg4}$ &$N=1$ &      $N=2$ &      $N=3$ \\ \hline
              2 &              2 &     2 &          3 &          4 \\ \hline
              3 &              4 &     5 &         15 &         35 \\ \hline
              4 &              6 &    14 &        105 &        560 \\ \hline
              5 &              8 &    42 &        903 &     13,244 \\ \hline
              6 &             10 &   132 &      8,778 &    392,084 \\ \hline
              7 &             12 &   429 &     92,235 & 13,251,095 \\ \hline
              8 &             14 & 1,430 &  1,023,165 &            \\ \hline
              9 &             16 & 4,862 & 11,821,953 &            \\ \hline
\end{tabular}
\end{center}
\protect\caption{\label{table3}Number $N_N^{\tt alg3,4}(L)$
  of basis states in the even sector of algorithm {\tt alg3} and {\tt alg4}.
  For $N=2,3$ coupled models these numbers
  are shown up to the largest strip width $L$ for which we were able
  to diagonalise the transfer matrix. With {\tt alg4} it is possible to
  access larger strip widths $L^{\tt alg4} = 2(L^{\tt alg3}-1)$ than
  with {\tt alg3}, using the {\em same} number of states. Since for a
  strip of width $L$, {\tt alg4} needs to employ intermediary states
  of ($L+2$)-point connectivities, the size of the basis given here
  pertains to these intermediary states.}
\end{table}

\subsubsection{Magnetic properties}

At the expense of increasing the size of the basis it is possible
to generalise {\tt alg3} to treat the case of a Potts model in a
uniform magnetic field $H$ \cite{Blote82}. To this end, note that the
interaction with the field can be accounted for through the inclusion
of a term $H \sum_i \delta_{\sigma_i,\sigma_0}$ in the Hamiltonian,
where a so-called ghost spin $0$ of fixed value $\sigma_0 \equiv 1$ has
been introduced. Taking each site of the lattice to be connected to
$0$ through a `ghost edge', this has the usual form of a
nearest-neighbour interaction. The mapping to the random cluster model
therefore goes through in exact analogy with
Sec.~\ref{sec:rand-cluster}. In specifying the connectivities one now
needs both to keep track of which sites are connected (directly or
indirectly) to the ghost site and, at the same time, to specify how
the remaining sites are interconnected.

The extended $L$-point connectivity states thus defined can be ordered
and enumerated \cite{Blote82}, and their number $d_L$ is found to grow
asymptotically as $5^L$. For $H \to 0$ the transfer matrix 
has the following block form
\begin{equation}
  {\bf T} = \left[ \begin{array}{cc}
                {\bf T}^{00} & {\bf T}^{01} \\
                {\bf 0}      & {\bf T}^{11}
             \end{array} \right],
\end{equation}
where superscript 1 (0) refers to the (non-)ghost connectivities, and
the magnetic scaling dimension $x_H$ is obtained via Eq.~(\ref{fx})
from the largest eigenvalues of the ${\bf T}^{00}$ and ${\bf T}^{11}$ blocks.
The physical content of this relation is that by acting
repeatedly with ${\bf T}^{11}$ on some initial (row) state
$|{\bf v}_0 \rangle \neq 0$ one measures the decay of clusters
extending back to row 0. This must have the same spatial dependence as
the spin-spin correlation function and hence be related to $x_H$
\cite{Blote82}.

For $L=1,2,3,\ldots$ the size of the magnetic block ${\bf T}^{11}$
is $d_L - c_L$ = 1,3,10,37,146,599,\ldots. {}For three coupled models
this means that computations with one, two or all three layers in the
magnetic sector are feasible for strip widths up to $L=6$. Since our
most refined algorithm {\tt alg4} can access $L=10$ we turn our
attention to this next.

\subsection{Algorithm {\tt alg4}}
\label{sec:alg4}

\begin{figure}
\begin{center}
 \leavevmode
 \epsfysize=80pt{\epsffile{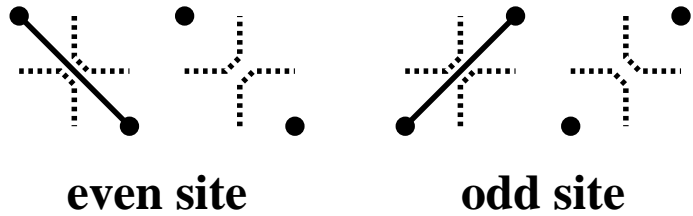}}
 \end{center}
 \protect\caption[3]{\label{fig:medial}The relation between the random
 cluster model on 
 the square lattice and the loop model on the corresponding medial
 graph. The clusters consist of connected components of coloured edges
 (thick lines) or isolated sites (filled circles). Loops on the medial
 graph (dashed lines) are defined by the convention that they wrap
 around the cluster boundaries.}
\end{figure}

The configurations of the random cluster model on some graph are in
a one-to-one correspondence with configurations of a fully-packed loop
model on the {\em medial graph} \cite{Baxter82}.
By definition the nodes of the medial graph
are situated at the mid-points of the edges of the original graph. In
our case the graph on which the Potts model is defined is simply the
square lattice, and the medial graph is then nothing but another
square lattice that has been rotated through $\pi/4$ with respect to
the original one, and rescaled by a factor of $1/\sqrt{2}$.
Bipartitioning the medial graph into even and odd sublattices the
precise correspondence is as shown in Fig.~\ref{fig:medial}.

The partition function for $N$ coupled Potts models can now be
rewritten in the loop picture by using Euler's relation
\cite{Baxter82}. Namely, for each layer the number of clusters ($C$),
loops ($l$), coloured edges ($e$) and sites ($s$) are related by
$2C = l + s - e$. Thus, in the case of two models, Eq.~(\ref{Z2new})
is turned into
\begin{equation}
  Z = q^s \sum_{\cal G} q^{l/2} k_0^{E_0}
      \left( \frac{k_1}{q^{1/2}} \right)^{E_1}
      \left( \frac{k_2}{q} \right)^{E_2}.
  \label{Z2loop}
\end{equation}
Note that the factor of $q^{-e}$ has been redistributed using
$e = E_1 + 2 E_2$.
Similarly, for the three-layered model Eq.~(\ref{Z3new}) is
transformed into
\begin{equation}
  Z = q^{3s/2} \sum_{\cal G} q^{l/2} k_0^{E_0}
      \left( \frac{k_1}{q^{1/2}} \right)^{E_1}
      \left( \frac{k_2}{q} \right)^{E_2}
      \left( \frac{k_3}{q^{3/2}} \right)^{E_3}.
  \label{Z3loop}
\end{equation}

When constructing the transfer matrix the advantage of this
representation is that less information is needed to keep track of the
loops than was the case in the cluster representation. Roughly
speaking this is because the loop model is {\em fully packed}, so that
one does not need to waste information to specify where is `the empty
space' in between the clusters.

The strip width $L$ is now defined as the number of `dangling ends'
resulting from cutting the loops in between two rows of sites on the
medial graph. A sparse-matrix decomposition can then be made by adding
one vertex at a time, rather than one edge as was the case in the
cluster representation. This has been illustrated in
Fig.~\ref{fig:seam} for a strip of width $L=6$. For each added vertex
there are two possible configurations of the loop segments that must
be summed over, {\em cfr.}~Fig.~\ref{fig:medial}.
When adding the first vertex of a new row the
number of dangling ends increases from $L$ to $L+2$, and it only goes
back to $L$ once the full row of $L$ vertices has been
completed. Therefore, we need to work interchangingly with bases
specifying the loop connectivities of $L$ and $L+2$ dangling ends
\cite{Blote89,jj_npb98}.

By definition of the loops, each of the dangling ends is connected to
exactly one other dangling end. In particular the strip width $L$ must
be {\em even}. Employing basically the same recursive argument as for
the clusters \cite{Blote82} it is found that the possible number of
connectivities among $L$ dangling ends is now only $c_{L/2}$
\cite{Blote89}. For $L \gg 1$ this increases as $2^L$, and in fact the
loop representation of the Potts model is even more efficient than the
spin representation of the Ising model!%
\footnote{Using Stirling's formula a more accurate estimation of the
 asymptotic behaviour of $c_{L/2}$ is found as 
 $2^L (L+2)^{-3/2} \sqrt{8/\pi}$. This approximation is asymptotically
 exact in a strict sense, and its relative precision is better than
 10 \% for $L \ge 6$.}
This is witnessed by Table~\ref{table3}, where we show some explicit values of
$N_N^{\tt alg4}(L) = N_N^{\tt alg3}\big( (L+2)/2 \big)$.

At this point a brief comment on the boundary conditions is in
order. Since the medial lattice is rotated through $\pi/4$ with
respect to the original one, the imposition of periodic boundary
conditions on the loops is not {\em a priori} equivalent to the
boundary conditions hitherto used for the clusters. Indeed, these two
possibilities are connected by a modular transformation, as should be
clear from Fig.~\ref{fig:seam}. The consistency between the results
obtained from {\tt alg3} and {\tt alg4} will thus serve as a useful
check of the modular invariance of the critical system under
investigation.

To implement Eqs.~(\ref{Z2loop}) and (\ref{Z3loop}) for several
coupled Potts models we need to keep track of the edge weights on the
original lattice as well as the number of closed loops on the medial
graph. As shown in Fig.~\ref{fig:medial} the loop configuration
suffices to determine the positions of the coloured edges on the
original lattice, and so the edge weights can easily be determined
from each vertex appended to the medial graph. By the same token the
single-layer algorithm furnishes an even more efficient way of
performing transfer matrix calculations for the random-bond Potts model
than the one presented in Ref.~\cite{cj}.

\subsubsection{Magnetic properties}

In {\tt alg4} the magnetic sector of the transfer matrix is constructed
by using the fact that the spin-spin correlator maps onto a disorder
operator under duality \cite{cj}. For a single Potts model at the
self-dual (critical) point, or for several coupled Potts models along
the self-dual lines described in Sec.~\ref{sec:duality}, this leads
to a relation between the magnetic scaling exponent and the largest
eigenvalue of a modified transfer matrix on which {\em twisted
boundary conditions} have been imposed.

Defining the local order parameter as
\begin{equation}
  M_a(r) = \left( \delta_{\sigma(r),a} - \frac{1}{q} \right),
  \ \ \ \ a=1,\ldots,q.
  \label{orderparameter}
\end{equation}
it is easily seen that the magnetic two-point correlator
$G_{aa}(r_1,r_2) = \langle M_a(r_1) M_a(r_2) \rangle$ is proportional
to the probability that the points $r_1$ and $r_2$ belong to the
same cluster. Let us briefly recall that any given configuration of
the random clusters is dual to one where each coloured edge in the
direct lattice is intersected by an uncoloured edge in the dual
lattice, and {\em vice versa}. Taking the two points $r_1$ and $r_2$
to reside at opposite ends of the cylinder, the graphs contributing to
$G_{aa}(r_1,r_2)$ thus correspond to dual graphs where clusters are
forbidden to wrap around the cylinder. This is equivalent to computing
the dual partition function with twisted boundary conditions
\begin{equation}
  \sigma \to (\sigma+1) \mbox{ mod } q
\end{equation}
across a {\em seam} running from $r_1$ to $r_2$.
By permuting the Potts spin states the shape of this seam
can be deformed at will as long as it connects $r_1$ and $r_2$.

A realisation of the Potts model transfer matrix in the presence of a
seam was first described in Ref.~\cite{cj} within the context of {\tt alg3}. 
Since there is a one-to-one correspondence between clusters and loops
that wrap around the cylinder these ideas can also be applied to
{\tt alg4}.

\begin{figure}
\begin{center}
 \leavevmode
 \epsfysize=400pt{\epsffile{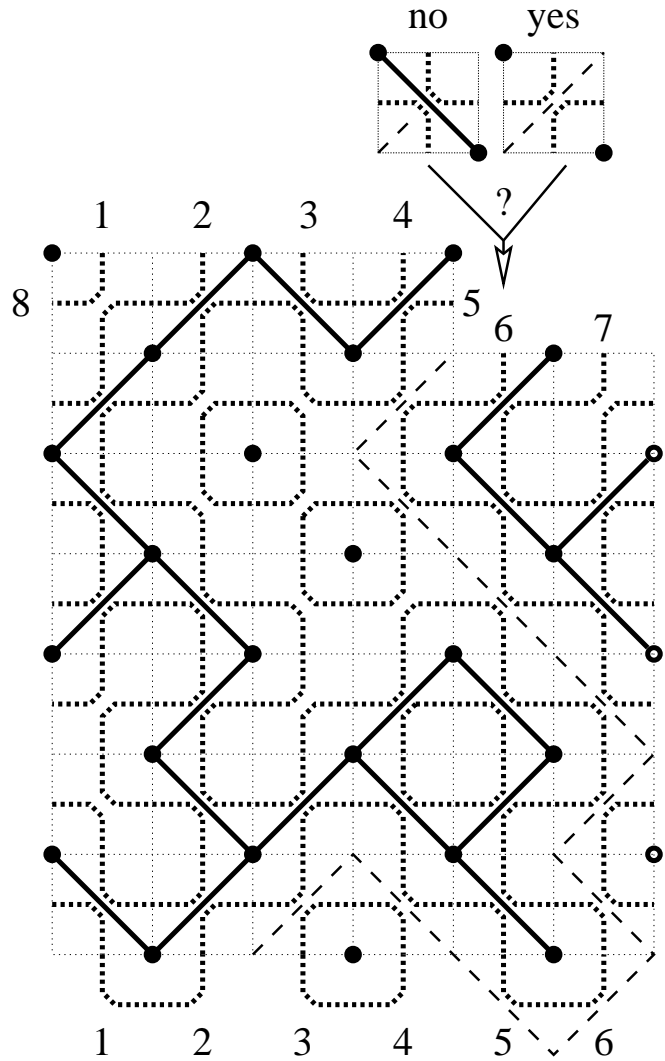}}
 \end{center}
 \protect\caption[3]{\label{fig:seam}Computation of the partition
 function with 
 twisted boundary conditions. Coloured edges (solid linestyle) connect
 the sites of the direct lattice (filled circles) so as to form
 clusters. The loops on the medial lattice (short boldface dashes) surround
 the clusters, and both cluster and loops are subject to periodic
 boundary conditions 
 (indicated by open circles) across the strip of width $L=6$. A seam
 (long dashes) connecting sites of the dual lattice prevents the
 clusters and loops from wrapping around the cylinder. By adding a
 single vertex at a time the transfer matrix can be decomposed as a
 product of sparse matrices. Each multiplication by a single-vertex
 matrix corresponds graphically to augmenting the `jigsaw puzzle' by
 one of the two `pieces' shown in the top. In the situation at hand, one
 of these is forbidden by the twisted boundary conditions.}
\end{figure}

In Fig.~\ref{fig:seam} we show a configuration contributing to the
partition function with twisted boundary conditions. As the square
lattice is self-dual we shall take the clusters and loops to live on
the direct lattice and its medial lattice respectively. The extended
connectivity state of the $L$ (resp.~$L+2$) dangling ends is the
direct product of the $c_{L/2}$ possible connectivities mediated by
the loops,
and an integer specifying the position of the seam. The seam is a path
inside the infinite {\em dual} cluster spanning the length of the
cylinder, and after the addition of each vertex it must be updated
according to the invariant that no loop closure take place across the seam.

Now consider the situation shown on Fig.~\ref{fig:seam} where we are
about to add the fifth vertex of the top row. Since this is an even site
the two possibilities for the loop configuration are those shown in
the left part of Fig.~\ref{fig:medial}. However, the first of these
would lead to a forbidden loop closure, as witnessed by the fact that
the seam is `trapped' and cannot be updated. The corresponding entry
in the transfer matrix is therefore zero, and only the graph realising
the second possibility contributes.

Using the numbering of the $L$ (resp.~$L+2$) dangling ends of an
(in)complete row shown at the bottom (top) part of Fig.~\ref{fig:seam}
it is easily seen that the seam position is always immediately to the
right of an {\em even} end in every other row, and to the
right of an {\em odd} end in the remaining rows. Therefore, in all
cases the number of permissible seam positions is equal to {\em half} the
number of dangling ends. The total number of extended $L$-point
connectivity states is therefore
\begin{equation}
  \frac{L c_{L/2}}{2} = \frac{L!}{\left( \frac{L}{2} - 1 \right)!
                                  \left( \frac{L}{2} + 1 \right)!}
                      \sim 2^L .
\end{equation}
Taking into account that for a strip of width $L$ we need intermediate
states of $(L+2)$-point connectivities, the size
$N^{\rm alg4}_{N_{\rm mag}}$ of the transfer matrix for three coupled
models  with $N_{\rm mag}=1,2,3$ layers in the odd sector is as shown
in Table \ref{tab:alg4mag}. Note that for $N_{\rm mag}=1,2$ two of the
layers are indistinguishable in the sense of Sec.~\ref{sec:indist},
whilst for $N_{\rm mag}=3$ all three layers are indistinguishable.

\begin{table}
\begin{center}
\begin{tabular}{|r||r|r|r||} \hline
  $L$ & $N_{\rm mag}=1$ & $N_{\rm mag}=2$ & $N_{\rm mag}=3$ \\ \hline
   2  &              12 &              20 &              20 \\ \hline
   4  &             225 &             600 &             680 \\ \hline
   6  &           5,880 &          22,344 &          30,856 \\ \hline
   8  &         189,630 &         930,510 &       1,565,620 \\ \hline
  10  &       6,952,176 &      41,451,696 &      83,112,744 \\ \hline
\end{tabular}
\end{center}
\protect\caption{\label{tab:alg4mag}Number
  $N^{\tt alg4}_{N_{\rm mag}}(L)$ of basis states needed by {\tt alg4}
  to access the scaling exponents of the three-layered system that
  correspond to the insertion of a
  magnetisation operator in $N_{\rm mag}=1,2,3$ of the layers.}
\end{table}

A very important remark pertains to the proper choice of the initial
vector $|{\bf v}_0 \rangle$ used for finding the largest
eigenvalue, {\em cfr.}~Eq.~(\ref{iterate}). Namely, when more than one
layer is in the odd sector the seam positions on all layers must
coincide at points $r_1$ and $r_2$. The reason why this is so can
readily be illustrated for the case of two coupled models.
With the initial vector (symbolically) chosen as
$|{\bf v}_0 \rangle = |1,1,1,\ldots\rangle \otimes |1,1,1,\ldots\rangle$
one would observe the asymptotic scaling of the correlator
\begin{equation}
 \left \langle \sum_{r_1,r_1',r_2,r_2'}
   M_a^{(1)}(r_1) M_a^{(2)}(r_1') M_a^{(1)}(r_2) M_a^{(2)}(r_2')
 \right \rangle
 \sim
 \left \langle \sum_{r_1,r_2} M_a(r_1) M_a(r_2) \right \rangle^2,
\end{equation}
where the summations run over the $L$ points at each extremity of the
cylinder. At large distances $|r_2 - r_1| \gg 1$ one would expect this to scale
with the exponent $2 x_H^{(1)}$, where $x_H^{(1)}$ is just the scaling
dimension of the usual magnetisation operator. However, with 
$|{\bf v}_0 \rangle = |1,0,0,\ldots\rangle \otimes |1,0,0,\ldots\rangle$
one would instead observe the scaling of
\begin{equation}
 \left \langle \sum_{r_1,r_2}
   M_a^{(1)}(r_1) M_a^{(2)}(r_1) M_a^{(1)}(r_2) M_a^{(2)}(r_2)
 \right \rangle,
\end{equation}
and this should decay as $|r_2 - r_1|^{-2 x_H^{(2)}}$, where $x_H^{(2)}$
is the sought scaling dimension of the {\em local} operator
$M_a^{(1)}(r) M_a^{(2)}(r)$. Indeed this is the simplest example of a
{\em multiscaling exponent} as discussed by Ludwig \cite{ludwig} in
the context of the random-bond Potts model ($N \to 0$).

The proper choice of $|{\bf v}_0 \rangle$ is tantamount to anchoring
all the seams at the point $r_1$. When applying Eq.~(\ref{iterate})
one should then theoretically project
${\bf T}^{|r_2-r_1|} |{\bf v}_0 \rangle$ out on a state with a
definite and identical seam position for all the layers before taking
the norm. However, since we expect all the non-zero entries of the iterated
vector to grow as $\exp \big( |r_2-r_1| \lambda(L) \big)$ this would
just correspond to multiplying by a constant before taking the
logarithm. Clearly such a constant would not contribute to the
computed value of $\lambda(L)$, and so the projection can be omitted.
%
% JLJ
% Maybe I should write a small note on why the string dimensions are
% non-local objects (and therefore uninteresting?)...

\section{Numerical results}
\label{sec:results}

Using the transfer matrix algorithms just described we are able to
find the effective values of the central charge and of the various
thermal and magnetic scaling dimensions, all as functions of the
coupling constant $a$, parametrising the position on the self-dual
lines identified in Sec.~\ref{sec:duality}. We are furthermore able to
monitor the scale dependence of these quantities by changing the
strip width $L$.

Let us briefly recapitulate the physical information that we hope to
extract from these data. First, we should like to provide compelling
evidence that, for $N>2$ coupled Potts models, a novel, non-trivial
critical fixed point is located on the self-dual line, in the limit
$a \to \infty$. We shall presently describe how such a conclusion may
be attained from our numerical data. Second, we aim at fixing
the values of the critical exponents at this fixed point. This is done
by taking the limit $a \to \infty$ explicitly in the transfer
matrices, in order to obtain high-precision data for rather large
strip widths (up to $L=12$ for $N=3$). Extrapolating these data to the
infinite system limit $L\to\infty$ we shall be able to identify the
$a \to \infty$ fixed point with the one obtained from perturbative CFT
in Sec.~\ref{sec:RG}, and to associate the numerically obtained
scaling dimensions with those of physical operators in the continuum
limit. Third, we provide an independent check of the criticality of
the fixed point under consideration. This is done by verifying the
existence of scaling laws, using Monte Carlo simulations directly in
the limit $a \to \infty$. As a by-product we extract values of the scaling
dimensions that agree with the more precise estimates obtained from
the transfer matrices. And fourth, we use our transfer matrices to
inquire further into the structure of the (presently unknown) CFT
governing the critical behaviour of the $a \to \infty$ model. To this
end we take a closer look at the higher eigenvalues in the even
sector. These data determine the scaling dimensions of less relevant
operators in the Verma module, and they give crucial information of 
the operator content and descendence structure of the sought
CFT. Finally, a comparison of the results obtained from algorithms
{\tt alg3} and {\tt alg4} serves as a check of the modular invariance
of the theory.

Although our main interest is $N=3$ coupled models we have also
produced numerical results for the case of two coupled $q$-state Potts
models, for several values of $q \in [2,4]$. For $q=2$ this is the
Ashkin-Teller model. This model is critical on the entire half line
$a \in [0,\infty]$, and it provides a useful check of our algorithms
since the critical exponents are known exactly as a function of
$a$ \cite{Kadanoff,Baxter82}. For $q>2$, on the other hand, Vaysburd
\cite{vays} has predicted a dynamical mass generation and thus a
non-critical model. Since this prediction was made under several
assumptions it is reassuring to see that our numerics confirms it. At
the same time, our $N=2$ data provide a clear illustration of the
difference between a first order phase transition scenario and a
second order one.

\subsection{Finding critical points from effective exponents}

The key to the search for critical fixed points is, of course,
Zamolodchikov's $c$-theorem \cite{zamo}. This theorem states that the
effective central charge (the $c$-function) decreases along the RG
flow and is stationary at its fixed points. In view of our definition
(\ref{c-function}) of the $c$-function, this is equivalent to the
familiar statement that the $\beta$-function vanishes at a fixed point.

A local minimum (maximum) of the effective central charge thus corresponds to
a (un)stable fixed point. When, as is the case here, the RG flow
takes place in a multi-dimensional space of coupling constants one can
also imagine the existence of saddle points, corresponding to a
partially stable fixed point. But generically we expect the flow to
take place {\em along} the self-dual line, and accordingly we can
limit the search to this one-dimensional self-dual submanifold. This
assumption is nicely corraborated by our numerical results. Indeed we
find that in general any motion perpendicular to the self-dual line
leads to a {\em decrease} of the central charge [see, {\em e.g.},
Fig.~\ref{cc2im}.b and Fig.~\ref{cc33pd} below]. Hence, invoking duality,
this line must serve as a mountain ridge for the central charge.

A notable exception to this scenario happens for the
Ashkin-Teller model, where the self-dual half line $a \in [0,\infty]$
bifurcates into two mutually dual lines of critical points at
$a=0$. However, this anomaly is very clearly signalled by the
vanishing of the $c$-function of the other self-dual half line,
$a \in [-\infty,0[$. We have not observed a similar situation for any
other value of $q$, or of $N$.

Once a candidate for a fixed point has been identified the question
arises whether it is critical or not. To address this point it is
useful to examine the dependence of $c_{\rm eff}$ on the system size
$L$. If a critical point is involved, the finite-size estimates
$c_{\rm eff}(L)$ should converge very rapidly, with increasing $L$,
to the true value of the
central charge at this point. On the other hand, if the point is not
critical a finite correlation length $\xi$ must be involved. As has
been shown by Bl{\"o}te and Nightingale \cite{bn} for the particular case
of a first order phase transition, Eq.~(\ref{fc}) for the finite-size
scaling of the free energy must then replaced by 
\beq
  f_0(L) = f_0(\infty) - {\rm const.}
  \frac{{\rm e}^{-L/\xi}}{L^2} + \cdots.
  \label{f1o}
\eeq
This in turn implies that the effective central charge, {\em i.e.},
the one measured assuming the scaling form (\ref{fc}), will vanish
exponentially as a function of the strip width $L$. We shall soon see that
our numerical data distinguish very easily between these two situations.

Complementary information about the location of possible fixed points
is provided by the effective values of the thermal and magnetic
scaling dimensions $x_{\rm eff}(L)$ along the self-dual line. In many
situations a fixed point is signalled by the fact that the curves
$x_{\rm eff}(L)$ and $x_{\rm eff}(L+1)$ intersect at some coupling
$a(L)$. The finite-size estimates $a(L)$ then converge very rapidly
towards the true value of the fixed-point coupling, as $L \to \infty$.
Indeed, this is nothing but the well-known technique of
phenomenological renormalisation \cite{Night76}.
Evidently a fixed point can also be characterised as a point of high
symmetry. It is thus hardly surprising to see from our numerics, that
in many cases local extrema of $x_{\rm eff}$ as a function of $a$
serve equally well to locate the position of the fixed point.

As was the case for the central charge, once a fixed point has been
located its precise nature (critical or not) can be inferred by
observing the size-dependence of $x_{\rm eff}(L)$. Since a finite
correlation length implies an exponential, rather than a power law,
decay of the various correlators, effective values of the scaling
dimensions extracted from Eq.~(\ref{fx}) will be observed to vanish
exponentially with $L$, if $\xi < \infty$.

\subsection{Two coupled models}

\subsubsection{Ashkin-Teller model}

The case of two coupled Ising models is nothing but the well-known
Ashkin-Teller model \cite{AT}. We shall nevertheless begin by
investigating this case in order to demonstrate the method outlined
above, and to see how we can reproduce known results.

\begin{figure}
\begin{center}
\leavevmode
\epsfysize=200pt{\epsffile{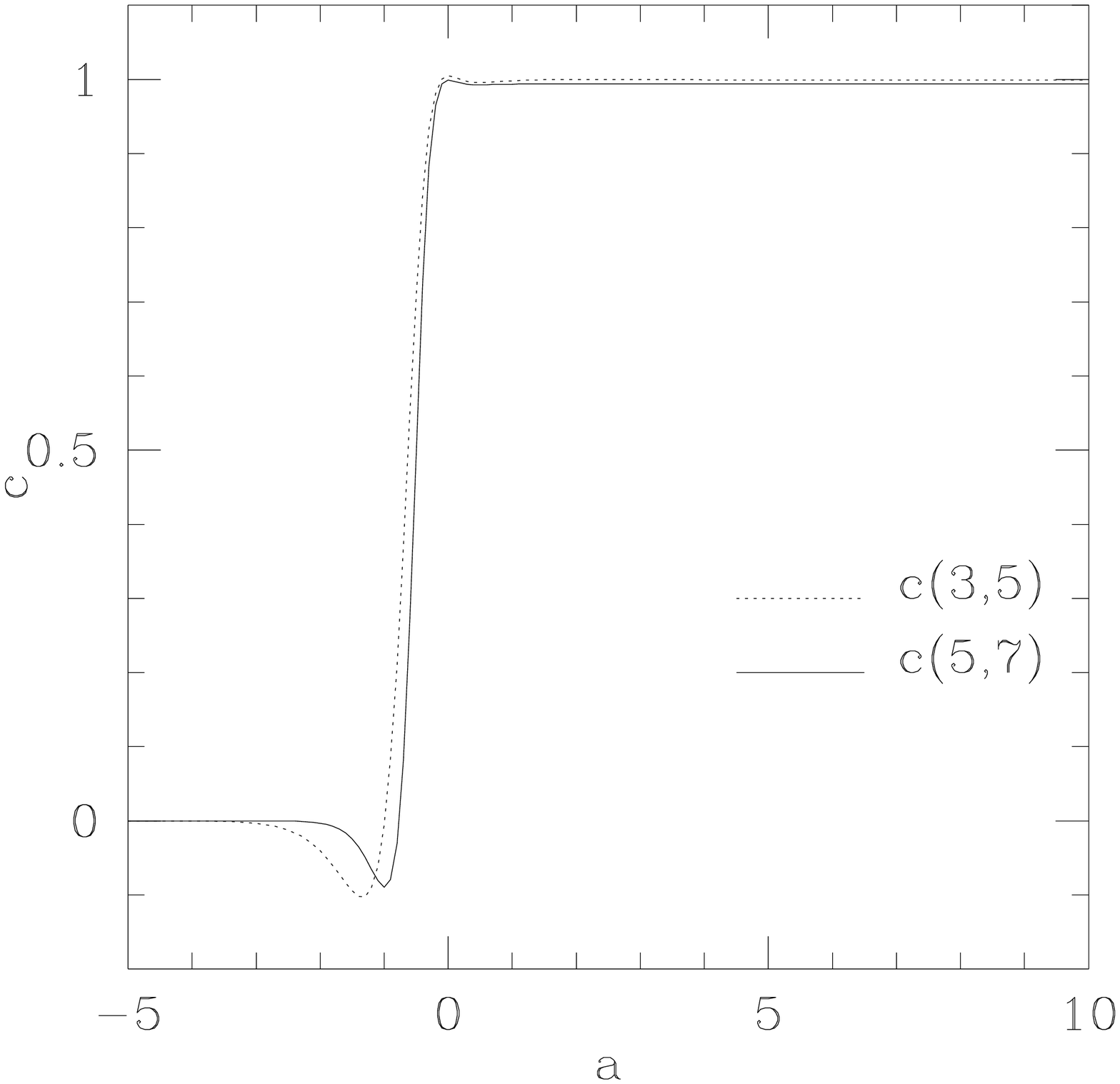}}
\epsfysize=200pt{\epsffile{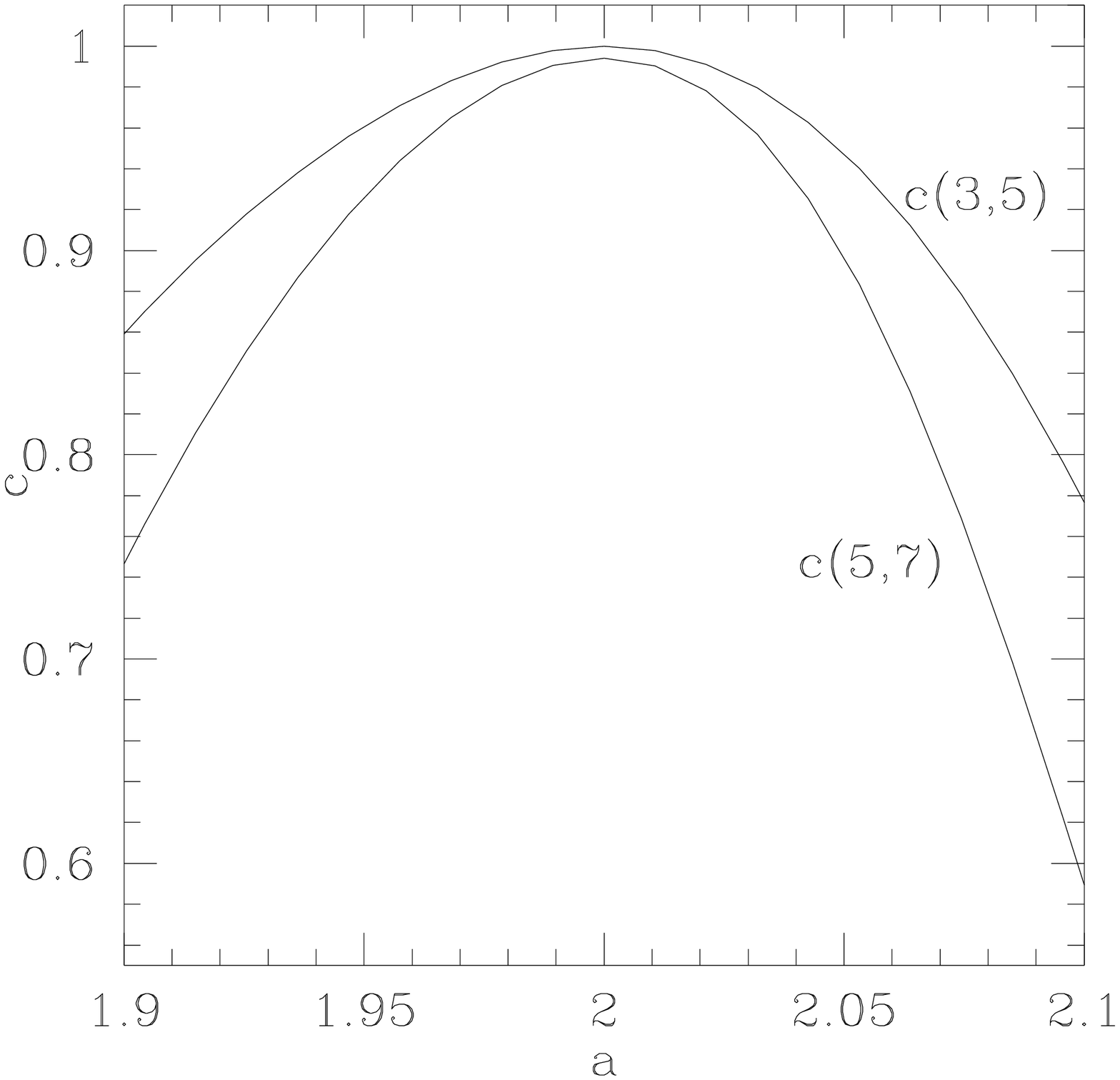}}
\end{center}
\protect\caption[1]{\label{cc2im}Central charge for two coupled Ising
 models. (a) shows the effective central charge as a function of the parameter
 $a$. (b) corresponds to moving perpendicularly off the self-dual line.}
\end{figure}

As has been shown by Baxter \cite{Baxter82} the isotropic
Ashkin-Teller model can be mapped to a staggered eight-vertex
model. This fact that this model is staggered seems to impede its
solvability, but on the self-dual line given by Eq.~(\ref{sd2m}),
$a \in [-\infty,\infty]$, the eight-vertex model becomes
{\em homogeneous} and hence soluble. For $a \ge 0$ the model is
critical, with the following exact values of the critical exponents
\cite{Kadanoff}:
\beq
  c         = 1, \ \ \ \
  x_t       = \frac{1}{2-y}, \ \ \ \
  x_H^{(1)} = \frac{1}{8}, \ \ \ \
  x_H^{(2)} = \frac{1}{8-4y},
  \label{AT-exact}
\eeq
where
\beq
  y = \frac{2}{\pi} {\rm arccos} \left(
  {\rm e}^{-a} + \frac12 {\rm e}^{-2a} - \frac12 \right).
\eeq

In Fig.~\ref{cc2im}.a we show the numerical values for the effective
central charge, in the form of the three-point fits,
{\em cfr.}~Eq.~(\ref{fc2}).
For every $a < 0$ the central charge tends to zero in the large-system
limit, indicating non-critical behaviour, and for $a \ge 0$ it
approaches unity, in complete agreement with Eq.~(\ref{AT-exact}).
The transfer matrix has also been diagonalised directly at $a=\infty$,
again with the result $c \simeq 1$.

\begin{figure}
\begin{center}
\leavevmode
\epsfysize=200pt{\epsffile{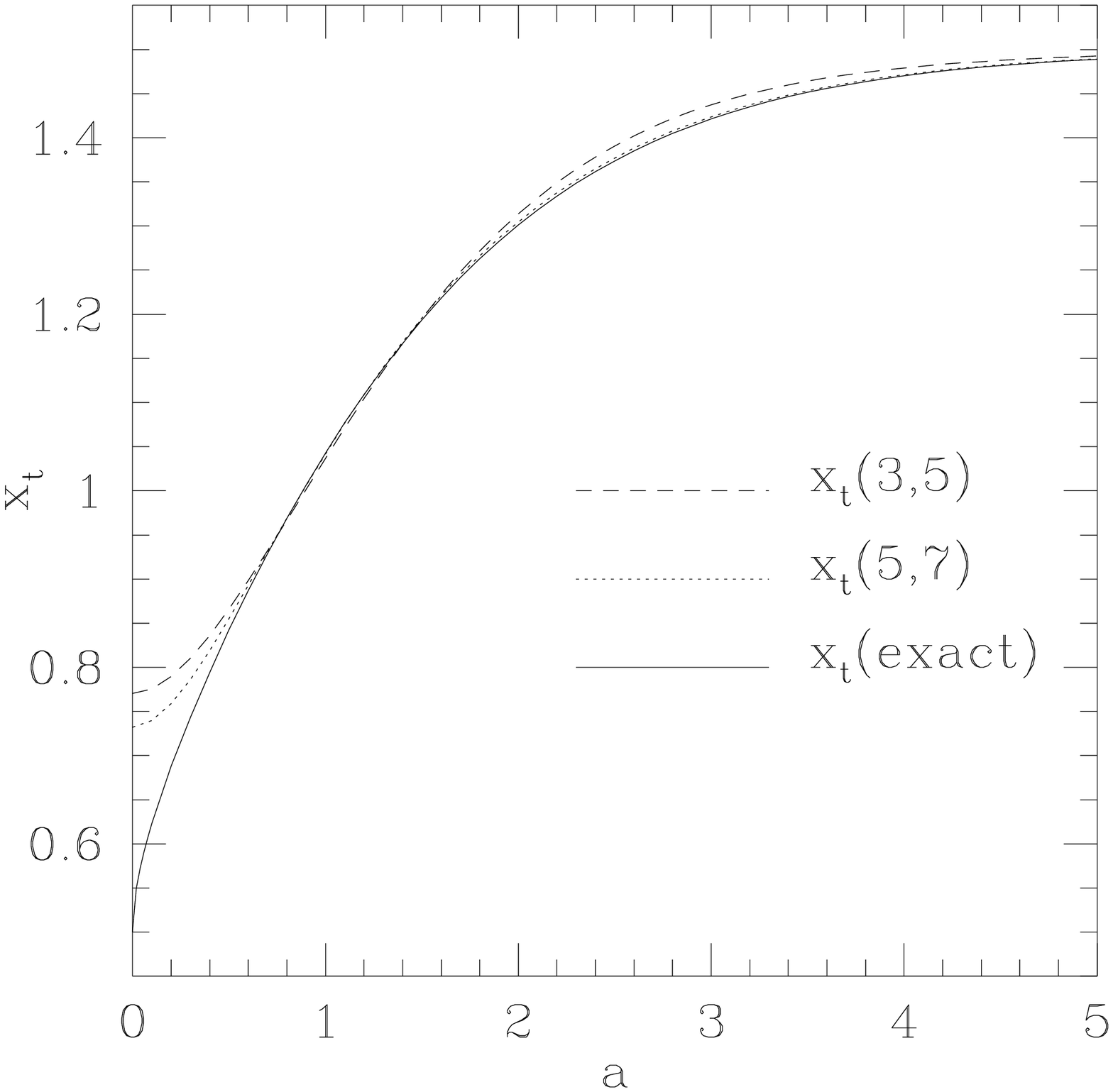}}
\epsfysize=200pt{\epsffile{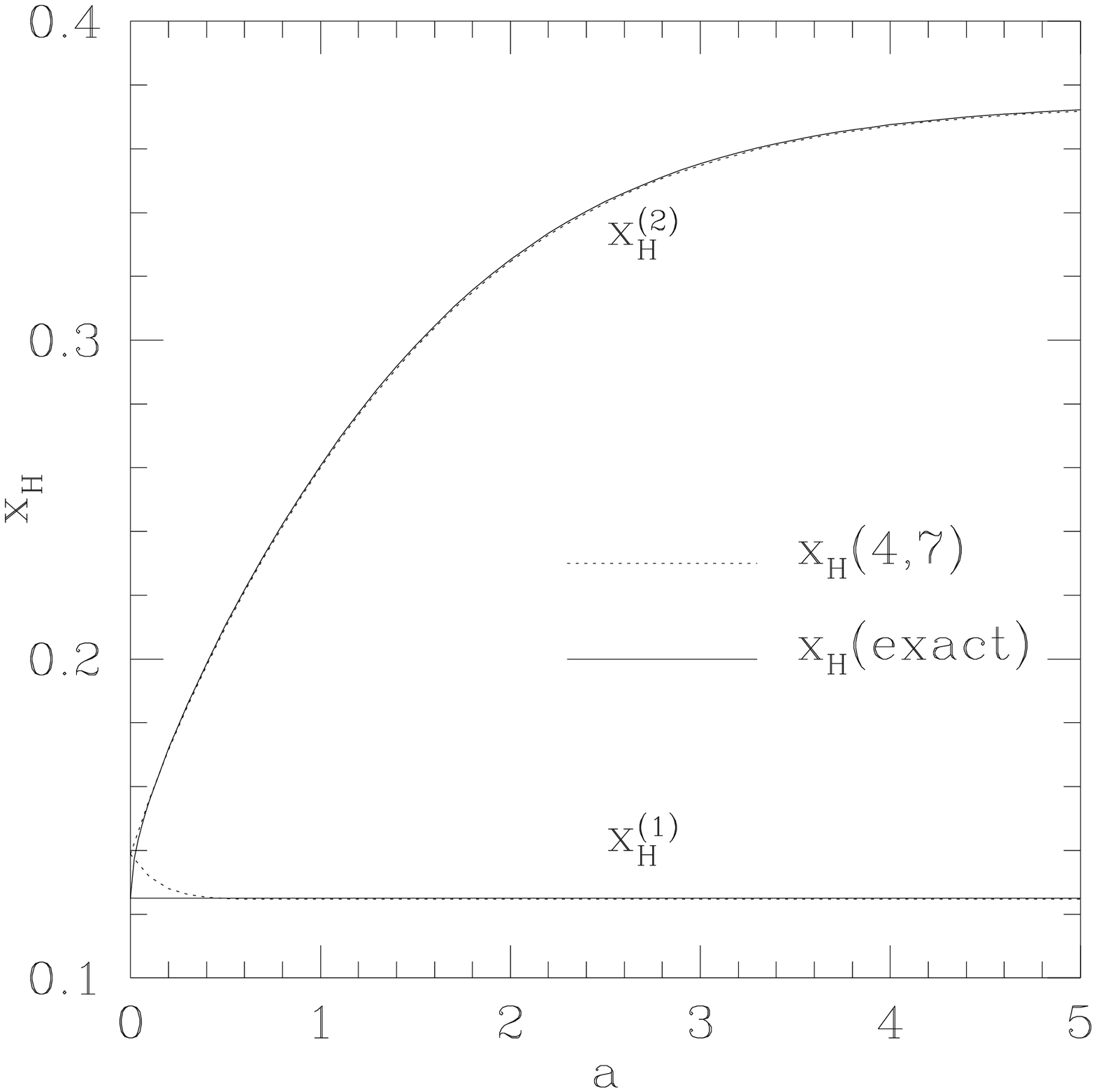}}
\end{center}
\protect\caption[2]{\label{x2im}Thermal dimension $x_t$ and magnetic
 dimensions $x_H^{(1)}$, $x_H^{(2)}$ for two coupled Ising models.}
\end{figure}

Fig.~\ref{cc2im}.b depicts the behaviour of the central charge along a
line perpendicular to the self-dual line at the point $a=2$. The value
$a=2$ is chosen arbitrary, and we observe similar curves for other
values of $a>0$. As we move into the non-self dual regime the central
charge decreases, and the larger the strip width the faster the decrease.
In the $L\to\infty$ limit we should have $c_{\rm eff}=0$ exactly.
We can therefore conclude that the self-dual line
indeed acts as a ridge of the central charge, thus confirming the
hypothesis that the RG flow is along that line.  We also observed
that for $a<0$ the critical line bifurcates into two mutually dual
Ising-like lines \cite{Baxter82}.

Next, in Fig.~\ref{x2im}, we present measured values of the thermal
scaling dimension, $x_t$, as well as the two magnetic ones,
$x_H^{(1)}$ and $x_H^{(2)}$. As the $\beta$-function is exactly zero
we can expect very accurate values, and as the comparison with the
analytical results (\ref{AT-exact}) shows this is indeed the case. The
only exception is the point $a=0$, which corresponds to the 4-state
Potts model. Due to the presence of a marginal operator there are
strong logarithmic corrections \cite{Cardy-log}, and the analytical
results $x_t = 1/2$ and $x_H^{(1)} = x_H^{(2)} = 1/8$ are not accurately
reproduced. Finally, by taking the explicit $a\to\infty$ limit in the
transfer matrices we have verified the results
$x_t = 3/2$, $x_H^{(1)} = 1/8$ and $x_H^{(2)} = 3/8$.

\subsubsection{Two coupled 3-state Potts models}

\begin{figure}
\begin{center}
\leavevmode
\epsfysize=200pt{\epsffile{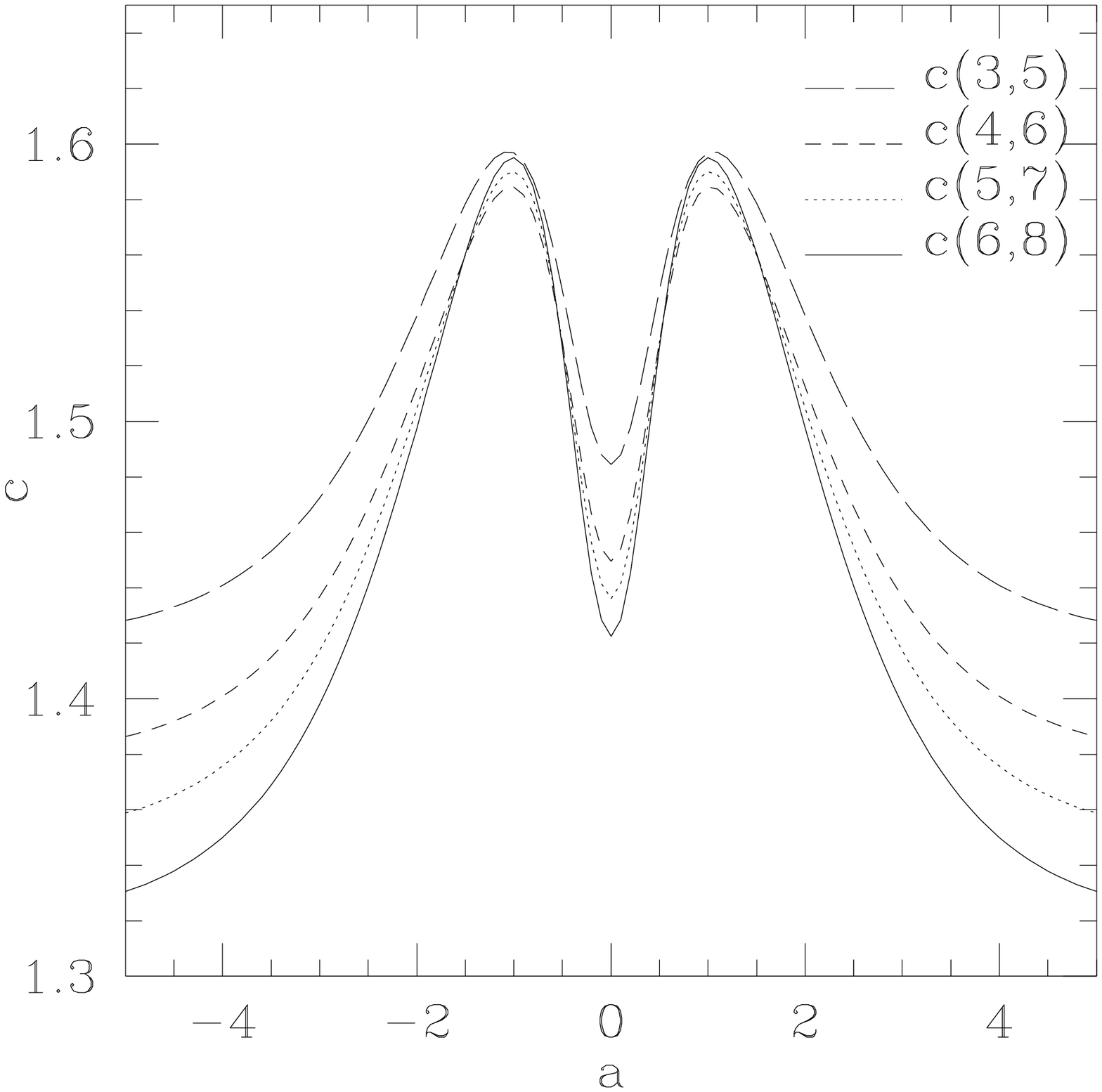}}
\epsfysize=200pt{\epsffile{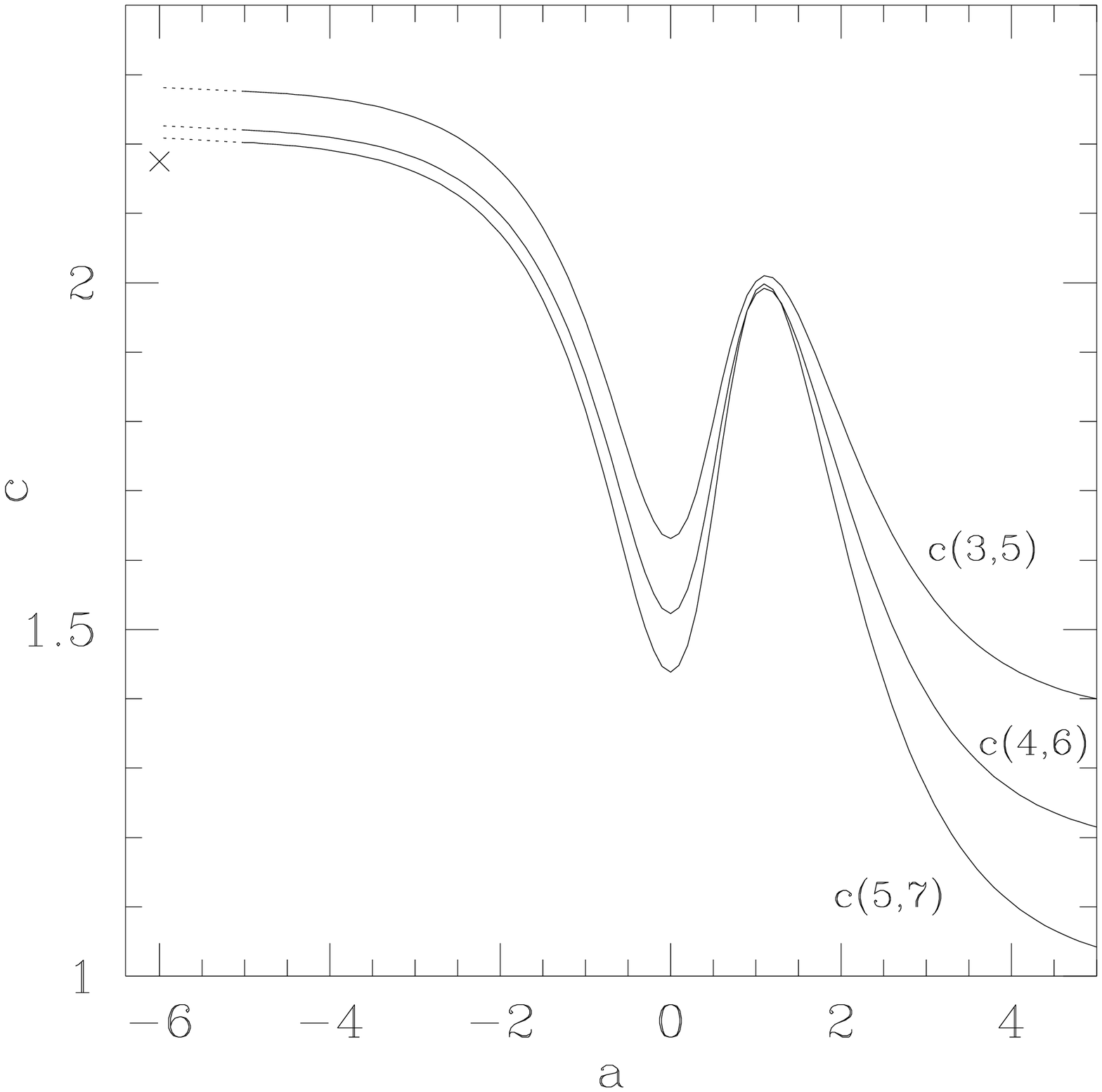}}
\end{center}
\protect\caption[3]{Central charge for two coupled $3$-state Potts
models (a) and two coupled $4$-state Potts models (b).
 The dots in (b) correspond to an interpolaton from the measurements made
 directly at $a=-\infty$ and $a=-5$. We also indicate by a cross a fit
 at $a=-\infty$ for strip widths $L$=10--14.}
\label{cc23p}
\end{figure}

In the case $N=2$, $q=3$ there is an $a \leftrightarrow -a$ symmetry
at the level of the free energy, as is reflected by the values of the
effective central charge shown in Fig.~\ref{cc23p}. The origin of this
symmetry becomes clear if one considers the possible local Boltzmann
weigths $W(L_{ij},a)$ [{\em cfr.}~Eq.~(\ref{local-Boltz})] associated
with the interactions along edge $(ij)$. For two coupled $q$-state
Potts models these are given by
\beq
 W(0,a) = 1, \ \ \ \
 W(1,a) = {\rm e}^a, \ \ \ \
 W(2,a) = {\rm e}^{2a+b} = 2 {\rm e}^a + q - 1.
\eeq
In the transfer matrix, starting from an arbitrary state, the first weight
will appear $(q-1)^2$ times, the second one $2(q-1)$ times,
and the last one just once. Specialising now to $q=3$, we see that
$W(0,a)$ and $W(1,a)$ have the same degeneracy, and that
\beq
 W(0,-a) = {\rm e}^{-a} W(1,a), \ \ \ \
 W(1,-a) = {\rm e}^{-a} W(0,a), \ \ \ \
 W(2,-a) = {\rm e}^{-a} W(2,a).
\eeq
Thus we have established an exact symmetry for the specific free
energy on the self dual line
\beq
  f(-a,q=3) = f(a,q=3) - 2a.
\eeq

Returning to the central charge values shown in Fig.~\ref{cc23p}.a we
see a local maximum at $a = \log(1+\sqrt{3})$, the fixed point
corresponding to two decoupled 3-state Potts models. At this point the
finite-size estimates converge very rapidly towards $c=2 \times \frac45$.
The other trivial fixed point is located at $a=0$ and corresponds to a
single 9-state Potts model. Here the estimates $c_{\rm eff}(L)$
decrease steadily with increasing strip width, as expected for a first
order phase transition. Any model with a bare coupling
$a_0 > \log(1+\sqrt{3})$ renormalises towards the $a\to\infty$ fixed
point, but this again corresponds to a first order transition
\cite{vays} as witnessed by the steady decrease of the estimates for
$c$.

\subsubsection{Two coupled 4-state Potts models}

\begin{table}
\begin{center}
\begin{tabular}{|r||c|c|} \hline
    $L$ & $c(L,L+4)$ & $x_t(L,L+2)$ \\ \hline
      4 &      2.251 &        1.032 \\ \hline
      6 &      2.231 &        0.856 \\ \hline
      8 &      2.203 &        0.808 \\ \hline
     10 &      2.175 &        0.762 \\ \hline
     12 &            &        0.720 \\ \hline
\end{tabular}
\end{center}
\protect\caption{\label{tab:24p}Effective exponents for two coupled
 4-state Potts models at $a=-\infty$.}
\end{table}

The effective central charge for two 4-state models can be inferred
from Fig.~\ref{cc23p}.b, and apart from the disappearence of the
$a \leftrightarrow -a$ symmetry conclusions are as in the case of two
3-state models. The range of $a$-values shown on the figure cannot
entirely exclude the possibility of a novel fixed point at
$a=-\infty$, but the data of Table \ref{tab:24p}, obtained by using
{\tt alg4} directly at this point, tell us that if $a=-\infty$ indeed
is a fixed point it must be unstable and non-critical.

\subsubsection{The point $a = \infty$}

As has already been stressed in Sec.~\ref{sec:numerics} the
topological algorithms {\tt alg3} and {\tt alg4} enable us to treat
$q$, the number of Potts spin states, as a continuously varying
parameter. We finish the discussion of the two-models case by
presenting some results for the effective exponents for several
fractional values of $q$, obtained directly in the $a \to \infty$ limit.

\begin{table}
\begin{center}
\begin{tabular}{|c||c|c|c|c|c|} \hline
 $q$  & $c(4,8)$ & $c(6,10)$ & $c(8,12)$ & $c(10,14)$ & $c(12,16)$ \\ \hline
 2.00 & 0.988504 & 0.995838  & 0.998256  & 0.999142   & 0.999528   \\ \hline
 2.25 & 1.142687 & 1.149660  & 1.151260  & 1.151271   & 1.150843   \\ \hline
 2.50 & 1.263404 & 1.264818  & 1.260899  & 1.255641   & 1.250201   \\ \hline
 2.75 & 1.351006 & 1.338262  & 1.319944  & 1.300188   & 1.280109   \\ \hline
 3.00 & 1.406234 & 1.367834  & 1.322338  & 1.274092   & 1.224246   \\ \hline
 3.25 & 1.430583 & 1.353605  & 1.266530  & 1.174515   & 1.079463   \\ \hline
 3.50 & 1.426433 & 1.298639  & 1.157785  & 1.011642   & 0.864787   \\ \hline
 3.75 & 1.396976 & 1.208954  & 1.008126  & 0.807868   &            \\ \hline
 4.00 & 1.346005 & 1.092727  & 0.833815  & 0.590962   &            \\ \hline
\end{tabular}
\end{center}
\caption{\label{tab:c2p}Effective central charge for two coupled
 $q$-state Potts models at $a\to\infty$, obtained from {\tt alg4}.}
\end{table}

First, in Table \ref{tab:c2p} we display the central charge
values. For $q=2$ these converge very nicely towards $c=1$, the exact
Ashkin-Teller value (\ref{AT-exact}). For higher $q$ the increasing
spacing between succesive estimates signals non-critical behaviour, or
in other words, the free energies are not well fitted by Eq.~(\ref{fc}).
It is convenient to have such data at one's disposal, since they
facilitate the distinguishing between first and second order phase
transitions in the sequel. Evidently, for $q=2.25$ it not {\em a priori}
easy to clearly point out the first order nature of the transition,
since the correlation length is very large. However, it is interesting
to notice that whilst for $q=2$ the estimates {\em increase}
monotonically as a function of $L$, for all $q>2$ they eventually
begin to {\em decrease} for large enough $L$.

\begin{table}
\begin{center}
\begin{tabular}{|c||c|c|c|c|c|} \hline
 $q$  &$x_t(6,8)$&$x_t(8,10)$&$x_t(10,12)$&$x_t(12,14)$&$x_t(14,16)$\\ \hline
 2.00 & 1.467238 & 1.475693  & 1.490527   & 1.495490   & 1.497566   \\ \hline
 2.25 & 1.745413 & 1.554317  & 1.588008   & 1.605994   & 1.618971   \\ \hline
 2.50 & 1.863461 & 1.768649  & 1.702778   & 1.739814   & 1.769780   \\ \hline
 2.75 & 1.780405 & 1.717780  & 1.672429   & 1.635165   & 1.603300   \\ \hline
 3.00 & 1.707436 & 1.630297  & 1.575072   & 1.532502   & 1.499100   \\ \hline
 3.25 & 1.648514 & 1.565197  & 1.510808   & 1.474971   & 1.452970   \\ \hline
 3.50 & 1.605054 & 1.524930  & 1.482309   & 1.464823   & 1.466231   \\ \hline
 3.75 & 1.576886 & 1.508815  & 1.487488   & 1.497813   &            \\ \hline
 4.00 & 1.562949 & 1.514505  & 1.521838   & 1.566459   &            \\ \hline
\end{tabular}
\end{center}
\caption{\label{tab:xt2p}Effective values of the thermal scaling
 dimension for two 
 coupled $q$-state Potts models at $a\to\infty$, obtained from {\tt alg4}.}
\end{table}

Table \ref{tab:xt2p} provides the analogous estimates for the thermal
scaling dimension. For $q=2$ they converge towards $x_t = 3/2$,
in accordance with Eq.~(\ref{AT-exact}). Otherwise their variation is
not monotonic in $L$, presumably signalling the presence of a finite
correlation length.

\subsection{Three coupled models}

We now turn to our primary interest, namely the case of $N=3$ coupled
Potts models. The self-dual line here starts at the point where
$b=-\infty$, that is $a \in [a_{\rm min},\infty]$ with
\beq
 a_{\rm min} = \log \left( \frac{\sqrt{q}+1}{\sqrt{q}+2} \right).
\eeq
Unlike the case of two models just discussed we shall find that the
$a\to\infty$ limit corresponds to a {\em critical} fixed
point for {\em all} values of $q \in [2,4]$. The determination of the
associated critical exponents thus becomes of paramount interest, and
we have dedicated considerable computational effort to this purpose.
Let us discuss the various cases in turn.

\subsubsection{Three coupled Ising models}

\begin{figure}
\begin{center}
\leavevmode
\epsfysize=200pt{\epsffile{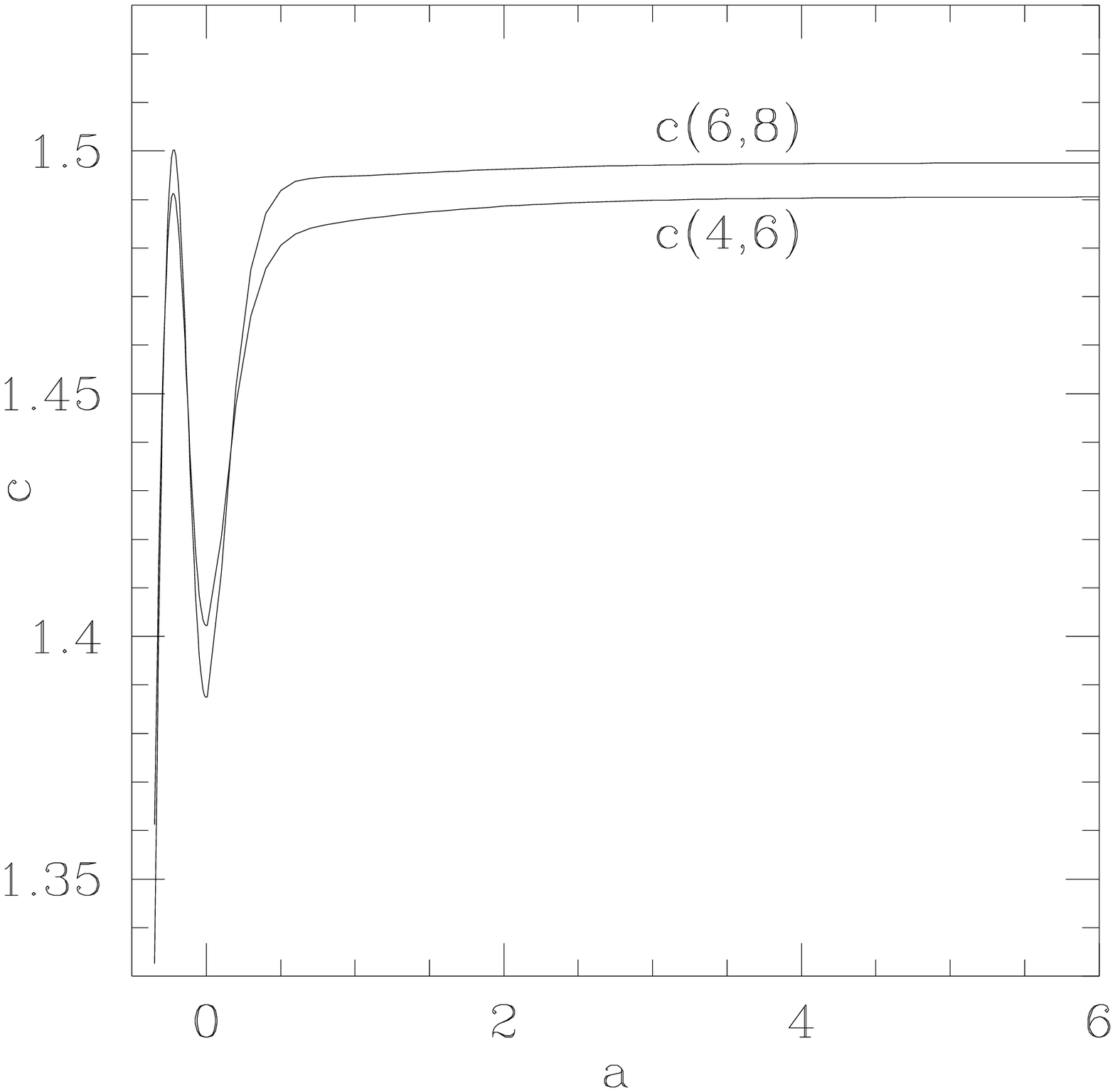}}
\epsfysize=200pt{\epsffile{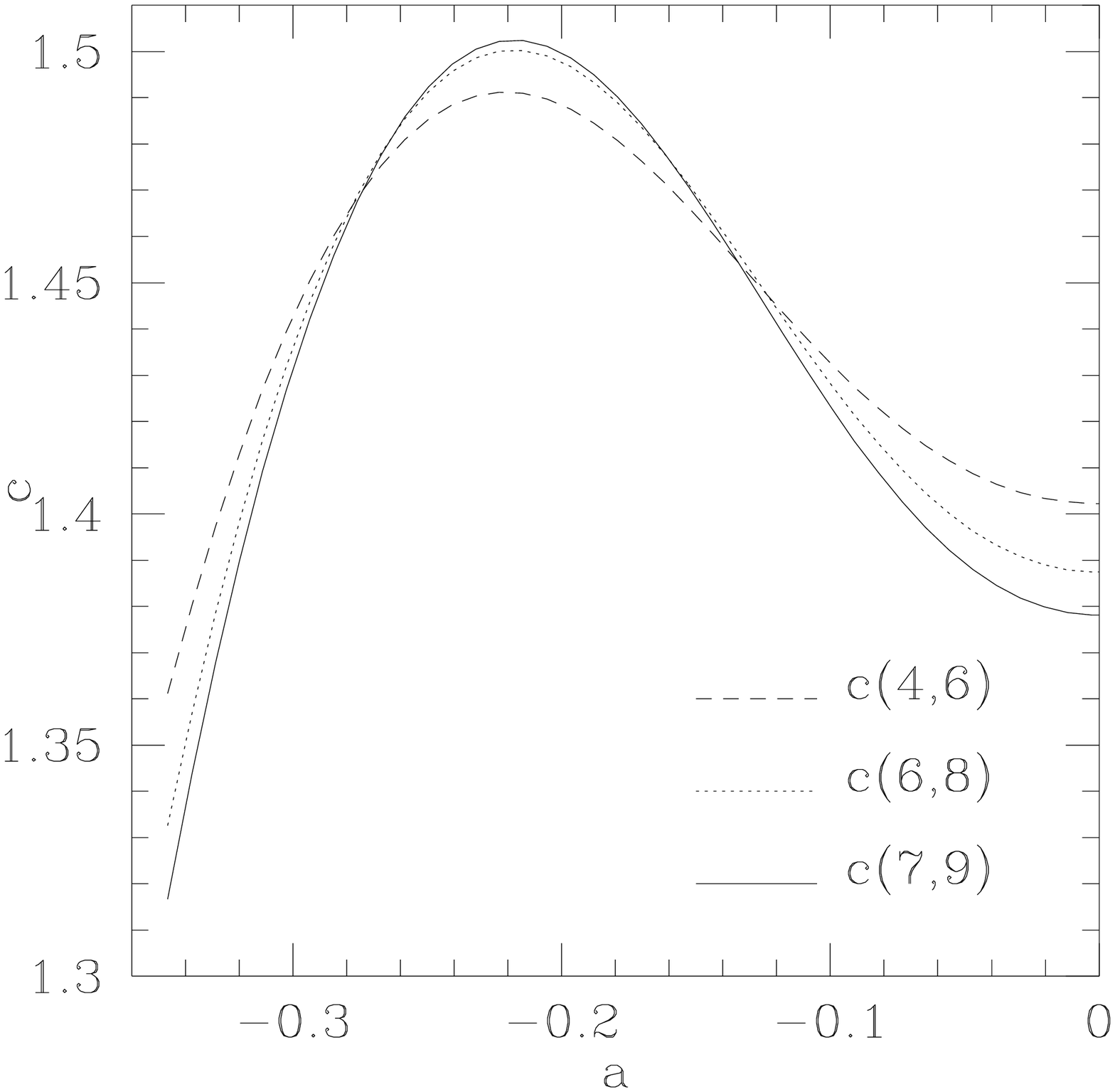}}
\end{center}
\protect\caption[3]{\label{cc3im}Central charge for three coupled
Ising models.}
\end{figure}

This is the three-colour Ashkin-Teller model. The effective central
charge is displayed in Fig.~\ref{cc3im}.a. At the trivial fixed point
at $a=0$ we find the usual dip in $c$, reflecting the first-order
nature of the transition in the 8-state Potts model.

For all $a>0$ the central charge is approximately constant,
$c \simeq 3/2$, and we have also verified this value in the limit
$a \to \infty$. This situation is very reminiscent of that of the
(standard) {\em two}-colour Ashkin-Teller model, and seems to allow
for a situation where the entire half line $a>0$ consists of critical
fixed points along which the critical exponents vary
continuously. Whether this is indeed the case will be the subject of a
separate publication \cite{futurepaper}.

\begin{figure}
\begin{center}
\leavevmode
\epsfysize=200pt{\epsffile{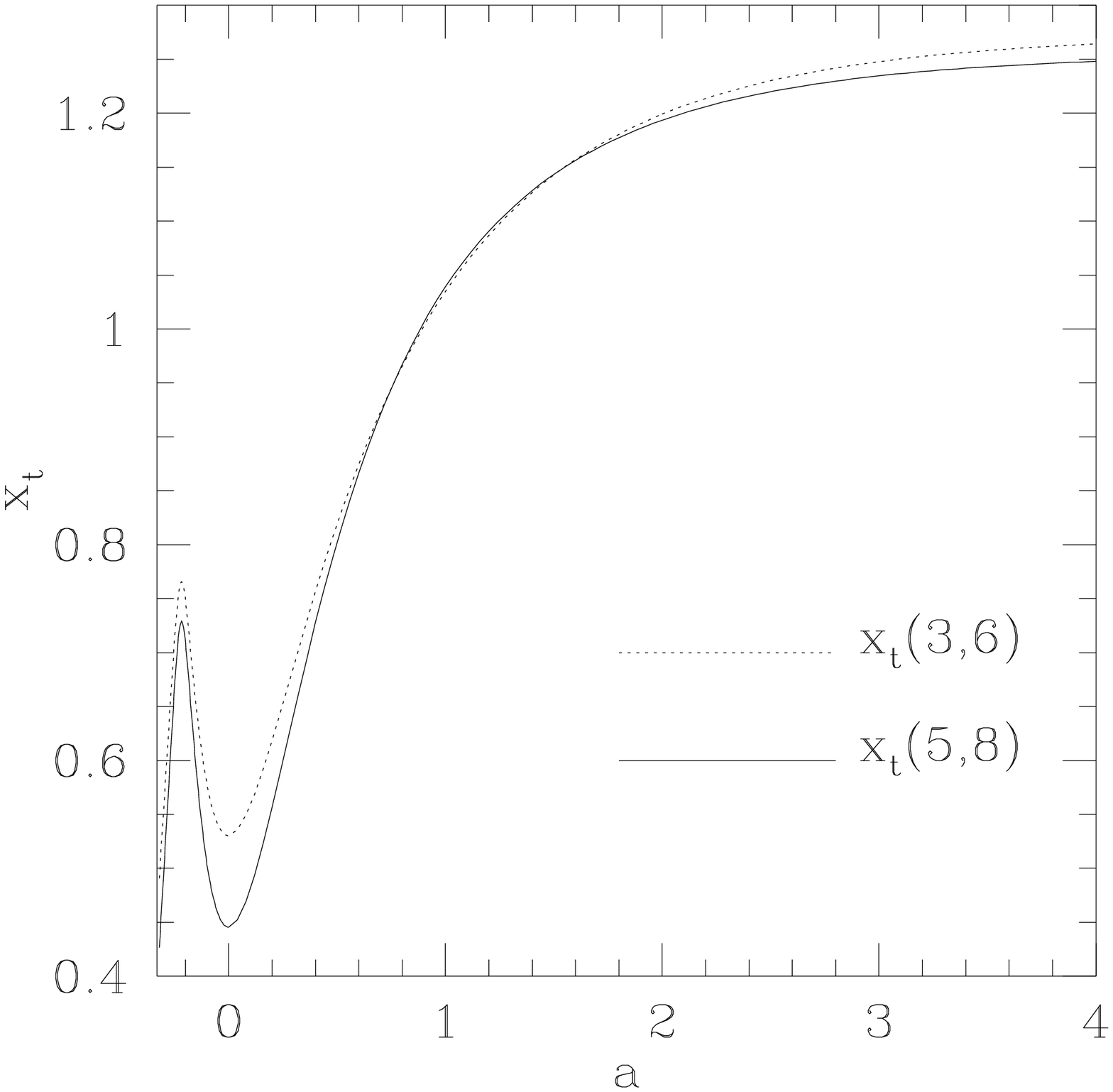}}
\epsfysize=200pt{\epsffile{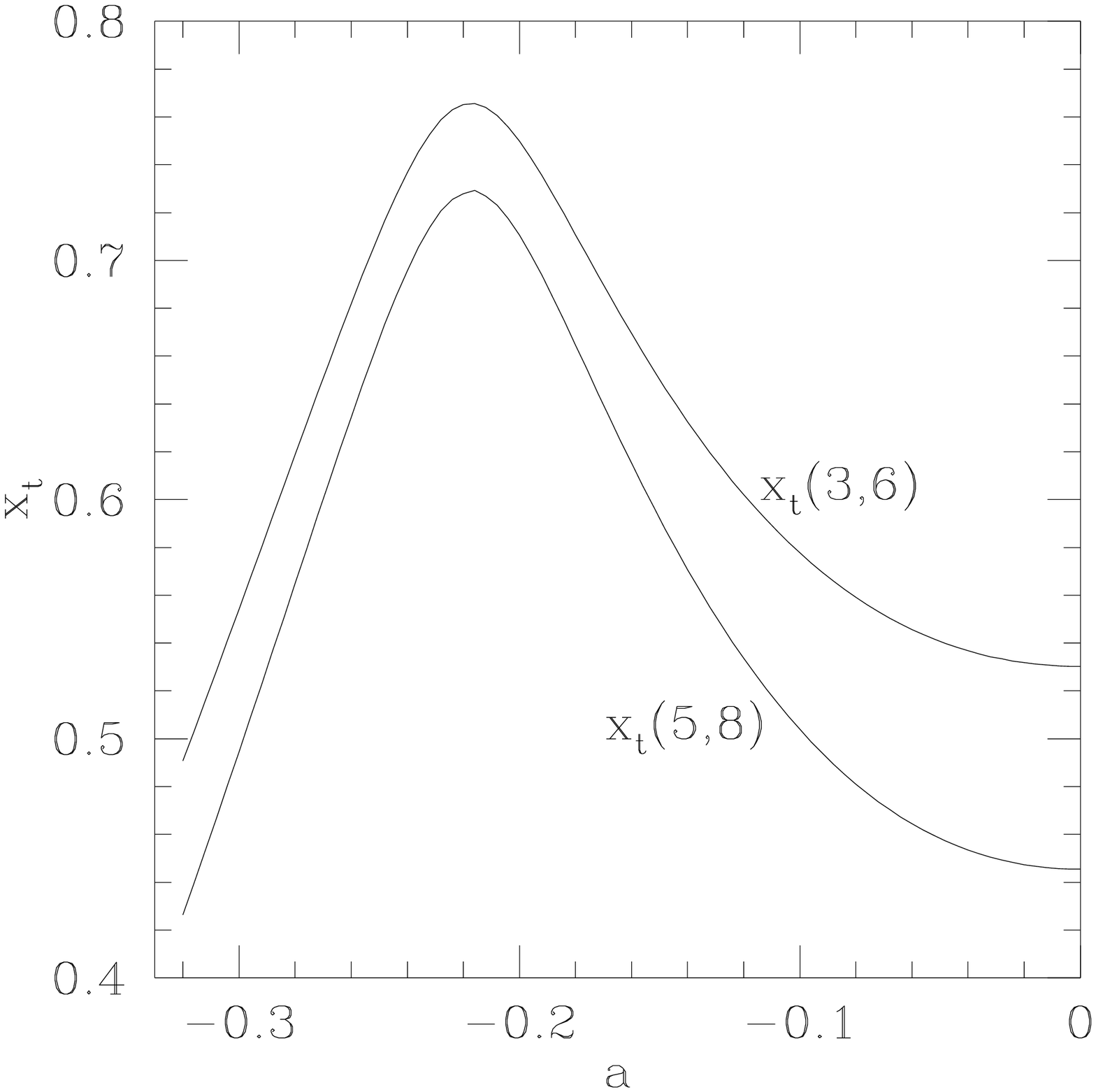}}
\end{center}
\protect\caption[4]{\label{xt3im}Thermal exponent for three coupled
Ising models.}
\end{figure}

Fig.~\ref{cc3im}.b presents a closer look at the antiferromagnetic
region $a<0$. We find here a surprising candidate for an unstable
critical fixed point at $a \simeq -0.22$ with a central charge of
$c \simeq 1.5$. Further support for this suspicion is found in the
maximum for the effective thermal exponent on Fig.~\ref{xt3im}.b, and
in the crossing of the curves for the magnetic exponents on
Fig.~\ref{xm3im}.b. More results along these lines will be published
elsewhere \cite{futurepaper}.

The results for the thermal exponent on the half line $a>0$ are shown
on Fig.~\ref{xt3im}.a. Everywhere the convergence seems to be very
rapid, except at $a=0$ where we expect an extrapolated effective
exponent of $x_t^{\rm eff}=0$ due to the first order transition. In
the $a\to\infty$ limit we estimate $x_t = 1.24 \pm 0.01$; more
detailed results are provided by Table \ref{tab:xt-3p} below.

Finally, in Fig.~{\ref{xm3im}  we present results for the three
different magnetic exponents, $x_H^{(1)}$, $x_H^{(2)}$ and $x_H^{(3)}$.
The first exponent is constant, $x_H^{(1)} \simeq 1/8$ for $a>0$, as in the
case of two coupled Ising models, whilst the other two depend on
$a$. In the $a\to\infty$ limit we find $x_H^{(2)} \simeq 0.27$ and
$x_H^{(3)} \simeq 0.46$ as witnessed by Tables \ref{tab:xh2-3p} and
\ref{tab:xh3-3p}.

\begin{figure}
\begin{center}
\leavevmode
\epsfysize=200pt{\epsffile{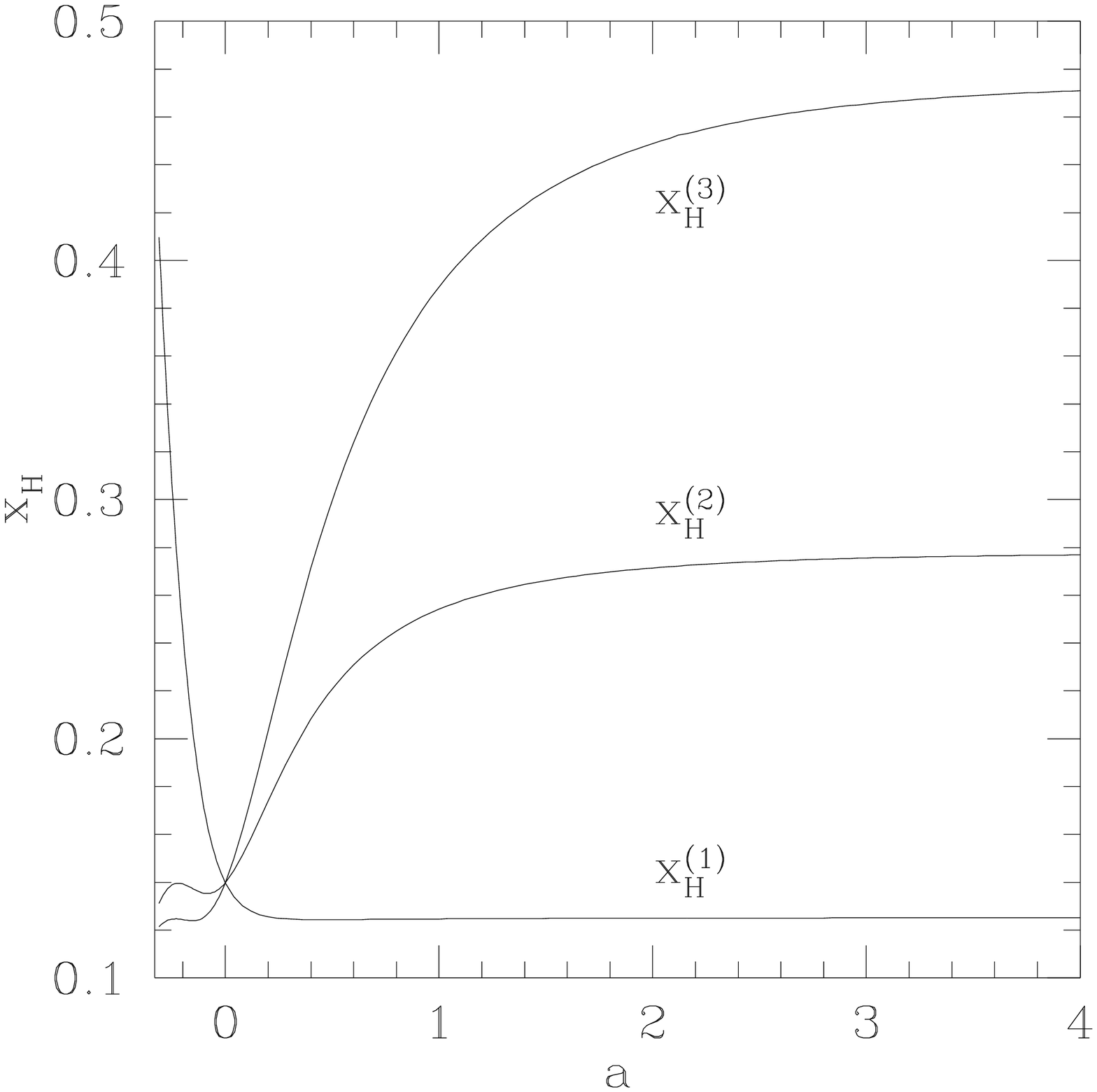}}
\epsfysize=200pt{\epsffile{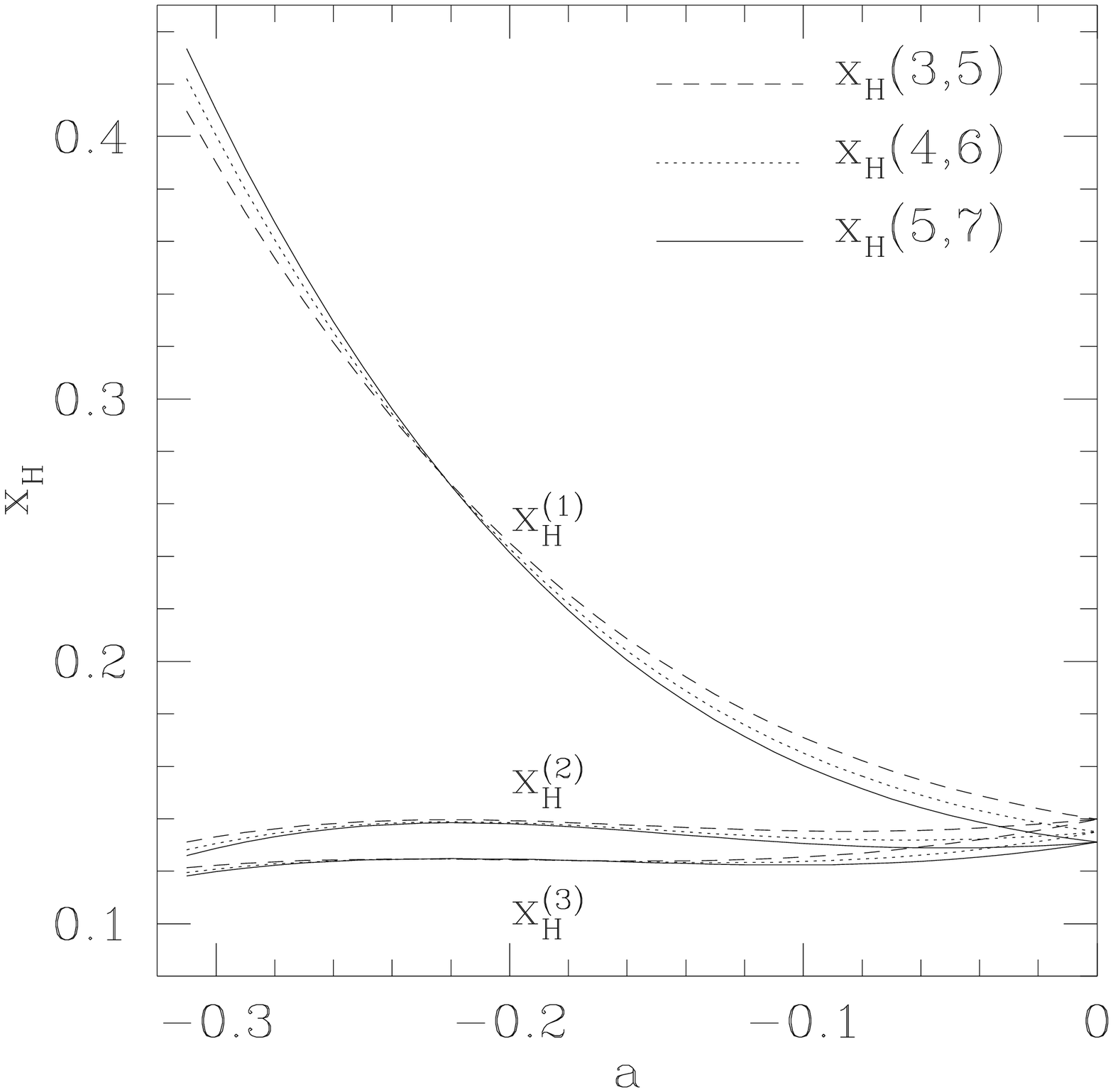}}
\end{center}
\protect\caption[5]{\label{xm3im}Magnetic exponents for three coupled
Ising models.}
\end{figure}

\subsubsection{Three coupled 3-state Potts models} 

Until now, the models for which we have presented numerical data are
believed to be somewhat atypical representatives for the general class
of $N$ coupled $q$-state Potts models. On one hand, for $N=2$ coupled
models the ${\cal O}(g^2)$-term in the $\beta$-function (\ref{betap})
vanishes, and neighter perturbative RG nor
numerics predict any non-trivial critical fixed points for $q \neq 2$.
On the other, for $q=2$ the energy-energy coupling is exactly
marginal, leading either to the Ashkin-Teller scenario with an entire line of
critical fixed points along which the exponents vary continuously, or
simply to a flow back towards the decoupled fixed point.
For the case of $N=3$ coupled 3-state Potts models, which we consider
now, the situation can thus be believed to be generic: Perturbation
theory predicts a novel critical fixed point, Eq.~(\ref{fpcoup}), with
unique critical exponents, and so does numerics, as we shall presently
see.

Consider first the effective central charge along the self-dual line
$a \in [a_{\rm min},\infty]$, where
$a_{\rm min} = \log \left( \frac{\sqrt{3}+1}{\sqrt{3}+2} \right)$.
On Fig.~\ref{cc33p}.a we recognise the familiar structure of a local
minimum and a local maximum, signaling the usual
two trivial fixed points. However, for
larger values of $a$ a novel feature emerges, as witnessed by
Fig.~\ref{cc33p}.b. The central charge is here a very slightly
decreasing function of $a$, indicating a flow towards the fixed point
at $a = \infty$. Furthermore, the finite-size dependence of the
estimates leads us to the conclusion that this point is now
{\em critical}.

\begin{figure}
\begin{center}
\leavevmode
\epsfysize=200pt{\epsffile{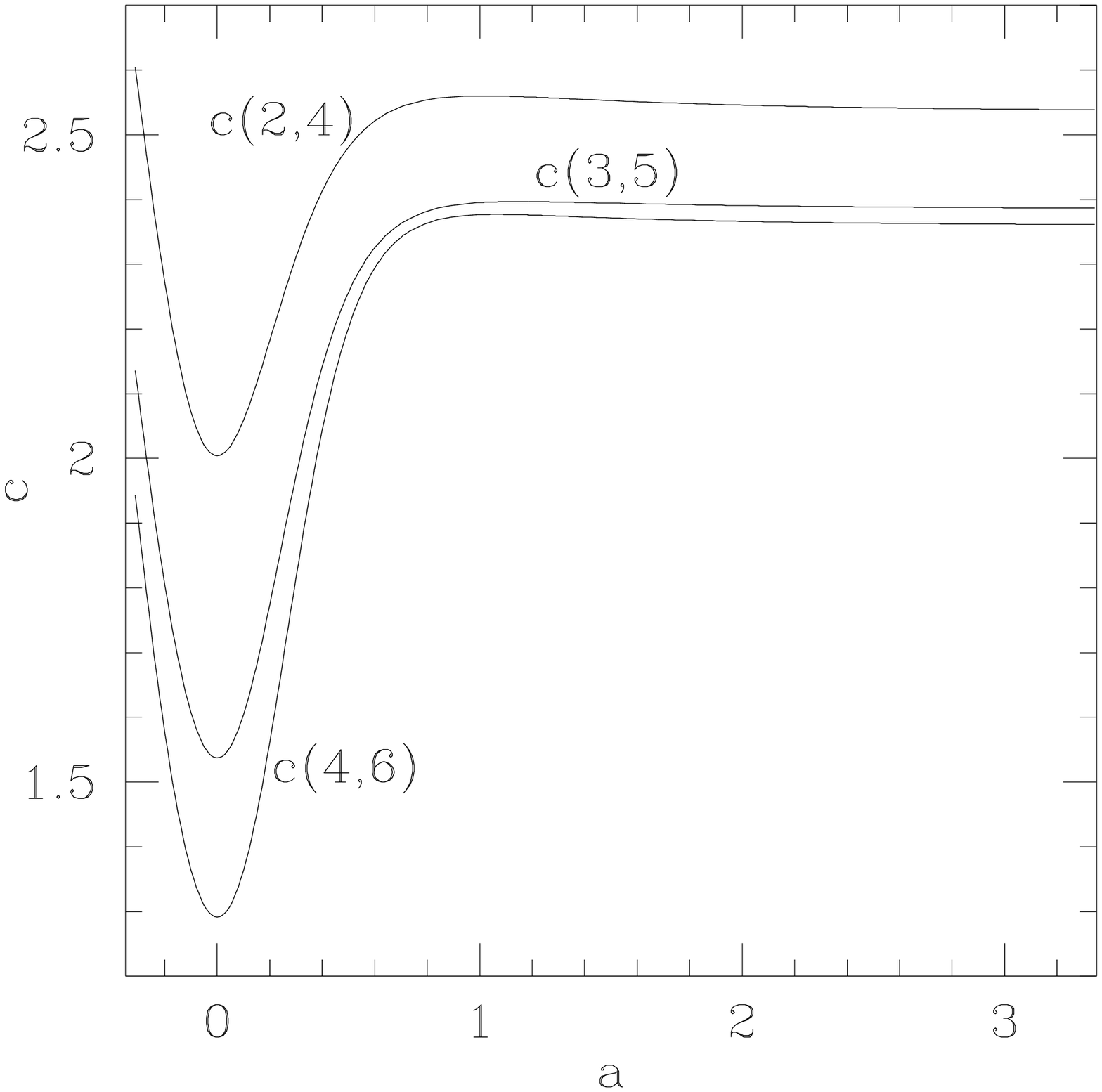}}
\epsfysize=200pt{\epsffile{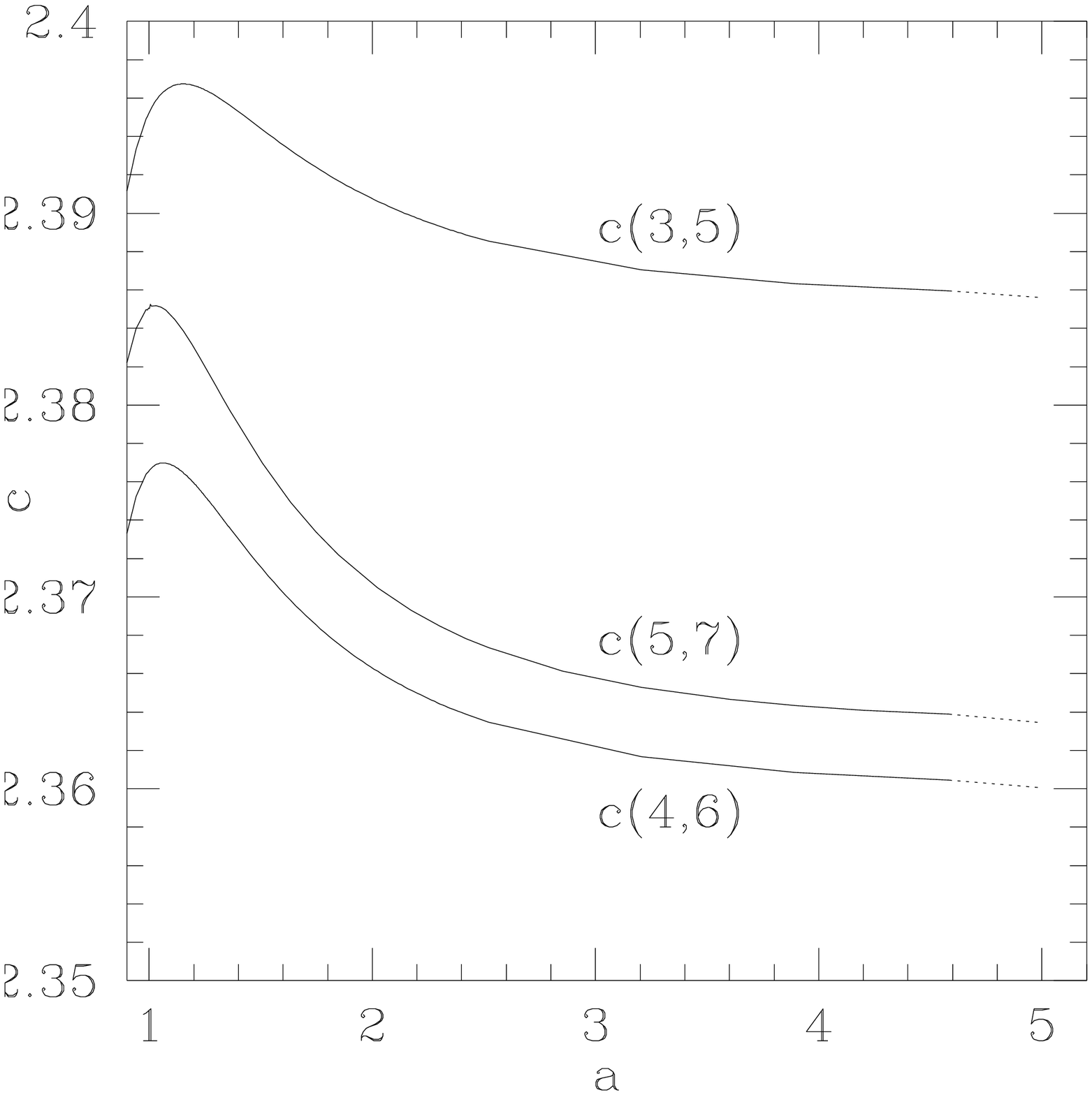}}
\end{center}
\protect\caption[3]{\label{cc33p}Central charge for three coupled
 3-state Potts models. The dots designate an interpolation from
 $a=4.5$ to $a=+\infty$.} 
\end{figure}

The more accurate data of Table~\ref{tab:c-3p} corroborate this
scenario. First, we see that the finite-size estimates converge very
rapidly to the value
\beq
  c = 2.377 \pm 0.003,
\eeq
% Geometric extrapolation of the two-point estimates, with a
% conservative error-bar. ---JLJ
in very good agreement with the perturbative prediction
$c_{\rm FP}=2.3808 + {\cal O}(\epsilon^5)$ from
Eq.~(\ref{c-correction}).%
\footnote{In fact, the latter prediction would be slightly {\em smaller} if
  we could go to even higher order in perturbation theory, since the
  series (\ref{c-correction}) is known to be alternating.}
Second, the convergence of the estimates in Table~\ref{tab:c-3p} is
now from {\em below}, convincingly excluding the possibility that
$c_{\rm eff}$ could eventually tend to zero, as would be the case for
a non-critical fixed point.
Third, the numerical value for $c$ is comfortably away from that of 
the decoupled fixed point, $c_{\rm pure} = 3 \times  \frac45 = 2.4$.
\begin{figure}
\begin{center}
\leavevmode
\epsfysize=200pt{\epsffile{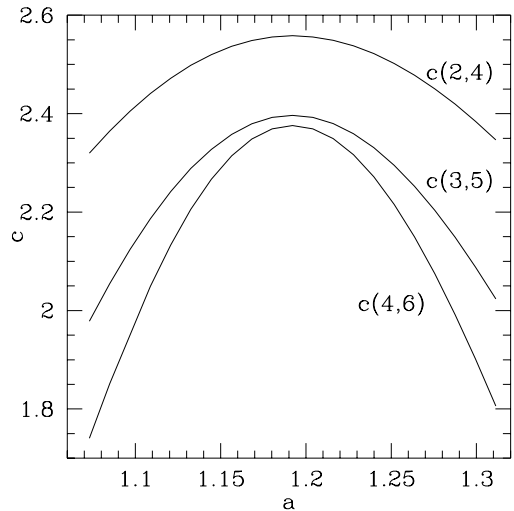}}
\end{center}
\protect\caption[3]{\label{cc33pd}Central charge for three coupled
 3-state Potts models when moving perpendicularly off the self-dual line.}
\end{figure}

\begin{figure}
\begin{center}
\leavevmode
\epsfysize=200pt{\epsffile{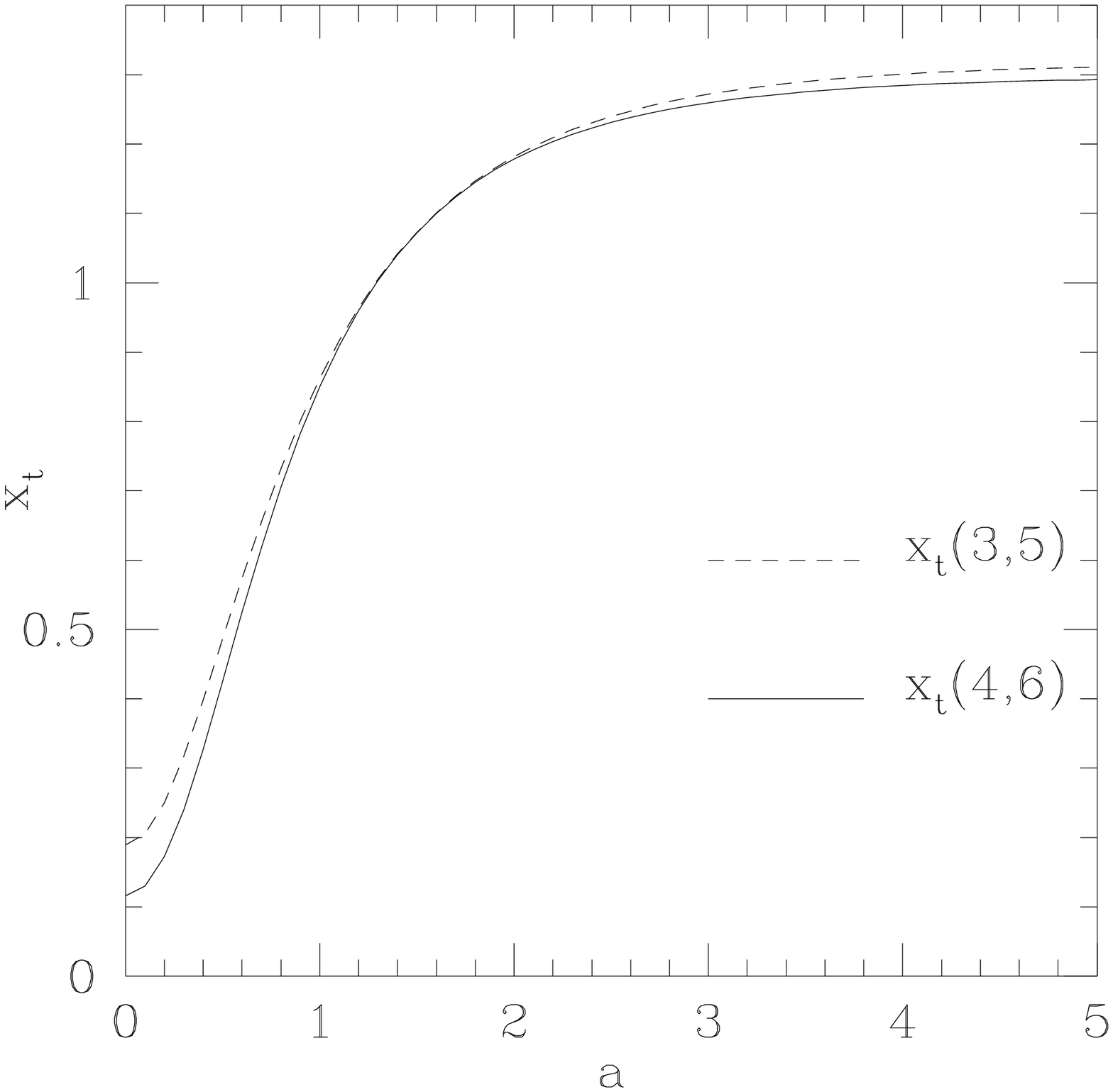}}
\epsfysize=200pt{\epsffile{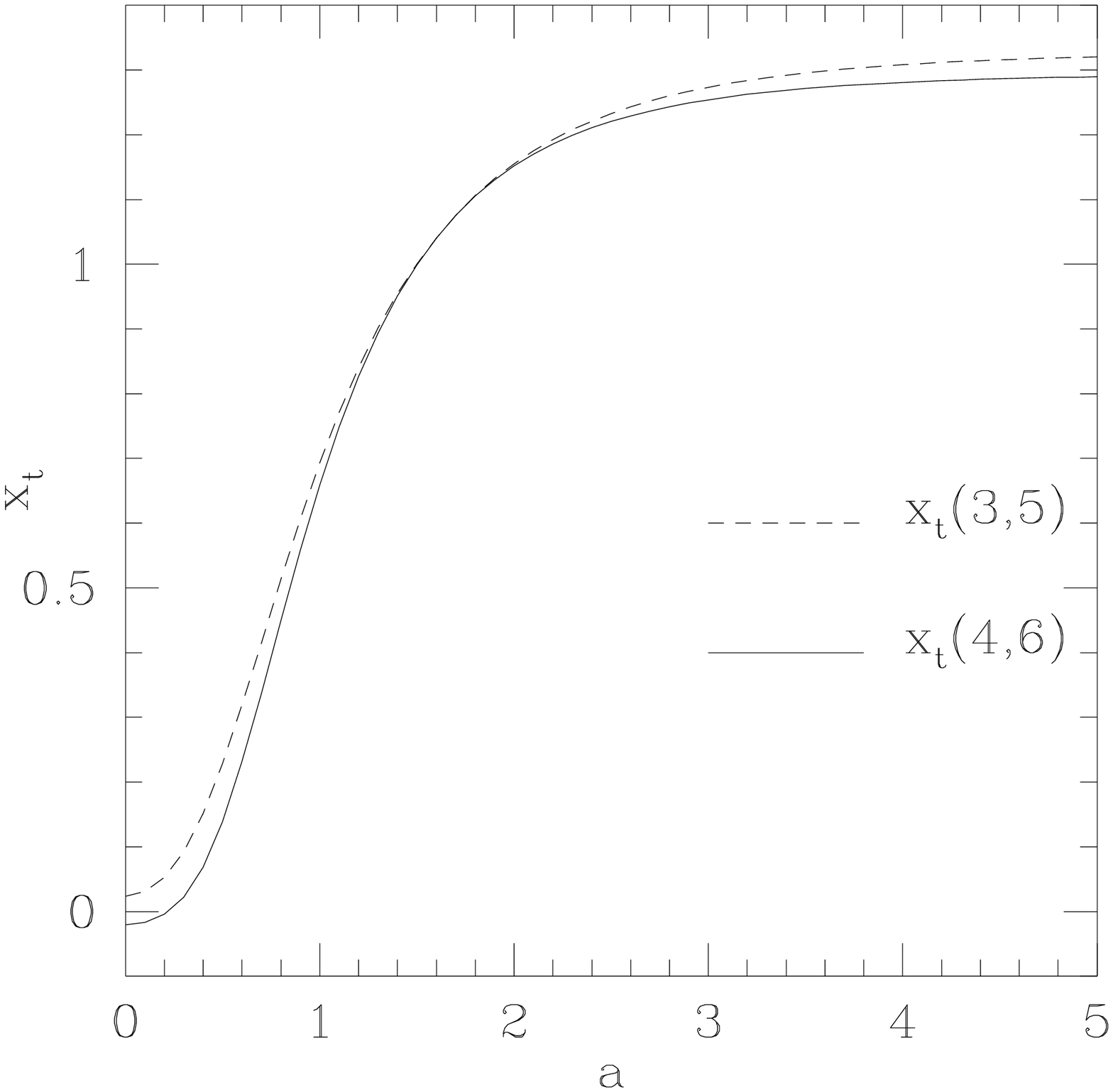}}
\end{center}
\protect\caption[3]{\label{xt33p}Thermal exponent for $x_t$ for three
 coupled $q$-state Potts models with $q=3$ and  $q=4$ respectively.}
\end{figure}

In Fig.~\ref{cc33pd} we show the behaviour of the central charge along a
line perpendicular to the self-dual line at the point $a=1.19217$. Once
again, this value of $a$ was chosen arbitrarily; similar curves are observed
for other values of $a>0$. 

Plots of the effective thermal and magnetic exponents are shown in
Figs.~\ref{xt33p}.a and \ref{xh33p}.a respectively. More accurate values
obtained directly at the $a=\infty$ critical point are given in
Table~\ref{tab:xt-3p}--\ref{tab:xh3-3p}.

\begin{figure}
\begin{center}
\leavevmode
\epsfysize=200pt{\epsffile{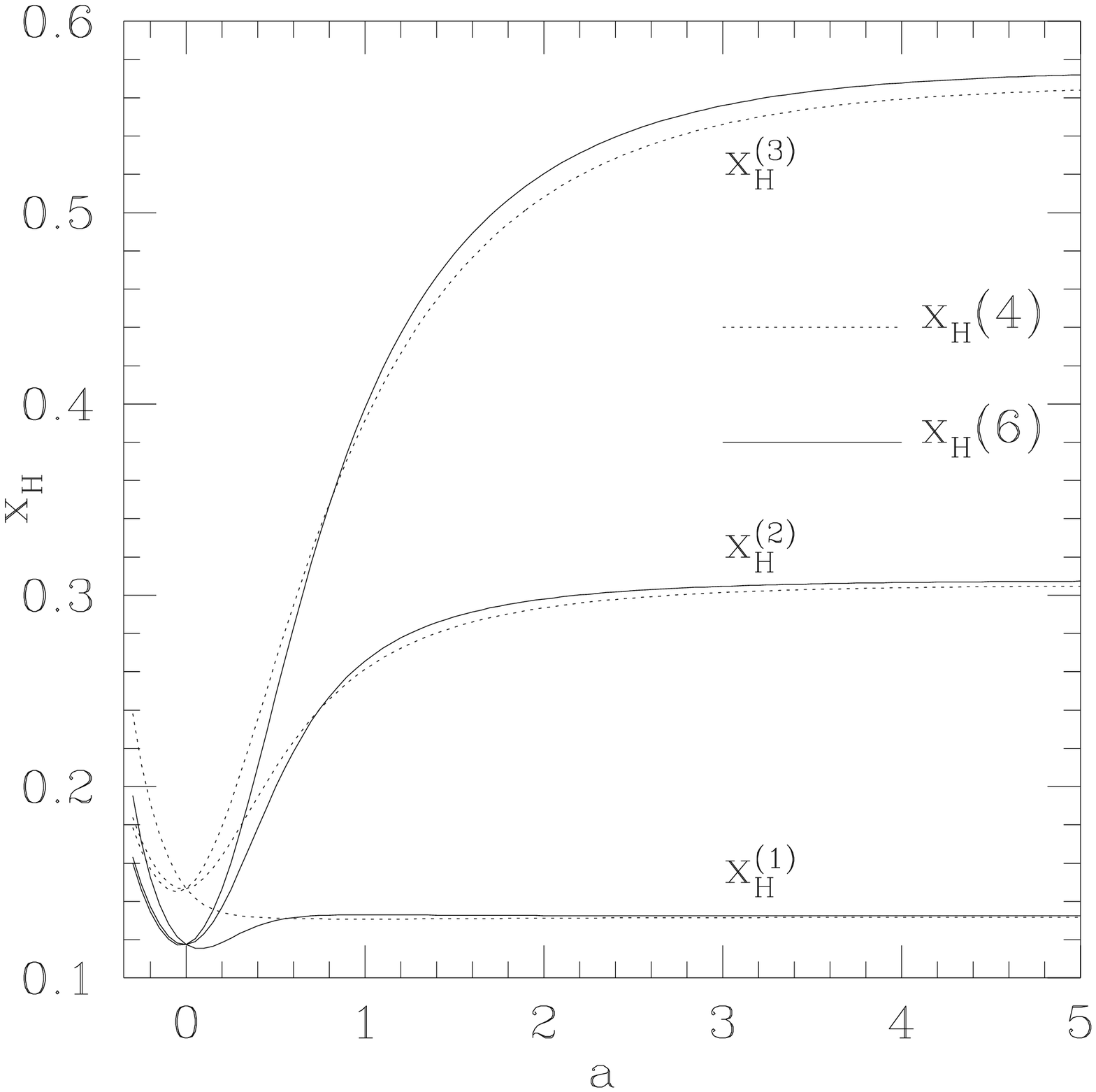}}
\epsfysize=200pt{\epsffile{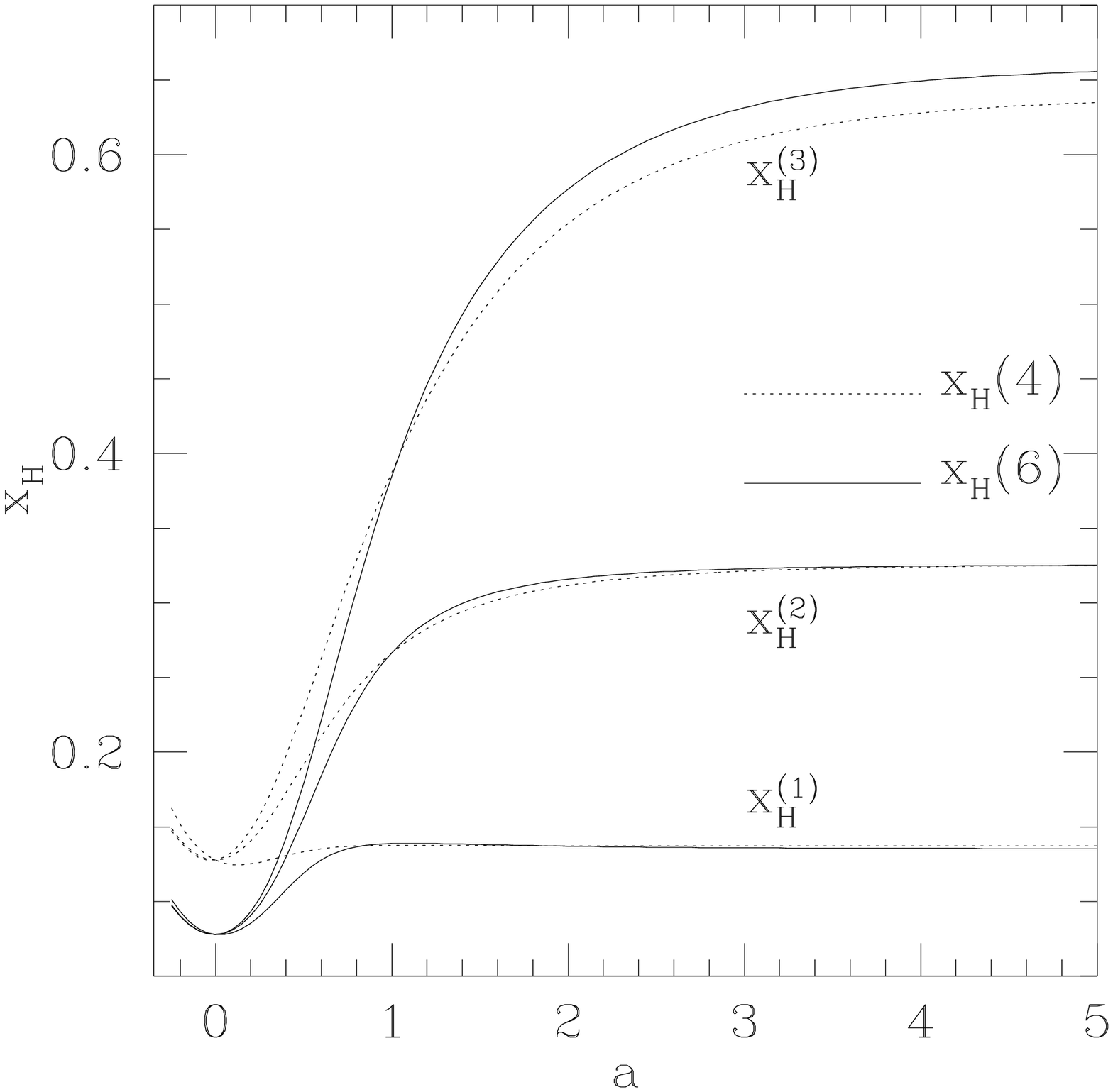}}
\end{center}
\protect\caption[3]{\label{xh33p}Magnetic exponents $x_H^{(1)}$,
 $x_H^{(2)}$ and $x_H^{(3)}$ for three coupled $q$-state Potts models
 with $q=3$ and  $q=4$ respectively.} 
\end{figure}

\subsubsection{Three coupled 4-state Potts models}

The situation for three coupled Potts models is very similar to the
case of $q=3$ treated above. The plots of the effective central charge
given in Fig.~\ref{cc34p} again give us strong reasons to believe that
the $a\to\infty$ limit represents the non-trivial critical fixed point
predicted by perturbation theory. Also the effective thermal and magnetic
exponents of Figs.~\ref{xt33p}.b and \ref{xh33p}.b respectively exhibit
a structure similar to the 
$q=3$ case, but the critical exponents at $a\to\infty$ are of course
distinct.

\begin{figure}
\begin{center}
\leavevmode
\epsfysize=200pt{\epsffile{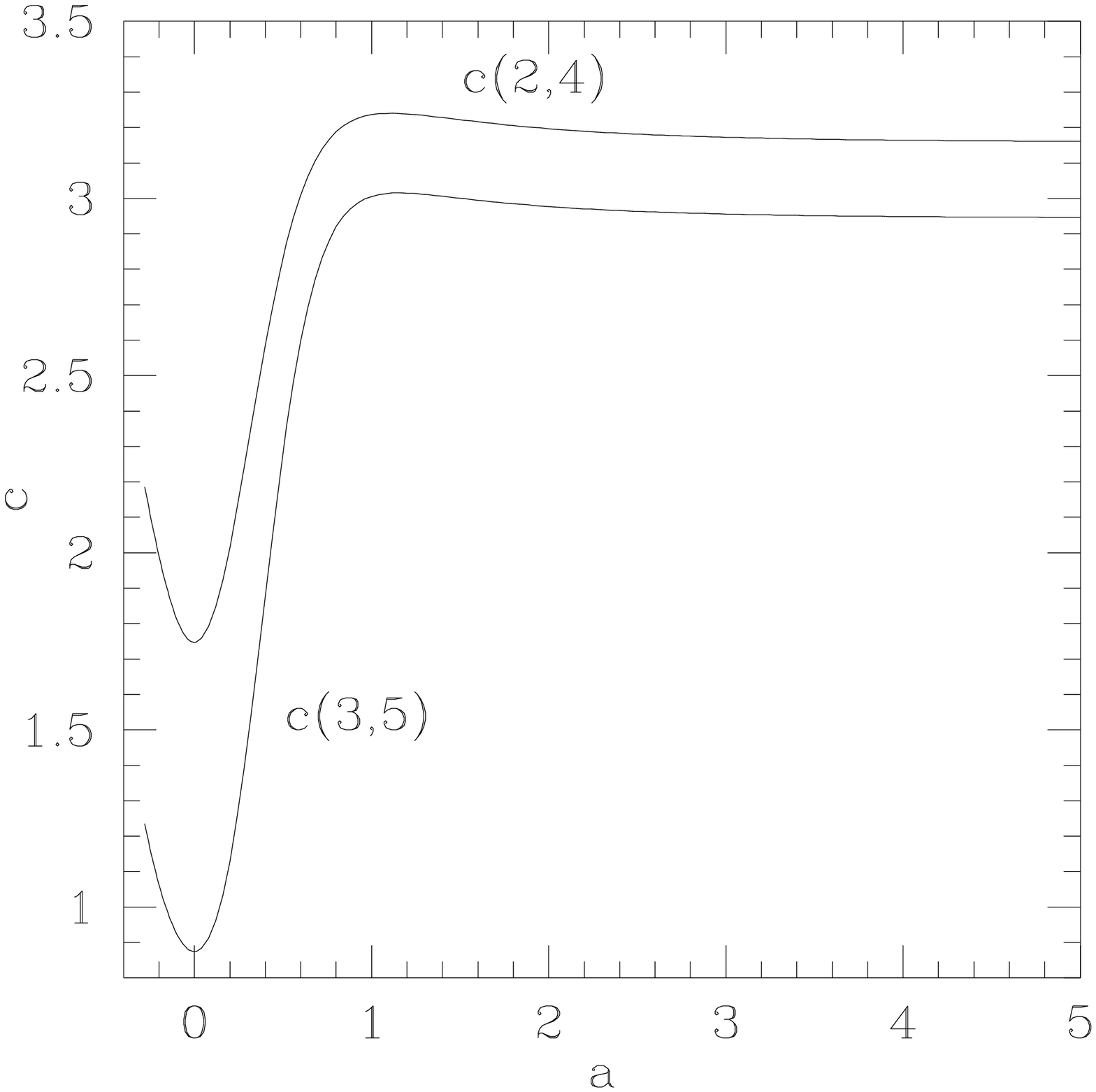}}
\epsfysize=200pt{\epsffile{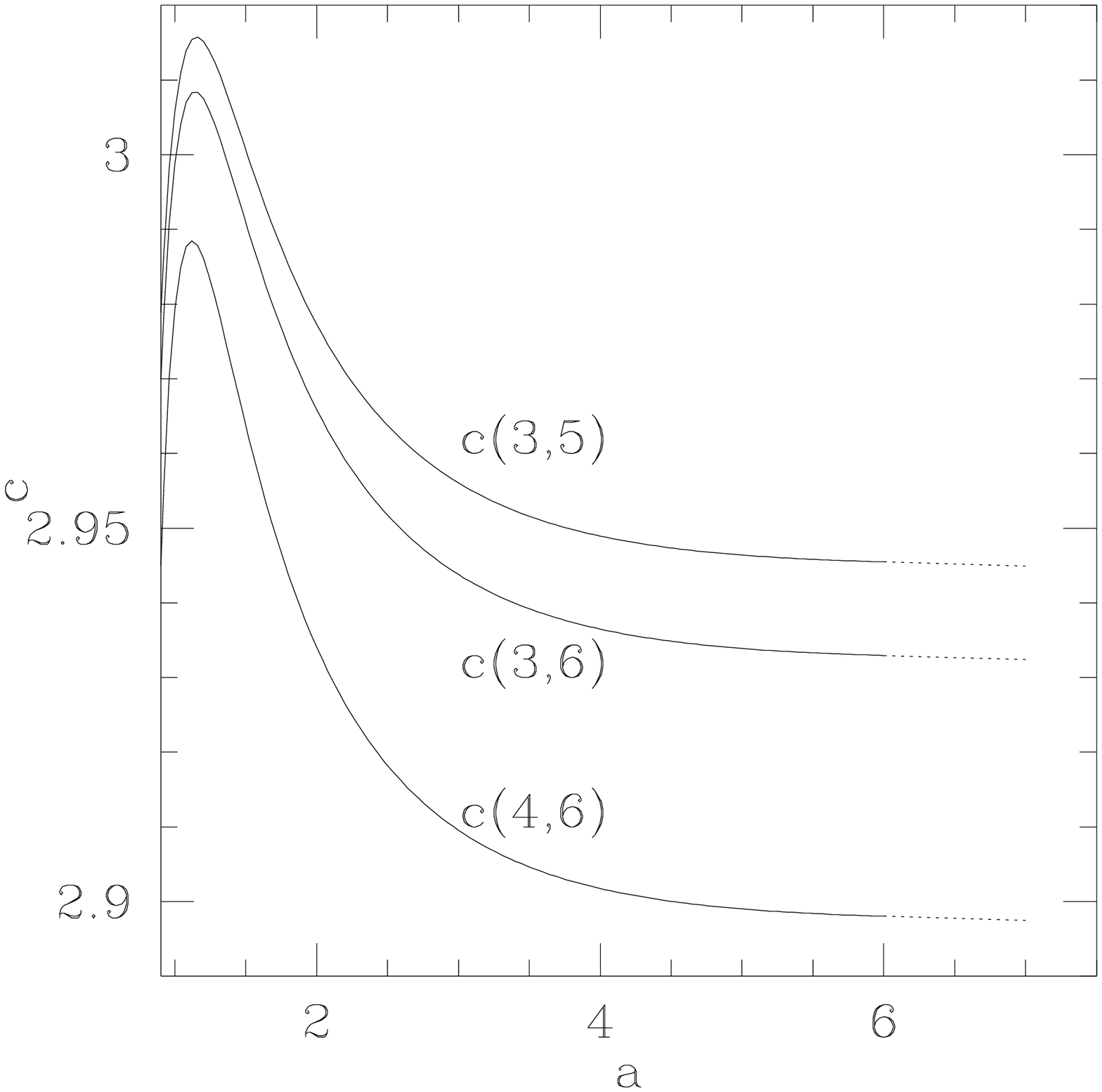}}
\end{center}
\protect\caption[7]{\label{cc34p}Central charge for three coupled
 $4$-state Potts models.}
\end{figure}

\subsubsection{The $a \to \infty$ limit for general $q \in [2,4]$}

After having numerically identified the non-trivial critical fixed
point for three coupled models, that was predicted by the perturbative
RG, we would like to accurately determine the associated critical
exponents as a function of $q$. This is done by diagonalising the
transfer matrix directly at $a=\infty$, using {\tt alg4}.

In Table~\ref{tab:c-3p} we show finite-size estimates of the central
charge, obtained from three-point fits of the form (\ref{fc2}). For
all $q \in [2,4]$ these estimates seems to converge as
$L\to\infty$. Of course, an explicit extrapolation to the infinite
system limit from just three data points would easily turn out to be
rather subjective. It should however be noted that the convergence is
in most cases monotonic, and the last estimate therefore provides a
quite accurate upper or lower bound for the extrapolated quantity.

\begin{table}
\begin{center}
\begin{tabular}{|c||c|c|c|} \hline
 $q$  & $c(4,8)$ & $c(6,10)$ & $c(8,12)$ \\ \hline
 2.00 & 1.489114 & 1.497443  & 1.500008  \\ \hline
 2.25 & 1.747538 & 1.757162  & 1.760185  \\ \hline
 2.50 & 1.976384 & 1.986719  & 1.989905  \\ \hline
 2.75 & 2.179933 & 2.190166  & 2.193005  \\ \hline
 3.00 & 2.361502 & 2.370594  & 2.372345  \\ \hline
 3.25 & 2.523740 & 2.530453  & 2.530147  \\ \hline
 3.50 & 2.668818 & 2.671752  & 2.668217  \\ \hline
 3.75 & 2.798557 & 2.796195  & 2.788092  \\ \hline
 4.00 & 2.914511 & 2.905265  & 2.891139  \\ \hline
\end{tabular}
\end{center}
\caption{\label{tab:c-3p}Results for the central charge of three
 coupled models at $a=\infty$ and for varying values of $q$.}
\end{table}

The results for $c$ are in very good agreement with the perturbative
RG in the range $q \in [2,3]$. For larger values of $q$ the
perturbation theory is expected to break down, and so the numerical
values are likely to be more reliable.

\begin{table}
\begin{center}
\begin{tabular}{|c||c|c|c|c|c|} \hline
 $q$  &$x_t(6,8)$&$x_t(8,10)$&$x_t(10,12)$ \\ \hline
 2.00 & 1.257900 & 1.248706  & 1.242186    \\ \hline
 2.25 & 1.263919 & 1.255942  & 1.250623    \\ \hline
 2.50 & 1.268895 & 1.262901  & 1.259341    \\ \hline
 2.75 & 1.272294 & 1.268558  & 1.266905    \\ \hline
 3.00 & 1.273618 & 1.271938  & 1.271901    \\ \hline
 3.25 & 1.272423 & 1.272143  & 1.272966    \\ \hline
 3.50 & 1.286448 & 1.268344  & 1.268400    \\ \hline
 3.75 & 1.261129 & 1.260138  & 1.258753    \\ \hline
 4.00 & 1.250663 & 1.247052  & 1.242082    \\ \hline
\end{tabular}
\end{center}
\caption{\label{tab:xt-3p}Results for the thermal scaling dimension of
three coupled models at $a=\infty$ using {\tt alg4}.}
\end{table}

The corresponding values of the thermal scaling dimension are given in
Table~\ref{tab:xt-3p}. In this case the comparison with the
perturbative results is more tricky. We recall that $x_t$ is obtained
by fitting the first gap in the even sector of the transfer matrix
to Eq.~(\ref{fx2}). This is expected to correspond to the scaling
dimension of the most relevant symmetric energy operator,
$\ve_1+\ve_2+\ve_3$. However, $\Delta_{\ve_1+\ve_2+\ve_3}$ has only
been determined to order ${\cal O}(\epsilon^3)$
[see Eq.~(\ref{symm-energy})], and so the discrepancy between the
values of Table~\ref{tab:xt-3p} and of Table~\ref{tablecrit} for large
$q$ should hardly come as a surprise. On the other hand,
for $q$ slightly larger than two numerics is supposed to be hampered
by large logarithmic corrections due to the near-marginality of the
energy operator \cite{Cardy-log}. In this range we thus expect the
perturbative results to be the more reliable.

\begin{table}
\begin{center}
\begin{tabular}{|c||c|c||c|c|} \hline
 $q$  & $c(3,5)$ & $c(4,6)$ & $x_t(4,5)$ & $x_t(5,6)$ \\ \hline
 2.00 & 1.5040   & 1.4906   & 1.2692     & 1.2600     \\ \hline
 2.25 & 1.7653   & 1.7490   & 1.2798     & 1.2691     \\ \hline
 2.50 & 1.9965   & 1.9775   & 1.2890     & 1.2774     \\ \hline
 2.75 & 2.2022   & 2.1801   & 1.2964     & 1.2842     \\ \hline
 3.00 & 2.3856   & 2.3601   & 1.3018     & 1.2890     \\ \hline
 3.25 & 2.5495   & 2.5198   & 1.3051     & 1.2913     \\ \hline
 3.50 & 2.6962   & 2.6615   & 1.3062     & 1.2907     \\ \hline
 3.75 & 2.8274   & 2.7869   & 1.3049     & 1.2868     \\ \hline
 4.00 & 2.9449   & 2.8975   & 1.3011     & 1.2796     \\ \hline
\end{tabular}
\end{center}
\caption{\label{tab:alg3-3p}Results for three models at $a\to\infty$
using {\tt alg3}.} 
\end{table}

In Table~\ref{tab:alg3-3p} we show some more results for $c$ and
$x_t$, but this time obtained from {\tt alg3}. The usefulness of these
data is twofold. First, they clearly demonstrate the
superiority of {\tt alg4} as regards the accuracy and the convergence
properties (monotonicity) of the finite-size estimates. Second, the
agreement with Table~\ref{tab:c-3p}--\ref{tab:xt-3p} serves as a check
of the modular invariance of the $a=\infty$ critical point, as
explained in Sec.~\ref{sec:alg4}.

\begin{table}
\begin{center}
\begin{tabular}{|c||c|c|c|c|c|} \hline
 $q$  &$x^{(1)}_H(4,6)$&$x^{(1)}_H(6,8)$&$x^{(1)}_H(8,10)$ \\ \hline
 2.00 &       0.124815 &       0.125054 &        0.125112  \\ \hline
 2.25 &       0.127712 &       0.127847 &        0.127851  \\ \hline
 2.50 &       0.129955 &       0.129915 &        0.129810  \\ \hline
 2.75 &       0.131635 &       0.131334 &        0.131050  \\ \hline
 3.00 &       0.132825 &       0.132170 &        0.131623  \\ \hline
 3.25 &       0.133588 &       0.132480 &        0.131577  \\ \hline
 3.50 &       0.133977 &       0.132318 &        0.130962  \\ \hline
 3.75 &       0.134040 &       0.131734 &        0.129831  \\ \hline
 4.00 &       0.133816 &       0.130778 &        0.128237  \\ \hline
\end{tabular}
\end{center}
\caption{\label{tab:xh1-3p}$x_H^{(1)}$ for three coupled Potts models
at $a=\infty$.}
\end{table}

We now turn our attention to the magnetic exponents $x_H^{(N_{\rm mag})}$
obtained by imposing twisted boundary conditions on $N_{\rm mag}=1,2$ or
3 of the $N=3$ coupled Potts models. The numerical results are
presented in Table~\ref{tab:xh1-3p}--\ref{tab:xh3-3p}. Comparing with
the analytic results of Table~\ref{tablecrit}--\ref{tablemom} we find,
as expected, that these scaling exponents must be those of the first
three moments of the magnetisation. Excluding the case of $q=2$ where
the remarks made above hold true, we find a very good agreement
between $x_H^{(1)}$ and $\Delta_{\sigma_1}$, and between
$x_H^{(2)}$ and $\Delta_{\sigma_1 \sigma_2}$ in the range $q \le 3$.
For larger $q$ the perturbative RG breaks down with a vengeance, as
witnessed by its predictions of negative scaling dimension.

Finally, the agreement between $x_H^{(3)}$ and
$\Delta_{\sigma_1 \sigma_2 \sigma_3}$, especially in the range $2.5<q<3.5$,
is quite striking, and is the first validation of the second order
perturbative computation proposed previously by one of us \cite{MAL}.

\begin{table}
\begin{center}
\begin{tabular}{|c||c|c|c|c|c|} \hline
 $q$  &$x^{(2)}_H(4,6)$&$x^{(2)}_H(6,8)$&$x^{(2)}_H(8,10)$ \\ \hline
 2.00 &       0.277424 &       0.276866 &                  \\ \hline
 2.25 &       0.287411 &       0.287143 &         0.286780 \\ \hline
 2.50 &       0.296026 &       0.295985 &         0.295839 \\ \hline
 2.75 &       0.303408 &       0.303470 &         0.303445 \\ \hline
 3.00 &       0.309671 &       0.309661 &         0.309600 \\ \hline
 3.25 &       0.314915 &       0.314615 &         0.314310 \\ \hline
 3.50 &       0.319226 &       0.318393 &         0.317599 \\ \hline
 3.75 &       0.322685 &       0.321061 &         0.319512 \\ \hline
 4.00 &       0.325367 &       0.322693 &         0.320119 \\ \hline
\end{tabular}
\end{center}
\caption{\label{tab:xh2-3p}$x_H^{(2)}$ for three coupled Potts models
at $a=\infty$.}
\end{table}

\begin{table}
\begin{center}
\begin{tabular}{|c||c|c|c|c|c|} \hline
 $q$  &$x^{(3)}_H(4,6)$&$x^{(3)}_H(6,8)$&$x^{(3)}_H(8,10)$ \\ \hline
 2.00 &       0.475424 &       0.471563 &                  \\ \hline
 2.25 &       0.503049 &       0.500930 &                  \\ \hline
 2.50 &       0.529688 &       0.529874 &                  \\ \hline
 2.75 &       0.555511 &       0.558541 &                  \\ \hline
 3.00 &       0.580638 &       0.587025 &         0.592223 \\ \hline
 3.25 &       0.605162 &       0.615388 &                  \\ \hline
 3.50 &       0.629150 &       0.643668 &                  \\ \hline
 3.75 &       0.652656 &       0.671885 &                  \\ \hline
 4.00 &       0.675719 &       0.700047 &         0.721371 \\ \hline
\end{tabular}
\end{center}
\caption{\label{tab:xh3-3p}$x_H^{(3)}$ for three coupled Potts models
at $a=\infty$.}
\end{table}

\subsubsection{Higher exponents in the even sector for $N=3$, $q=3$}

In an attempt to numerically identify more of the operators predicted
by perturbation theory, we have looked at the scaling dimensions
extracted from the finite-size scaling of higher gaps in the even
sector of the transfer matrix. These computations are extremely
time-consuming, since, in order to study the first $k$ gaps, we need
to iterate and orthogonalise $k+1$ vectors, {\em cfr.}~Eq.~(\ref{iterate}).
We have therefore focused exclusively on the case of three coupled
3-state Potts models.

The primary operators we expect to find in the even sector must all be
energetic, since the $Z_q$ symmetry associated with the permutation of
the Potts spin labels has been factored out. Furthermore, since
{\tt alg4} by construction treats all three layers symmetrically, we
only expect to find such operators that are symmetric under any
permutation of the layer indices. According to Sec.~\ref{sec:RG} the
permissible operators are thus
\beq
 I, \ \ \ \
 \ve_1+\ve_2+\ve_3, \ \ \ \
 \ve_1 \ve_2 + \ve_2 \ve_3 + \ve_3 \ve_1, \ \ \ \
 \ve_1 \ve_2 \ve_3.
 \label{primaries}
\eeq
If the (presently unknown) conformal field theory of the system can
be assumed to have the usual structure \cite{BPZ} we can, in addition
to these primaries, in general expect to find descendent operators
with their appropriate degeneracies, reflecting the 
null vector structure of the Verma module. The identity operator $I$
can be thought of as the primary corresponding to the zeroth gap, thus
having scaling dimension $\Delta_I = 0$ by definition.

\begin{table}
\begin{center}
\begin{tabular}{|c||c|c|c|c|c||c|c|} \hline
Gap & $x(4)$ & $x(6)$ & $x(8)$ & $x(10)$ & $x(12)$ &
      Extrapolation   & Operator \\ \hline \hline
 1  & 1.694  & 1.471  & 1.385  & 1.344   & 1.322 & 1.27 &
      $\ve_1+\ve_2+\ve_3$ \\ \hline
 2  & 2.603  & 2.380  & 2.272  & 2.219   & 2.148 & $\approx 2.0$
    & $T = L_{-1}I$ \\ \hline
 3  & 2.922  & 2.775  & 2.398  & 2.225   & 2.148 & $\approx 2.0$
    & $\overline{T} = \overline{L}_{-1}I$ \\ \hline
 4  & ---    & 3.104  & 2.398  & 2.225   & 2.174 & $\approx 2.1$
    & $\ve_1 \ve_2 + \ve_2 \ve_3 + \ve_3 \ve_1$ \\ \hline
 5  & ---    & 3.104  & 2.685  & 2.633   & 2.599 & $\approx 2.3$
    & $L_{-1}(\ve_1 + \ve_2 + \ve_3)$ \\ \hline
 6  & ---    & 3.626  & 3.279  & 2.785   & 2.598 & $\approx 2.3$
    & $\overline{L}_{-1}(\ve_1 + \ve_2 + \ve_3)$ \\ \hline
 7  & ---    & 3.991  & 3.279  & 2.785   & 2.598 & $\approx 2.4$
    & $\ve_1 \ve_2 \ve_3$ \\ \hline
 8  & ---    & 3.991  & 3.289  & 3.146   & 3.066 & $\approx 3.0$
    & $T' = L_{-3}I$ \\ \hline
\end{tabular}
\end{center}
\caption{\label{tab:gaps}Scaling dimensions extracted from the higher
gaps in the even sector as well as their identification with physical
operators.}
\end{table}

In Table~\ref{tab:gaps} we show the results for the first eight gaps,
obtained by employing {\tt alg4}. Since we do
not wish to make any {\em a priori} assumptions on the existence of
the stress tensor we report only the unadorned one-point fits
(\ref{fx}), for strip widths up to $L=12$. For $L=4$ we were only able
to iterate four orthogonal vectors, and consequently there are some
empty entries in the table.

The extrapolation to the $L\to\infty$ limit becomes increasingly
important for the higher gaps, but unfortunately also increasingly
difficult, since the convergence is slower and there are fewer data
points. The reported values of the extrapolated scaling dimensions are
therefore only indicative.

Let us recall from Sec.~\ref{sec:RG} that perturbation theory predicts
\beq
  \Delta_{\ve_1+\ve_2+\ve_3} \simeq 1.44, \ \ \ \
  \Delta_{\ve_1 \ve_2 + \ve_2 \ve_3 + \ve_3 \ve_1} \simeq 2.08, \ \ \ \
  \Delta_{\ve_1 \ve_2 \ve_3} \simeq 2.28.
\eeq
These values cannot be believed to be very precise, since at $q=3$ we
expect the perturbative expansion to be at the verge of breaking
down. Nevertheless, they are supposed to be of the right order of
magnitude, and we thus venture to make the identification shown in
Table~\ref{tab:gaps}.

Note that for $L \ge 8$ the 3rd and the 4th gap, and also the 6th and
the 7th, are exactly degenerate with full 16-digit machine
precision. One would therefore expect them to be connected by a simple
symmetry, as reflected by our conjectured identification.
However, for $L=12$ a crossover takes place so that henceforth it is
the second and the third gap that are degenerate rather than the third
and the fourth. This could have been anticipated by noticing that for
$L<12$ gaps 2--4 have quite different finite-size dependence. A
similar crossover is expected to take place amongst gaps 5--7 for
$L=14$. Accordingly our operator identifications pertain to the
expected arrangement of the gaps in the $L\to\infty$ limit.

Apart from the primaries (\ref{primaries}) we are able to identify the
holomorphic and antiholomorphic components of the stress tensor, both
with scaling dimension two. This provides further evidence for the
criticality of the system, and is completely independent of the
methods used this far. In addition we observe the level-one descendents
of the operator $\ve_1+\ve_2+\ve_3$ with the expected scaling
dimension.

Unfortunately we have not been able to iterate enough vectors to see
descendents at level two. This is a pity, since the degeneracy of
these scaling dimensions would enable us to make predictions about the
null vector structure of the Verma module.

\subsubsection{Monte Carlo Results}

Showing the existence of scaling operators, whose correlators obey scaling
laws, would confirm that the $a\to\infty$ limit point on the self-dual line
for three coupled models can be identified with a second order phase
transition. To do so, we performed Monte Carlo (MC) simulations with the
simplified Boltzmann weights (\ref{esd3r}) pertaining to this point. A
direct measure of correlators for the physically significant operators
({\em i.e.} those which are diagonal in the perturbation scheme)
clearly shows that the 
model is indeed critical, exhibiting scaling laws with associated exponents
sufficiently close to those obtained either by perturbative CFT or transfer
matrix techniques.  Since the critical exponents can be measured with much
more accuracy using transfer matrices techniques, MC was used solely to
establish the criticality of the fixed point model. 

We chose to measure explicitly the correlation functions for the spin
and energy operators. Whilst the definition of energy correlators is
straightforward, the $Z_q$ symmetry of spin variables has to be taken
into account, and one should rather consider the operators
\beq
 \Sigma_i \equiv \sigma_i - \overline{\sigma_i} \qquad \mbox{mod } q,
\eeq
where $\overline{\sigma_i}$ is the instantaneous average value of the
spin $\sigma$ on lattice $i$. The physically meaningful operators,
for which we computed correlators, are thus the following: 
\bea
 \ve_S     &\equiv& \ve_1+\ve_2+\ve_3 \\
 \ve_A     &\equiv& \ve_1-\ve_2\\
 \Sigma_S  &\equiv& \Sigma_1+\Sigma_2+\Sigma_3 \\
 \Sigma_A  &\equiv& \Sigma_1-\Sigma_2.
\eea
These operators have well-behaved distributions and are the ones
obeying scaling laws. We performed simulations on a $160 \times 160$
lattice with periodic boundary conditions in both directions. Sixteen
runs of 10,000 sweeps each were made, with a
preliminary thermalisation stage of 500 times the autocorrelation time 
(which is around 25 updates for our lattice size). We also checked that
the values of critical exponents were rather close to those predicted using
this lattice size.  A larger lattice would evidently provide a better
description of critical behaviour, but computation time grows rapidly with
the number of spins and, as we said, a precise measurement of critical
exponents is not what was sought here.  
  
Simulations were performed using a modified version of the Wolff one-cluster
algorithm \cite{wolff}. Let us consider the model given by 
Eq.~(\ref{dua3}), which we recall: 
\bea
 {\cal H}_{ij} &=& a\,(\delta_{\si_i,\si_j} +\delta_{\tau_i,\tau_j}
 +\delta_{\eta_i,\eta_j} ) 
 + b\,(\delta_{\si_i,\si_j}\delta_{\tau_i,\tau_j}
 +\delta_{\si_i,\si_j}\delta_{\eta_i,\eta_j}
 +\delta_{\tau_i,\tau_j}\delta_{\eta_i,\eta_j}) \nn\\
 &+& c\,\,\delta_{\si_i,\si_j} \delta_{\tau_i,\tau_j}
 \delta_{\eta_i,\eta_j}.
\eea
We first choose randomly one of the three models, say the one with the
indices $\sigma$ (the model is invariant under permutation of any pair of
layers). Then, we consider the model defined by 
\bea
 {\cal H}_{ij} &=& (a+b\,(\delta_{\tau_i,\tau_j}+\delta_{\eta_i,\eta_j})
 +c\,\,\delta_{\tau_i,\tau_j} \delta_{\eta_i,\eta_j}) \delta_{\si_i,\si_j}
 + {\rm const.} \\
 &=& a' \delta_{\si_i,\si_j} + {\rm const.}
\eea
Next we update the lattice with the usual Wolff algorithm but with the local
$a'=a+b\,(\delta_{\tau_i,\tau_j}+\delta_{\eta_i,\eta_j}) 
+c\,\,\delta_{\tau_i,\tau_j} \delta_{\eta_i,\eta_j}$.  Repeated three
times, this operation defines one update of the system.   

The measurement procedure is quite straightforward: For a given operator
${\cal O}$, all correlation functions $G(|x-y|) \equiv \langle {\cal
O}(x){\cal O}(y)\rangle$, which are manifestly translationally
invariant due to the choice of boundary conditions, are computed for
$0<|x-y|<80$.  Compiling results, one finds that the region where
scaling laws exist is limited to $|x-y|<20$ for all operators, except
for the symmetric energy, whose scaling law behaviour is observed only
for $|x-y|<8$, this being due to the fact that the critical exponent for
this operator is much larger than for the others.  This is quite deceiving,
but not unexpected.  It is clear that for critical correlations to have
some sense, they must be between points $x$ and $y$ such that $|x-y|\ll L$.
This, combined with the fact that our statistics were rather low, explains
why we had to restrict our attention to such narrow regions.  Doing so, one 
observes nice scaling properties with associated exponents close to the
ones computed by other analytical and numerical methods. 
Figures \ref{MCE} and \ref{MCS} present the
Log-Log plot of correlation functions along with statistical error bars.
Linear fits of these graphs give $2\Delta_{\cal O}$, and thus leads to the
following values for the critical exponents: 
\bea
 \Delta_{\ve_S} &\equiv&  1.27 \pm 0.13 \\
 \Delta_{\ve_A} &\equiv&  0.63 \pm 0.04 \\
 \Delta_{\Sigma_S} &\equiv&  0.13 \pm 0.03\\
 \Delta_{\Sigma_A} &\equiv&  0.10 \pm 0.03.
\eea
Altough not precise, the MC data clearly establish the existence of
scaling operators.  This confirms that the model under study is critical. 

\begin{figure}
\begin{center}
\leavevmode
\epsfysize=200pt{\epsffile{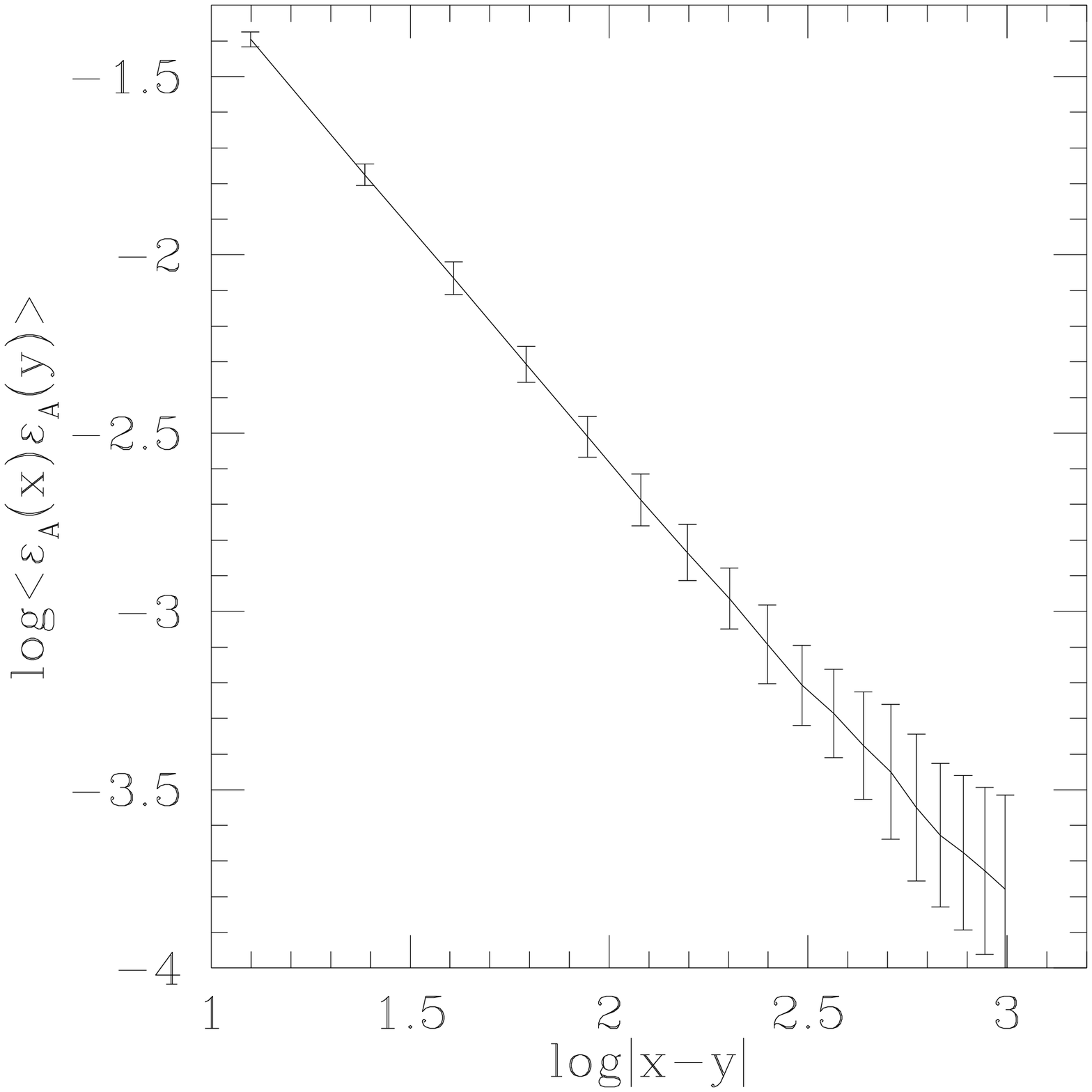}}
\epsfysize=200pt{\epsffile{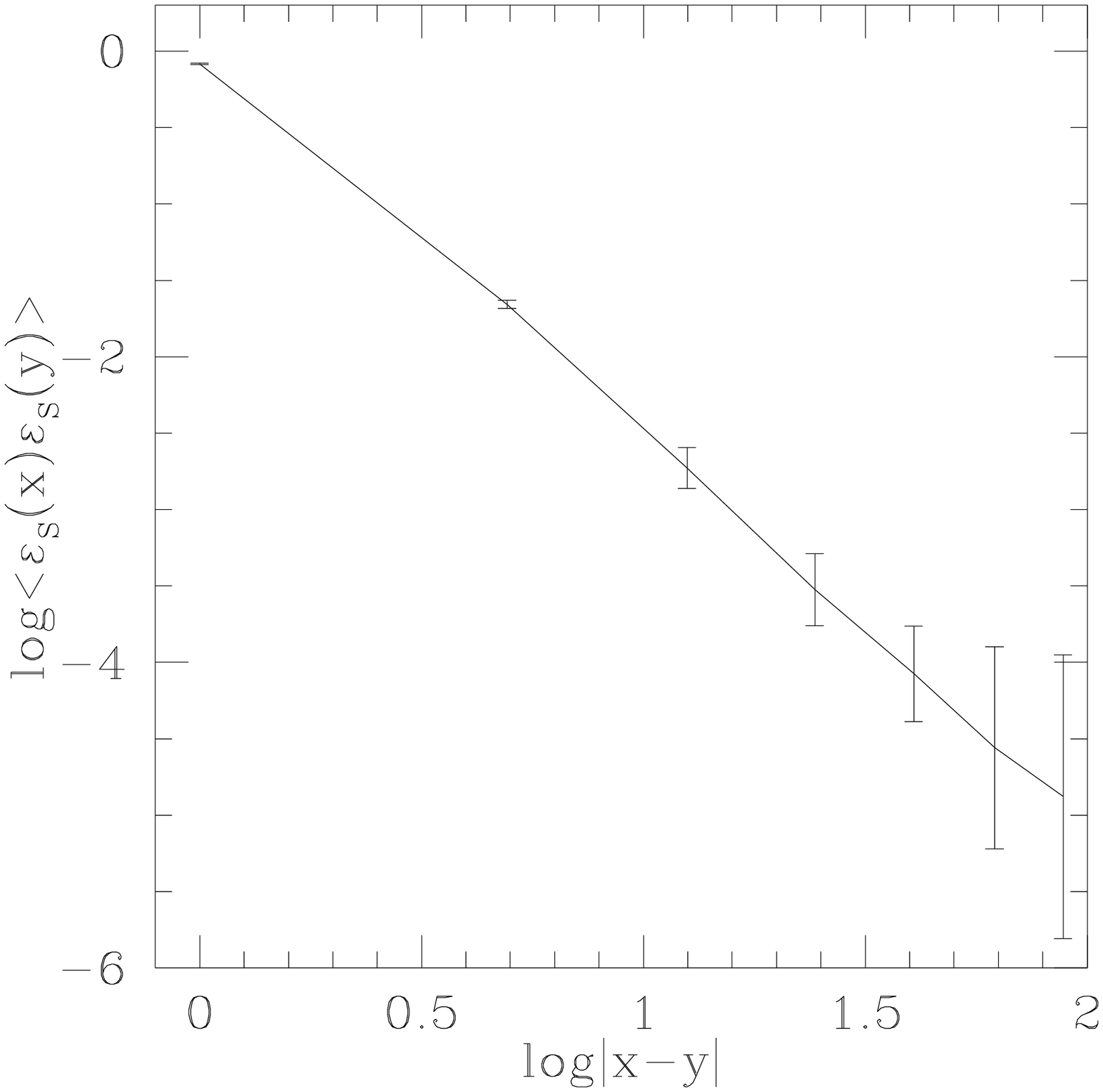}}
\end{center}
\protect\caption{(a) Monte Carlo results for $\log \langle
 \ve_A(x)\ve_A(y) \rangle$ as a function of $\log |x-y|$. (b) $\log
 \langle \ve_S(x)\ve_S(y) \rangle$ as a function of $\log |x-y|$.
\label{MCE}}  
\end{figure}

\begin{figure}
\begin{center}
\leavevmode
\epsfysize=200pt{\epsffile{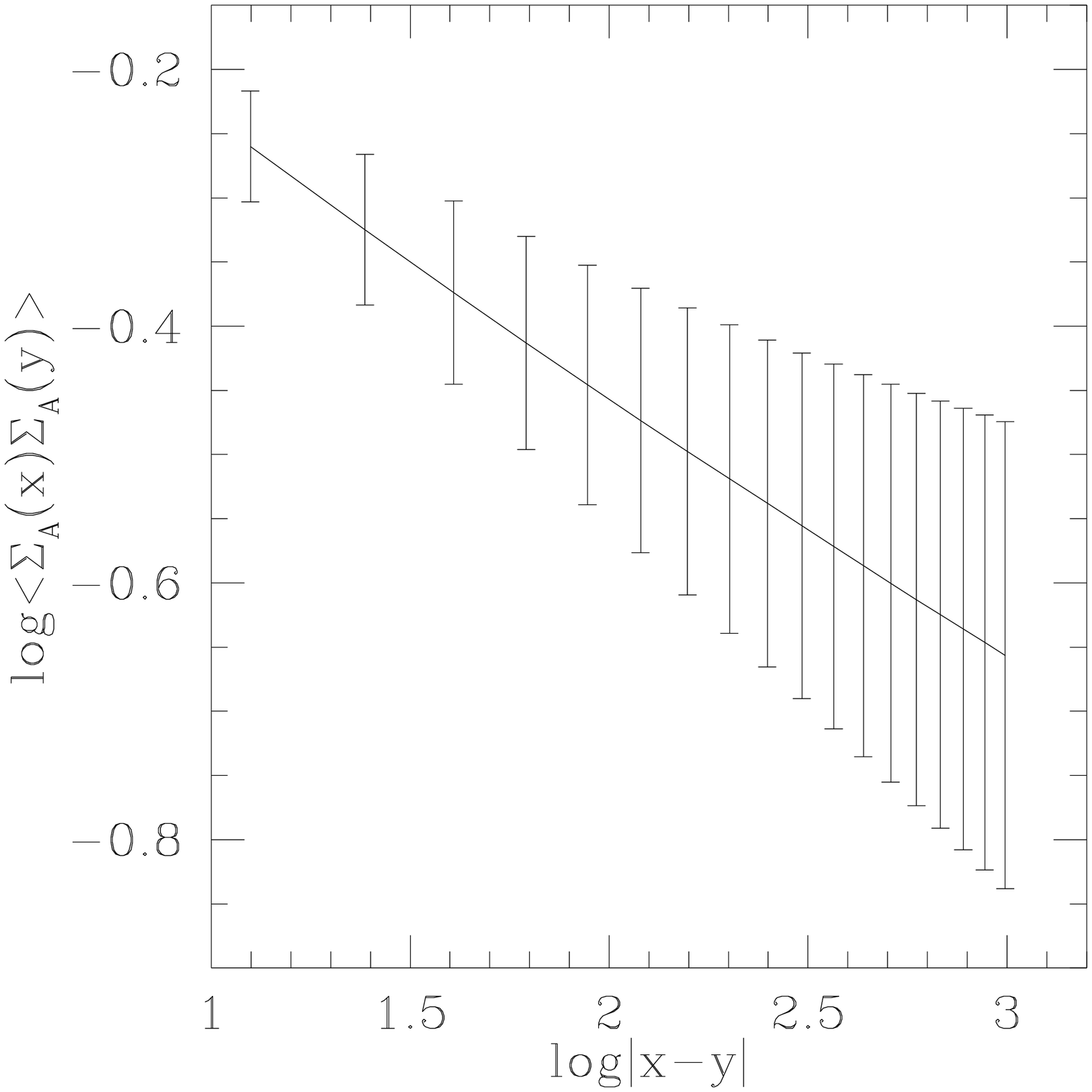}}
\epsfysize=200pt{\epsffile{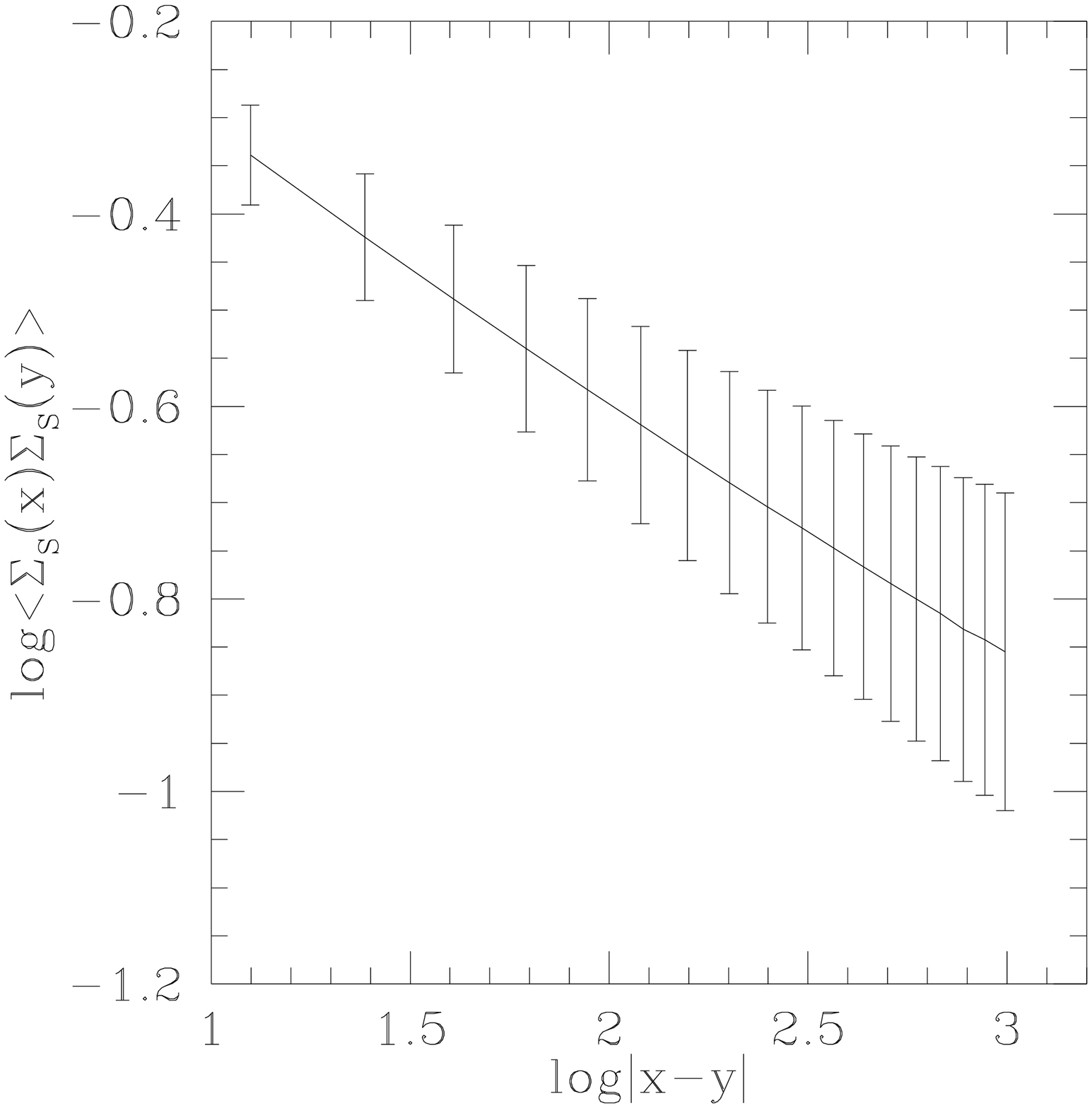}}
\end{center}
\protect\caption{(a) Monte Carlo results for $\log \langle
\Sigma_A(x)\Sigma_A(y) \rangle$ as a function of $\log |x-y|$.
(b) $\log \langle \Sigma_S(x)\Sigma_S(y) \rangle$ as a function of $\log
|x-y|$. \label{MCS}}   
\end{figure}

\section{Concluding remarks}
\label{sec:discussion}

Besides the long-reaching goal of describing exactly disordered
systems, the extensive study presented here is, we hope, interesting
in many aspects.  First, it introduces a new variant of the Potts
model transfer matrix method which proves to be the most accurate and
efficient up to now. Second, we have identified non-trivial fixed points both
numerically and analytically and have presented evidence that these
two types of fixed points are indeed the same and correspond to
non-trivial critical models. The universality classes of these
critical models are new, and several critical exponents were computed
for the first time.

The next step in our long-term project is to solve the critical model,
which was proven to be the end-point of the self-dual line.  At this point, as we have shown, the Hamiltonian simplifies greatly, and it might be
possible to map the system to a loop model, whose continuum limit is a
Liouville field theory~\cite{jj_npb98}. We believe that the results
presented here will be of the outmost importance when searching for
such a loop formulation.

\noindent{\large\bf Acknowledgments}

We would like to thank P.~Pujol for valuable
discussions. M.-A.L. acknowledges financial support from
NSERC Canada.

\end{document}